\documentclass[11pt]{article}
\usepackage{tikz}
\usetikzlibrary{patterns}
\usepackage[normalem]{ulem}
\usepackage{subfigure}
\usepackage{graphicx}  
\usepackage{hyperref}
\usepackage{epsfig,subfigure}  
\usepackage{color,soul}
\usepackage{enumerate}   
 
\usepackage{xcolor,cancel}
 
\usepackage{epsfig,subfigure}
\usepackage[cmex10,fleqn]{amsmath}
\usepackage{graphicx}
\usepackage{color}
\usepackage{amssymb}
\usepackage{amsfonts}
\usepackage{hyperref}
\usepackage{verbatim}
\usepackage{stfloats}
\usepackage[bookmarks=false]{}

\makeatletter
\newcommand{\crossout}[1]{%
  \begingroup
  \settowidth{\dimen@}{#1}%
  \setlength{\unitlength}{0.05\dimen@}%
  \settoheight{\dimen@}{#1}%
  \count@=\dimen@
  \divide\count@ by \unitlength
  \count0=20 \count4=\count@
  \loop
  \count2=\count0 % keep a copy
  \divide\count2\count4 \multiply\count2\count4
  \ifnum\count2<\count0
    \advance\count0 -\count2 % the remainder
    \count2=\count0
    \count0=\count4
    \count4=\count2
  \repeat
  \count0=20 \divide\count0\count4
  \count2=\count@ \divide\count2\count4
  \begin{picture}(0,0)
  \put(0,0){\line(\count0,\count2){20}}
  \put(0,\count@){\line(\count0,-\count2){20}}
  \end{picture}%
  #1%
  \endgroup
}
\makeatother

\newtheorem{theorem}{Theorem}

\newtheorem{definition}{Definition}
\newtheorem{lemma}{Lemma}

\newtheorem{remark}{Remark}

\begin{document}
\date{}
\title{Degrees of Freedom Region of the $(M,N_1,N_2)$ MIMO Broadcast Channel with  Partial CSIT: An Application of Sum-set Inequalities Based on Aligned Image Sets}
\author{ \normalsize Arash Gholami Davoodi and Syed A. Jafar \\
{\small Center for Pervasive Communications and Computing (CPCC)}\\
{\small University of California Irvine, Irvine, CA 92697}\\
{\small \it Email: \{gholamid, syed\}@uci.edu}
}
\maketitle

\begin{abstract}
The degrees of freedom (DoF) region is characterized for the  $2$-user multiple input multiple output (MIMO) broadcast channel (BC), where the transmitter is  equipped with $M$ antennas, the two receivers are equipped with $N_1$ and $N_2$  antennas, and the levels of channel state information at the transmitter (CSIT) for the two users are parameterized by $\beta_1, \beta_2$, respectively. The achievability of the DoF region was established by Hao, Rassouli and Clerckx, but no proof of optimality was heretofore available. The proof of optimality is provided in this work with the aid of  sum-set inequalities based on the aligned image sets (AIS) approach.
\end{abstract}

\section{Introduction}
The availability of channel state information at the transmitter(s) (CSIT) greatly affects the capacity of wireless networks, so much so that even the coarse degrees of freedom (DoF) metric is significantly impacted. Under perfect CSIT a  $K$-user interference channel has $K/2$ DoF  \cite{Cadambe_Jafar_int} and the corresponding $K$-user MISO BC has $K$ DoF almost surely  \cite{Caire_Shamai}. However, if CSIT is limited to finite precision then the DoF collapse to unity in both cases. The large gap between the two extremes underscores the  importance of studying partial CSIT settings. A key obstacle for these studies tends to be the proof of optimality once an achievable DoF region has been established based on the best known achievable schemes. For instance, the conjecture by Lapidoth, Shamai and Wigger  \cite{Lapidoth_Shamai_Wigger_BC}, that the DoF collapse under finite precision CSIT, remained open for nearly a decade, until it was finally settled using an unconventional (combinatorial) argument, called the aligned image sets (AIS) approach in \cite{Arash_Jafar_PN}. The AIS approach seeks to directly bound the number of codewords that can be resolved at one receiver while aligning at another receiver, under arbitrary levels of CSIT. Since its introduction in \cite{Arash_Jafar_PN},  the AIS approach has been successfully applied to construct proofs of optimality for a number of basic broadcast and interference channel settings. With each application the AIS approach has been further generalized, broadening its utility and scope. The unconventional nature of the AIS approach, in particular its reliance on combinatorial reasoning from first principles to bound the sizes of the aligned image sets, makes these generalizations quite challenging. Particularly relevant to this work is the recent effort in \cite{Arash_Jafar_sumset} to derive a new class of sumset inequalities based on the AIS approach, to serve as a toolkit for future DoF studies. In this work we demonstrate the utility of these sumset inequalities by providing the proof of optimality for a DoF region for the $2$-user MIMO BC under partial CSIT, that was shown to be achievable by by Hao, Rassouli and Clerckx in \cite{Hao_Rassouli_Clerckx}, but whose optimality was heretofore  open.

The setting of interest is a $2$-user MIMO BC where the transmitter is equipped with $M$ antennas,  the two receivers are equipped with $N_1$ and $N_2$  antennas, and the levels of CSIT for the two users are parameterized by $\beta_1, \beta_2 \in[0,1]$, respectively, such that $\beta_i=0$ represents no CSIT, $\beta_i=1$ represents perfect CSIT, and the intermediate values represent corresponding levels of partial CSIT. Existing results for this channel focus primarily on the two extremes of perfect CSIT and no CSIT. Exact capacity is known for the MIMO BC if the CSIT is perfect \cite{Weingarten_Steinberg_Shamai}. The collapse of DoF under no CSIT has been shown for this channel under restrictive assumptions such as isotropic fading that essentially appeal to the degraded BC perspective  \cite{Caire_Shamai, Jafar_scalar, Huang_Jafar_Shamai_Vishwanath, Vaze_Varanasi_MIMOBC}. The particular setting of  the MISO BC, where each user is equipped with only one antenna, i.e., $N_1=N_2=1$, has recently seen much progress based on the AIS approach in \cite{Arash_Jafar_PN, Arash_Jafar_TC, Arash_Bofeng_Jafar_BC}, leading ultimately to its full GDoF characterization in \cite{Arash_Bofeng_Jafar_BC} with arbitrary channel strengths and arbitrary channel uncertainty levels. For arbitrary antenna configurations and arbitrary levels of partial CSIT, an achievable DoF region is established by Hao, Rassouli and Clerckx in \cite{Hao_Rassouli_Clerckx}. The optimality of this achievable region has  been shown in \cite{Hao_Rassouli_Clerckx} for certain parameter regimes (mainly $N_1\leq N_2, M\leq N_2$), based on existing bounds, as well as AIS arguments. However, the general DoF region characterization remains open. Our main goal in this work is to provide a complete  DoF region characterization by providing the  proof of optimality that was heretofore missing for the remaining parameter regime. Remarkably, the proof makes  use of the sumset inequalities recently developed in \cite{Arash_Jafar_sumset}.

\section{Notation and Definitions}
For $n\in\mathbb{N}$, define the notation $[n]=\{1,2,\cdots,n\}$. The cardinality of a set $A$ is denoted as $|A|$. The notation~ $X^{[n]}$ stands~ for $\{X(1), X(2), \cdots, X(n)\}$. Moreover,  $X_{i}^{[n]}$ also stands for $\{X_i(t): \forall t\in[n]\}$. The
support of a random variable $X$ is denoted as supp$(X)$. The sets $\mathbb{R}$, $\mathbb{Q}$, $\mathbb{R}^n$ and $\mathbb{Q}^n$ stand for the sets of real numbers, rational numbers, all $n$-tuples of real numbers and all $n$-tuples of rational numbers, respectively. Moreover, the set $\mathbb{R}^{2+}$ is defined as the set of all pairs of non-negative numbers.  For any set $S$, we define the set $S^c$ as the complement of the set $S$. If $A$ is a set of random variables, then $H(A)$ refers to the joint entropy of the random variables in $A$. Conditional entropies, mutual information and joint and conditional probability densities of sets of random variables are similarly interpreted. Moreover, we use the Landau $O(\cdot)$ and $o(\cdot)$ notations as follows. For  functions $f(x), g(x)$ from $\mathbb{R}$ to $\mathbb{R}$, $f(x)=O(g(x))$ denotes that $\limsup_{x\rightarrow\infty}\frac{|f(x)|}{|g(x)|}<\infty$.  $f(x)=o(g(x))$ denotes that $\limsup_{x\rightarrow\infty}\frac{|f(x)|}{|g(x)|}=0$. We use the notation $A\doteq B$ to indicate that the difference $|A-B|$ is negligible in the DoF sense. 
We use $\mathbb{P}(\cdot)$ to denote the probability function $\mbox{Prob}(\cdot)$. For any real number $x$ we define $\lfloor x\rfloor$ as the largest integer that is smaller than or equal to $x$ when $x>0$,  the smallest integer that is larger than or equal to $x$ when $x<0$, and $x$ itself when $x$ is an integer.  The number $X_{r,s}$ may be represented as $X_{rs}$ if there is no cause of ambiguity. For any vector $V=\begin{bmatrix}v_1&\cdots&v_k\end{bmatrix}^T$ and non-negative integer numbers $m$ and $n$ less than $k$, let us define the  notation $V_{m\rightarrow n}$ as follows,
\begin{eqnarray}
V_{m\rightarrow n}&\triangleq&\left\{
\begin{array}{ll}
\begin{bmatrix}v_{m+1}&\cdots&v_{m+n}\end{bmatrix}^T,& m+n\le k\\
\begin{bmatrix}v_{m+1}&\cdots&v_k&v_1&\cdots&v_{m+n-k}\end{bmatrix}^T,& k<m+n
\end{array}
\right.
\end{eqnarray}
For any two vectors $V=\begin{bmatrix}v_1&\cdots&v_{k_1}\end{bmatrix}^T$ and $W=\begin{bmatrix}w_1&\cdots&w_{k_2}\end{bmatrix}^T$ define their concatenation as
$$V\bigtriangledown W\triangleq \begin{bmatrix}v_1&\cdots&v_{k_1}&w_1&\cdots&w_{k_2}\end{bmatrix}^T$$
Finally, for any $m\times n$ matrix $V=\left[\begin{matrix}
v_{11}&\cdots&v_{1n}\\
\ddots&\vdots&\ddots\\
v_{m1}&\cdots&v_{mn}\\
\end{matrix}\right]
$ and $a,b,c,d\in\mathbb{N}$ where $a+b\le m$ and $c+ d\le n$, define 
\begin{eqnarray}
V_{(a\rightarrow b):(c\rightarrow d)}&\triangleq&\left[\begin{matrix}
v_{a+1,c+1}&\cdots&v_{a+1,c+d}\\
\ddots&\vdots&\ddots\\
v_{a+b,c+1}&\cdots&v_{a+b,c+d}\\
\end{matrix}\right]\end{eqnarray}

%Motivated by \cite{Arash_Bofeng_Jafar_BC}, we characterize the sum GDoF of symmetric MIMO IC for general values of $\beta$, $\alpha$, $M$ and $N$. From the outer bound perspective, we first introduce a Lemma where we define two multi-letter random variables as finite precision linear combination of arbitrary numbers of dependent random variables with different power levels. Difference of the entropies of  these two multi-letter random variables are bounded with summation of difference of the levels. This Lemma and Fano's inequality together concludes the outer bound.  On the other hand, the achievability is based on the zero-forcing and interference management (power level management) where treating interference as noise together with multiple access decoding achieves optimal GDoF. Interestingly, profiting from both approaches simultaneous seems to be infeasible for some positive values of $\beta$.

 The following definitions, inherited from \cite{Arash_Jafar_sumset}, are replicated here for the sake of completeness.

\begin{definition}[Power Levels] Consider integer valued  variables $X_i$ over alphabet $\mathcal{X}_{\lambda_i}$,
\begin{eqnarray}
\mathcal{X}_{\lambda_i}&\triangleq&\{0,1,2,\cdots,\bar{P}^{\lambda_i}-1\}
\end{eqnarray}
where $\bar{P}^{\lambda_i}$ is a compact notation for $\left\lfloor\sqrt{P^{\lambda_i}}\right\rfloor$. We refer to $P\in\mathbb{R}_+$ as \emph{power}, and are primarily interested in limits as  $P\rightarrow\infty$. Quantities that do not depend on $P$ will be referred to as constants. The constant $\lambda_i\in\mathbb{R}_+$ denotes the \emph{power level} of $X_i$. 
\end{definition}

\begin {definition}\label{powerlevel} For  non-negative real numbers $X$,  $\lambda_1$ and $\lambda_2$, define  $(X)_{\lambda_1}$ and $(X)^{\lambda_2}_{\lambda_1}$  as,
 \begin{eqnarray}
(X)_{\lambda_1}&\triangleq& X-\bar{P}^{\lambda_1} \left \lfloor \frac{X}{\bar{P}^{\lambda_1}} \right \rfloor\\
(X)^{\lambda_2}_{\lambda_1}&\triangleq&\left \lfloor \frac{X-\bar{P}^{\lambda_2}\left \lfloor\frac{X}{ {\bar{P}}^{\lambda_2}}\right \rfloor }{{\bar{P}}^{\lambda_1}} \right \rfloor\label{mid}
\end{eqnarray}
\end {definition}

\begin{figure}[h]
\begin{center}
\begin{tikzpicture}
\draw [black, thick] (1,0) rectangle (2,6) node[midway] {$X$};
\draw[thick, <->] (2.18,0)--(2.18,2.5) node[midway, right]{$\lambda_1$};
\draw[thick, <->] (2.18,2.5)--(2.18,4) node[midway, right]{$\lambda_2$};
\draw[thick, <->] (2.18,4)--(2.18,6) node[midway, right]{$\lambda_3$};
\draw [help lines] (2,0)--(3,0);
\draw [help lines] (2,2.5)--(5,2.5);
\draw [help lines] (2,4)--(7,4);
\draw [help lines] (2,6)--(7,6);
\draw [thick](3,0)  -| (4,2.5) 
    node[pos=0.75,right] {$(X)_{\lambda_1}$}
    -| (3,0);

\draw [thick](5,2.5)  -| (6,4) 
    node[pos=0.75,right] {$(X)_{\lambda_1}^{\lambda_1+\lambda_2}$}
    -| (5,2.5);

\draw [thick](7,4)  -| (8,6) 
    node[pos=0.75,right] {$(X)_{\lambda_1+\lambda_2}^{\lambda_1+\lambda_2+\lambda_3}$}
    -| (7,4);
\end{tikzpicture}
\caption[]{Conceptual depiction of an arbitrary variable $X\in\mathcal{X}_{\lambda_1+\lambda_2+\lambda_3}$, and its power-level partitions $(X)_{\lambda_1}$, $(X)_{\lambda_1}^{\lambda_1+\lambda_2}$ and $(X)_{\lambda_1+\lambda_2}^{\lambda_1+\lambda_2+\lambda_3}$.}\label{tg}
\end{center}
\end{figure}
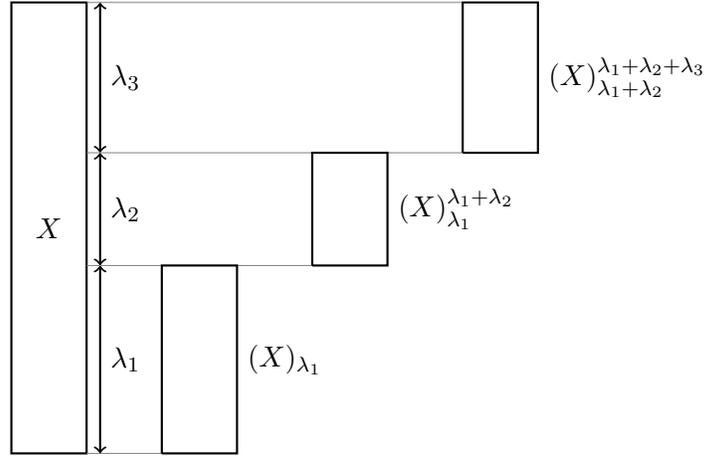

In words, for any $X\in\mathcal{X}_{\lambda_1+\lambda_2}$, $(X)^{\lambda_1+\lambda_2}_{\lambda_1}$ retrieves the top $\lambda_2$ power levels of $X$, while $(X)_{\lambda_1}$ retrieves the bottom $\lambda_1$ levels of $X$.  $(X)^{\lambda_3}_{\lambda_1}$ retrieves only the part of $X$ that lies between power levels $\lambda_1$ and $\lambda_3$.  Note that $X\in \mathcal{X}_\lambda$ can be expressed as $X={\bar{P}^{\lambda_1}}{(X)}_{\lambda_1}^{\lambda}+{(X)}_{\lambda_1}$ for $0\leq\lambda_1<\lambda$.  Equivalently, suppose $X_1\in\mathcal{X}_{\lambda_1}$, $X_2\in\mathcal{X}_{\lambda_2}$, $0<\lambda_2$ and $X=X_1+X_2\bar{P}^{\lambda_1}$. Then $X_1={(X)}_{\lambda_1}$, $X_2={(X)}^{\lambda_1+\lambda_2}_{\lambda_1}$.  A conceptual illustration of power level partitions is shown in 
Figure \ref{tg}.

Since expressions of the form $(X)^1_{1-\lambda}$ appear frequently in this particular work, let us define a compact notation for this as follows.
\begin{eqnarray}
(X)^\lambda\triangleq (X)^1_{1-\lambda}.
\end{eqnarray}

%\begin{figure}[h] 
%\centering
%\includegraphics[width=0.9\textwidth]{p5Themm.pdf}
%\caption[]{Illustration of an example of Definition \ref{powerlevel}. For an arbitrary random variable $X\in\mathcal{X}_{\lambda_1+\lambda_2+\lambda_3}$, the random variables $X_{\lambda_1}$, $X_{\lambda_1}^{\lambda_1+\lambda_2}$ and $X_{\lambda_1+\lambda_2}^{\lambda_1+\lambda_2+\lambda_3}$ are sketched.}\label{tg}
%\end{figure}

\begin {definition}
 For  the vector ${\bf V}=\begin{bmatrix}v_1&v_2&\cdots&v_k\end{bmatrix}^T$, we define  $({\bf V})_{\lambda_1}$ and $({\bf V})^{\lambda_2}_{\lambda_1}$ as,
 \begin{eqnarray}
%({\bf V})_{\lambda_1}&\triangleq& \begin{bmatrix}w'_1&w'_2&\cdots&w'_k\end{bmatrix}^T,\text{~~where,~~}w'_r=(v_r)_{\lambda_1}, \forall r\in[k]\\
({\bf V})_{\lambda_1}&\triangleq& \begin{bmatrix}(v_1)_{\lambda_1}&(v_2)_{\lambda_1}&\cdots&(v_k)_{\lambda_1}\end{bmatrix}^T\\
({\bf V})^{\lambda_2}_{\lambda_1}&\triangleq& \begin{bmatrix}(v_1)^{\lambda_2}_{\lambda_1}&(v_2)^{\lambda_2}_{\lambda_1}&\cdots&(v_k)^{\lambda_2}_{\lambda_1}\end{bmatrix}^T
\end{eqnarray}
\end{definition}

\begin{definition}[Bounded Density Channel Set $\mathcal{G}$]  \label{def:bd}
Let  $\mathcal{G}$ be a set of real-valued random variables, which satisfies both of the following conditions.
\begin{enumerate}
\item The magnitudes of all the random variables in $\mathcal{G}$ are bounded away from infinity, i.e., there exists a constant $\Delta<\infty$ such that for all $g\in\mathcal{G}$ we have $|g|\leq\Delta$.
\item There exists a finite positive constant $f_{\max}$, such that for all finite cardinality disjoint subsets $\mathcal{G}_1, \mathcal{G}_2$ of $\mathcal{G}$, the joint probability density function of all random variables in $\mathcal{G}_1$, conditioned on all random variables in $\mathcal{G}_2$, exists and is bounded above by $f_{\max}^{|\mathcal{G}_1|}$.
\end{enumerate}
 Without loss of generality we will assume that $f_{\max}\geq 1, \Delta\geq 1$. 
\end{definition}

\begin{definition}[Arbitrary Channel Set $\mathcal{H}$]  Let  $\mathcal{H}$ be a set of real-valued constants with magnitudes  bounded away from infinity, i.e.,  for all $h\in\mathcal{H}$ we have $|h|\leq\Delta$.
\end{definition}

%To distinguish channel coefficients that are known to the transmitter from those that are not known, we define the following two sets.
%\begin{definition}[Set of Channel Coefficients $\mathcal{G}$]
%Let $\mathcal{G}$ be a set of random variables that satisfy the bounded density assumption. 
%The set $\mathcal{G}$ contains channel coefficients whose realizations are not known to the transmitter, i.e., the transmitted symbols must be independent of channel coefficients that appear in $\mathcal{G}$, whereas the transmitted symbols are allowed to depend on channel coefficients that appear in $\mathcal{H}$. 
%\end{definition}

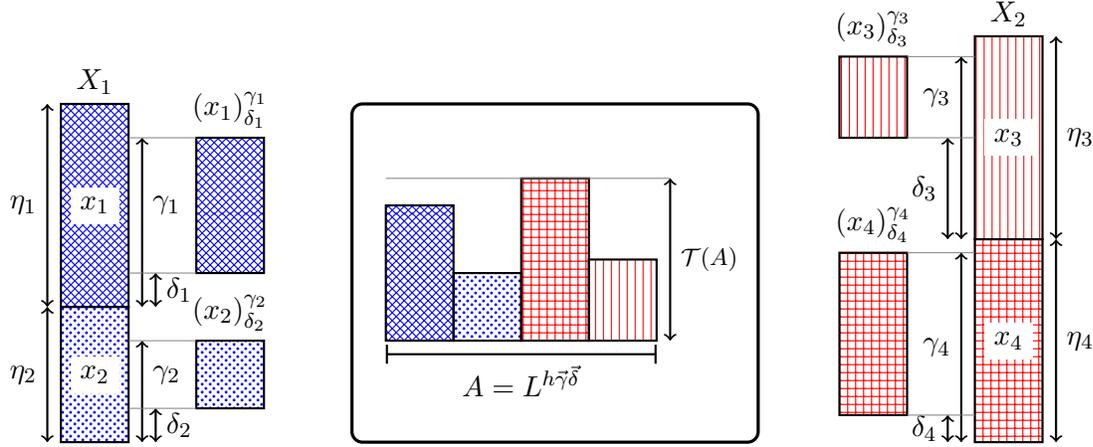
\begin{figure}[!h]
\begin{tikzpicture}[scale=0.9, baseline=(current bounding box.center)]

\begin{scope}[shift={(-0.5,0)}]
\draw[ thick, pattern = crosshatch dots, pattern color =blue] (0,0) rectangle (1,2) node[fill=white, midway]{$x_2$};
\draw[ thick, pattern = crosshatch, pattern color =blue] (0,2) rectangle (1,5)  node[fill=white, midway]{$x_1$};
\draw (0.5, 5) node[above]{$X_1$};
\draw[thick, <->](-0.2,0)--(-0.2,2) node[midway, left]{$\eta_2$};
\draw[thick, <->](1.2,0)--(1.2,0.5) -- node[ right]{$\gamma_2$} (1.2,1.5);
\draw[thick, <->](1.4,0)--(1.4,0.5) node[midway, right]{$\delta_2$};
\draw[thick, pattern = crosshatch dots, pattern color=blue](2,0.5) rectangle (3,1.5);
\draw (2.5,1.5) node[above]{$(x_2)^{\gamma_2}_{\delta_2}$};
\draw[help lines] (1,0.5)--(2,0.5);
\draw[help lines] (1,1.5)--(2,1.5);
\draw[thick, <->](-0.2,2)--(-0.2,5) node[midway, left]{$\eta_1$};
\draw[thick, <->](1.2,2)--(1.2,2.5) -- node[ right]{$\gamma_1$} (1.2,4.5);
\draw[thick, <->](1.4,2)--(1.4,2.5) node[midway, right]{$\delta_1$};
\draw[thick, pattern = crosshatch, pattern color=blue](2,2.5) rectangle (3,4.5);
\draw[help lines] (1,2.5)--(2,2.5);
\draw[help lines] (1,4.5)--(2,4.5);
\draw (2.5,4.5) node[above]{$(x_1)^{\gamma_1}_{\delta_1}$};
\end{scope}

\begin{scope}[shift={(4.3,1.5)}]
\draw[very thick, rounded corners] (-0.5, -1.5) rectangle (5.5, 3.5);

\draw[thick, pattern = crosshatch, pattern color=blue](0,0) rectangle (1,2);
\draw[thick, pattern = crosshatch dots, pattern color=blue](1,0) rectangle (2,1);
\draw[thick, pattern = grid, pattern color=red](2,0) rectangle (3,2.4);
\draw[thick, pattern = vertical lines, pattern color=red](3,0) rectangle (4,1.2);
\draw[thick, |-|] (0,-0.2)--(4,-0.2) node[midway, below]{$A=L^{h\vec{\gamma}\vec{\delta}}$};
\draw[thick,<->] (4.2,0)--(4.2,2.4) node[midway,right]{\footnotesize $\mathcal{T}(A)$};
\draw[help lines] (0,2.4)--(4.2,2.4);
\end{scope}

\begin{scope}[shift={(11,0)}]
\draw (2.5, 6) node[above]{$X_2$};

\draw[ thick, pattern = vertical lines, pattern color =red] (2,3) rectangle (3,6) node[fill=white, midway]{$x_3$};
\draw[thick, <->](3.2,3)--(3.2,6) node[midway, right]{$\eta_3$};

\draw[ thick, pattern = vertical lines, pattern color =red] (0,4.5) rectangle (1,5.7);
\draw[thick, <->](1.8,3)--(1.8,4.5) -- node[left]{$\gamma_3$} (1.8,5.7);
\draw[thick, <->](1.6,3)-- node[left]{$\delta_3$} (1.6,4.5);
\draw[help lines] (1,4.5)--(2,4.5);
\draw[help lines] (1,5.7)--(2,5.7);
\draw (0.5, 5.7) node[above]{$(x_3)^{\gamma_3}_{\delta_3}$};

\draw[ thick, pattern = grid, pattern color =red] (2,0) rectangle (3,3) node[fill=white, midway]{$x_4$};
\draw[thick, <->](3.2,0)--(3.2,3) node[midway, right]{$\eta_4$};

\draw[ thick, pattern = grid, pattern color =red] (0,0.4) rectangle (1,2.8);
\draw (0.5, 2.8) node[above]{$(x_4)^{\gamma_4}_{\delta_4}$};
\draw[thick, <->](1.6,0)--(1.6,0.4) node[midway, left]{$\delta_4$};
\draw[thick, <->](1.8,0)--(1.8,2.8) node[midway, left]{$\gamma_4$};
\draw[help lines] (1,0.4)--(2,0.4);
\draw[help lines] (1,2.8)--(2,2.8);
\end{scope}
\end{tikzpicture}
\caption{Visual illustration of $L^{\vec{\gamma}\vec{\delta}}$ and $\mathcal{T}(A)$. In this example, $x_1\in \mathcal{X}_{\eta_1}$ and $x_2\in\mathcal{X}_{\eta_2}$ are obtained as partitions of $X_1\in\mathcal{X}_{\eta_1+\eta_2}$. Similarly, $x_3\in \mathcal{X}_{\eta_3}$ and $x_4\in\mathcal{X}_{\eta_4}$ are  obtained as partitions of $X_2\in\mathcal{X}_{\eta_3+\eta_4}$. Note that $(\gamma_i, \delta_i)$ are only used to further trim the size of $x_i$, yielding $(x_i)_{\delta_i}^{\gamma_i}$ as the trimmed versions. These trimmed variables are then combined with arbitrary coefficients to produce $A=L^{\vec{\gamma}\vec{\delta}}$. Finally, note that  $\mathcal{T}(A)$ represents the size (power level) of the largest trimmed variable involved in $L^{\vec{\gamma}\vec{\delta}}$.}\label{fig:TA}
\end{figure}

\begin {definition}\label{deflc} For {real}  numbers {$x_1\in\mathcal{X}_{\eta_1},x_2\in\mathcal{X}_{\eta_2},\cdots,x_k\in\mathcal{X}_{\eta_k}$} and the vectors $\vec{\gamma}=(\gamma_1,\gamma_2,\cdots, \gamma_k)$ and $\vec{\delta}=(\delta_1,\delta_2,\cdots, \delta_k)$ define the notations $L_j^g(x_i,1\le i\le k)$, $L_j^h(x_i,1\le i\le k)$, {$L_j^{g\vec{\gamma}\vec{\delta}}(x_i,1\le i\le k)$ and $L_j^{h\vec{\gamma}\vec{\delta}}(x_i,1\le i\le k)$ } to represent,
\begin {eqnarray}
L^g_j(x_1,x_2,\cdots,x_k )&=&\sum_{1\le i\le k} \lfloor g_{j_i}x_i\rfloor\\
L_j^h(x_1,x_2,\cdots,x_k)&=&\sum_{1\le i\le k} \lfloor h_{j_i}x_i\rfloor\\
L_j^{g\vec{\gamma}\vec{\delta}}(x_1,x_2,\cdots,x_k)&=&\sum_{1\le i\le k} \lfloor g_{j_i}(x_i)^{\gamma_i}_{\delta_i}\rfloor\\
L_j^{h\vec{\gamma}\vec{\delta}}(x_1,x_2,\cdots,x_k)&=&\sum_{1\le i\le k} \lfloor h_{j_i}(x_i)^{\gamma_i}_{\delta_i}\rfloor
\end{eqnarray}
for  distinct random variables $g_{j_i}\in\mathcal{G}$,  arbitrary real valued constants {$h_{j_i}\in\mathcal{H}$}, and   arbitrary real valued constants $0\le\delta_i\le\gamma_i\le\eta_i$. For  the vector $V=\begin{bmatrix}v_1&v_2&\cdots&v_k\end{bmatrix}^T$ we similarly define the notations $L^g_j(V)$ and $L_j^h(V)$ to represent,
\begin {eqnarray}
L^g_j(V)=\sum_{1\le i\le k} \lfloor g_{j_i}v_i\rfloor\\
L^h_j(V)=\sum_{1\le i\le k} \lfloor h_{j_i}v_i\rfloor
\end{eqnarray}
\end {definition}
 Noting that these functions are \emph{approximately} (because of the $\lfloor\cdot\rfloor$ operations) linear, for simplicity we refer to them  as linear combinations. In particular,  we will refer to $L^{g}$ functions as \emph{random} linear combinations and to $L^h$ functions as \emph{arbitrary} linear combinations. The variables $x_i, v_i$ will generally be used to represent different parts of transmitted signals. Note that the subscripts, such as $L_j$, will be used to distinguish among different linear combinations, and may be dropped if there is no potential for ambiguity. 
{
\begin {definition}\label{def:length} For the linear combinations $A=L^{g\vec{\gamma}\vec{\delta}}(x_i,1\le i\le k)$ and $B=L^{h\vec{\gamma}\vec{\delta}}(x_i,1\le i\le k)$  where $x_1\in\mathcal{X}_{\eta_1},x_2\in\mathcal{X}_{\eta_2},\cdots,x_k\in\mathcal{X}_{\eta_k}$ we define $\mathcal{T}(A)$, $\mathcal{T}(B)$, $\mathcal{T}((A)^{\lambda}_\mu)$ and $\mathcal{T}((B)^{\lambda}_\mu)$  as,
\begin {eqnarray}
\mathcal{T}(A)=\mathcal{T}(B)&=&\max_{j\in[k]}(\gamma_{j}-\delta_{j})^+\label{total}\\
\mathcal{T}((A)^{\lambda}_\mu)=\mathcal{T}((B)^{\lambda}_\mu)&=&(\min(\lambda,\max_{j\in[k]}(\gamma_{j}-\delta_{j})^+)-\mu)^+\label{total}\\
\end{eqnarray}
\end {definition}}
Note that the terminology from Definition \eqref{deflc} is invoked in Definition \eqref{def:length}. Figure \ref{fig:TA} provides a visual illustration of $L^{h\vec{\gamma}\vec{\delta}}$ and $\mathcal{T}(A)$. From the definition of $\mathcal{T}(A)$ and $\mathcal{T}(B)$ in \eqref{total}, it follows that,
\begin {eqnarray}
A&\in&\{a:a\in\mathbb{Z}, |a| \le k\Delta\bar{P}^{\mathcal{T}(A)}\}\label{re1}\\
B&\in&\{a:a\in\mathbb{Z}, |a| \le k\Delta\bar{P}^{\mathcal{T}(B)}\}\label{re2}\\
(A)^{\lambda}_\mu&\in&\{a:a\in\mathbb{Z}, |a| \le \bar{P}^{\mathcal{T}((A)^{\lambda}_\mu)}\}\\
(B)^{\lambda}_\mu&\in&\{a:a\in\mathbb{Z}, |a| \le \bar{P}^{\mathcal{T}((B)^{\lambda}_\mu)}\}
\end{eqnarray}
 This is because the magnitudes of all\footnote{
Consider the terms $A=L^{g\vec{\gamma}\vec{\delta}}(x_i,1\le i\le k)$ and $(B)^{\lambda}_\mu=(L^{h\vec{\gamma}\vec{\delta}}(x_i,1\le i\le k))^{\lambda}_\mu$ and let us bound them as follows.
\begin {eqnarray}
|A|&=&|\sum_{1\le i\le k} \lfloor g_{i}(x_i)^{\gamma_i}_{\delta_i}\rfloor|\le k\Delta \max_{1\le i\le k}(x_i)^{\gamma_i}_{\delta_i}\le k\Delta \bar{P}^{\max_{1\le i\le k}(\gamma_i-\delta_i)^+}\le k\Delta\bar{P}^{\mathcal{T}(A)}.\\
|(B)^{\lambda}_\mu|&=&\left|\left \lfloor \frac{B-\bar{P}^{\lambda}\left \lfloor\frac{B}{ {\bar{P}}^{\lambda}}\right \rfloor }{{\bar{P}}^{\mu}} \right \rfloor\right| \le \left|  \frac{B-\bar{P}^{\lambda}\left \lfloor\frac{B}{ {\bar{P}}^{\lambda}}\right \rfloor }{{\bar{P}}^{\mu}}  \right|\le |  \frac{\bar{P}^{\min(\lambda,\max_{j\in[k]}(\gamma_{j}-\delta_{j})^+)}}{{\bar{P}}^{\mu}}  |\le\bar{P}^{\mathcal{T}((B)^{\lambda}_\mu)}
\end{eqnarray}} elements of $\mathcal{G},\mathcal{H}$ are bounded from above by $\Delta$.

\section{Sum-set Inequalities: Previous Results in \cite{Arash_Jafar_sumset}} {\label{sec-sys}}
Let us recall Theorem  $4$ of  \cite{Arash_Jafar_sumset}.

\begin{figure}[!h] 
\begin{eqnarray*}
H\left(\begin{tikzpicture}[scale=0.7, baseline=(current bounding box.center)]
\draw[ thick, pattern=crosshatch dots, pattern color=blue] (0,0+11) rectangle (1,3+11);
\draw[ thick, pattern=crosshatch, pattern color=blue] (0,3+11) rectangle (1,5+11);
\draw[ thick, pattern=grid, pattern color=blue] (0,5+11) rectangle (1,7.5+11);
\draw[ thick, pattern=vertical lines, pattern color=blue] (0,7.5+11) rectangle (1,9+11);
\draw[ thick, pattern=crosshatch dots, pattern color=red] (1,0+11) rectangle (2,3+11);
\draw[ thick, pattern=crosshatch, pattern color=red] (1,3+11) rectangle (2,5+11);
\draw[ thick, pattern=grid, pattern color=red] (1,5+11) rectangle (2,7.5+11);
\draw[ thick, pattern=vertical lines, pattern color=red] (1,7.5+11) rectangle (2,9+11);
\draw[ thick, pattern=crosshatch dots, pattern color=green] (2,0+11) rectangle (3,3+11);
\draw[ thick, pattern=crosshatch, pattern color=green] (2,3+11) rectangle (3,5+11);
\draw[ thick, pattern=grid, pattern color=green] (2,5+11) rectangle (3,7.5+11);
\draw[ thick, pattern=vertical lines, pattern color=green] (2,7.5+11) rectangle (3,9+11);
\draw[thick, <->] (3.2,0+11)--(3.2,3+11) node[midway, right]{$\lambda_{11}$};
\draw[thick, <->] (3.2,3+11)--(3.2,5+11) node[midway, right]{$\lambda_{12}$};
\draw[thick, <->] (3.2,5+11)--(3.2,7.5+11) node[midway, right]{$\lambda_{13}$};
\draw[thick, <->] (3.2,7.5+11)--(3.2,9+11) node[midway, right]{$\lambda_{14}$};
\draw[thick, |-|] (0,-0.2+11)--(3,-0.2+11) node[midway, below]{$Z_1=L_1^b$};
\draw (0.5, 9+11) node[above]{$X_1$};
\draw (1.5, 9+11) node[above]{$X_2$};
\draw (2.5, 9+11) node[above]{$X_3$};
\path (3, 10) node[right]{\Huge ,};
\draw[ thick, pattern= dots, pattern color=blue] (0,0) rectangle (1,3.8);
\draw[ thick, pattern=north east lines, pattern color=blue] (0,3.8) rectangle (1,5.2);
\draw[ thick, pattern=north west lines, pattern color=blue] (0,5.2) rectangle (1,8.1);
\draw[ thick, pattern=horizontal lines, pattern color=blue] (0,8.1) rectangle (1,9);
\draw[ thick, pattern= dots, pattern color=red] (1,0) rectangle (2,3.8);
\draw[ thick, pattern=north east lines, pattern color=red] (1,3.8) rectangle (2,5.2);
\draw[ thick, pattern=north west lines, pattern color=red] (1,5.2) rectangle (2,8.1);
\draw[ thick, pattern=horizontal lines, pattern color=red] (1,8.1) rectangle (2,9);
\draw[ thick, pattern= dots, pattern color=green] (2,0) rectangle (3,3.8);
\draw[ thick, pattern=north east lines, pattern color=green] (2,3.8) rectangle (3,5.2);
\draw[ thick, pattern=north west lines, pattern color=green] (2,5.2) rectangle (3,8.1);
\draw[ thick, pattern=horizontal lines, pattern color=green] (2,8.1) rectangle (3,9);
\draw[thick, <->] (3.2,0)--(3.2,3.8) node[midway, right]{$\lambda_{21}$};
\draw[thick, <->] (3.2,3.8)--(3.2,5.2) node[midway, right]{$\lambda_{22}$};
\draw[thick, <->] (3.2,5.2)--(3.2,8.1) node[midway, right]{$\lambda_{23}$};
\draw[thick, <->] (3.2,8.1)--(3.2,9) node[midway, right]{$\lambda_{24}$};
\draw[thick, |-|] (0,-0.2)--(3,-0.2) node[midway, below]{$Z_2=L_2^b$};
\draw (0.5, 9) node[above]{$X_1$};
\draw (1.5, 9) node[above]{$X_2$};
\draw (2.5, 9) node[above]{$X_3$};
\end{tikzpicture}
\right)
&\geq&
H\left(
\begin{tikzpicture}[scale=0.8, baseline=(current bounding box.center)]
\draw[ thick, pattern=vertical lines, pattern color=blue] (-1,0.1+9) rectangle (0,0.1+9.6);
\draw[ thick, pattern=vertical lines, pattern color=red] (0,0.1+9) rectangle (1,0.1+9.4);
\draw[ thick, pattern=vertical lines, pattern color=green] (1,0.1+9) rectangle (2,0.1+9.7);
\draw[thick,|-|](-1,0.1+8.8)--(2,0.1+8.8) node[midway, below]{ $Z_{11}=L_{11}^{\vec{\gamma_{11}}\vec{\delta_{11}}}$};
\draw[thick,<->](-1.2,0.1+9)--(-1.2,0.1+10.5) node[midway, left]{$\lambda_{14}$};
\draw[help lines](-1,0.1+9.7)--(2,0.1+9.7);
\draw[thick,<->](2.2,0.1+9)--(2.2,0.1+9.7) node[midway, right]{ $\mathcal{T}(Z_{11})$};
\path (3.6, 0.1+9) node[right]{\Huge ,};
\draw[ thick, pattern=horizontal lines, pattern color=blue] (4.9+0,0.1+9) rectangle (4.9+1,0.1+9.4);
\draw[ thick, pattern=horizontal lines, pattern color=red] (4.9+1,0.1+9) rectangle (4.9+2,0.1+9.2);
\draw[thick,|-|](4.9+0,0.1+8.8)--(4.9+2,0.1+8.8) node[midway, below]{ $Z_{21}=L_{21}^{\vec{\gamma_{21}}\vec{\delta_{21}}}$};
\draw[thick,<->](4.9+-0.1,0.1+9)--(4.9+-0.1,0.1+9.9) node[midway, left]{$\lambda_{24}$};
\draw[help lines](4.9+0,0.1+9.4)--(4.9+2,0.1+9.4);
\draw[thick,<->](4.9+2.2,0.1+9)--(4.9+2.2,0.1+9.4) node[midway, right]{ $\mathcal{T}(Z_{21})$};
\path (4.9+3.8, 0.1+9) node[right]{\Huge ,};

\draw[ thick, pattern=vertical lines, pattern color=blue] (0-0.5,5) rectangle (1-0.5,6);
\draw[ thick, pattern=vertical lines, pattern color=red] (1-0.5,5) rectangle (2-0.5,5.5);
\draw[ thick, pattern=crosshatch, pattern color=blue] (2-0.5,5) rectangle (3-0.5,6.9);
\draw[thick,<->](-0.2-0.5,5)--(-0.2-0.5,6.5) node[right]{$\lambda_{14}$};
\draw[thick,<->](-0.4-0.5,5)--(-0.4-0.5,7) node[left]{$\lambda_{12}$};
\draw[thick,<->](3.2-0.5,5)--(3.2-0.5,6.9) node[midway, right]{ $\mathcal{T}(Z_{12})$};
\draw[help lines](0,6.9)--(3-0.5,6.9);
\path (3.6-0.5, 5) node[right]{\Huge ,};
\draw[thick,|-|](0-0.5,4.8)--(3-0.5,4.8) node[midway, below]{ $Z_{12}=L_{12}^{\vec{\gamma_{12}}\vec{\delta_{12}}}$};
\draw[ thick, pattern=horizontal lines, pattern color=green] (4.8+0,5) rectangle (4.8+1,5.3);\draw[ thick, pattern=north west lines, pattern color=blue] (4.8+1,5) rectangle (4.8+2,6.2);
\draw[ thick, pattern=north west lines, pattern color=red] (4.8+2,5) rectangle (4.8+3,7.2);
\draw[thick,<->](4.8+-0.2,5)--(4.8+-0.2,5.9) node[right]{$\lambda_{24}$};
\draw[thick,<->](4.8+-0.4,5)--(4.8+-0.4,7.5) node[left]{$\lambda_{23}$};
\draw[thick,<->](4.8+3.2,5)--(4.8+3.2,7.2) node[midway, right]{ $\mathcal{T}(Z_{22})$};
\draw[help lines](4.8+0,7.2)--(4.8+3,7.2);
\path (4.8+3.5, 5) node[right]{\Huge ,};
\draw[thick,|-|](4.8+0,4.8)--(4.8+3,4.8) node[midway, below]{ $Z_{22}=L_{22}^{\vec{\gamma_{22}}\vec{\delta_{22}}}$};

\draw[ thick, pattern=vertical lines, pattern color=red] (0,0) rectangle (1,1);
\draw[ thick, pattern=vertical lines, pattern color=blue] (1,0) rectangle (2,0.8);
\draw[ thick, pattern=grid, pattern color=green] (2,0) rectangle (3,2.1);
\draw[ thick, pattern=crosshatch, pattern color=blue] (3,0) rectangle (4,1.3);
\draw[ thick, pattern=crosshatch, pattern color=red] (4,0) rectangle (5,1.6);
\draw[ thick, pattern=crosshatch dots, pattern color=green] (5,0) rectangle (6,2.8);
\draw[ thick, pattern=crosshatch dots, pattern color=red] (6,0) rectangle (7,2.6);
\draw[thick,<->](-0.2,0)--(-0.2,1.5) node[right]{$\lambda_{14}$};
\draw[thick,<->](-0.4,0)--(-0.4,3) node[left]{$\lambda_{11}$};
\draw[thick,<->](-0.6,0)--(-0.6,2.5) node[left]{$\lambda_{13}$};
\draw[thick,<->](-0.8,0)--(-0.8,2) node[left]{$\lambda_{12}$};
\draw[thick,<->](7.2,0)--(7.2,2.8) node[midway, right]{ $\mathcal{T}(Z_{13})$};
\draw[help lines](0,2.8)--(7,2.8);
\path (6, -1) node[right]{\Huge ,};
\draw[thick,|-|](0,-0.2)--(7,-0.2) node[midway, below]{ $Z_{13}=L_{13}^{\vec{\gamma_{13}}\vec{\delta_{13}}}$};

\draw[ thick, pattern=horizontal lines, pattern color=red] (0,-5) rectangle (1,-4.4);
\draw[ thick, pattern=horizontal lines, pattern color=green] (1,-5) rectangle (2,-4.6);
\draw[ thick, pattern=north west lines, pattern color=green] (2,-5) rectangle (3,-3.9);
\draw[ thick, pattern=north east lines, pattern color=blue] (3,-5) rectangle (4,-4.7);
\draw[ thick, pattern=north east lines, pattern color=green] (4,-5) rectangle (5,-3.8);
\draw[ thick, pattern= dots, pattern color=blue] (5,-5) rectangle (6,-2.6);
\draw[ thick, pattern= dots, pattern color=green] (6,-5) rectangle (7,-2.9);
\draw[thick,<->](-0.2,-5)--(-0.2,-4.1) node[right]{$\lambda_{24}$};
\draw[thick,<->](-0.4,-5)--(-0.4,-1.2) node[left]{$\lambda_{21}$};
\draw[thick,<->](-0.6,-5)--(-0.6,-2.1) node[left]{$\lambda_{23}$};
\draw[thick,<->](-0.8,-5)--(-0.8,-3.6) node[left]{$\lambda_{22}$};
\draw[thick,<->](7.2,-5)--(7.2,-2.6) node[midway, right]{ $\mathcal{T}(Z_{23})$};
\draw[help lines](0,-2.6)--(7,-2.6);
\draw[thick,|-|](0,-5.2)--(7,-5.2) node[midway, below]{ $Z_{23}=L_{23}^{\vec{\gamma_{23}}\vec{\delta_{23}}}$};

\end{tikzpicture}
\right)
\end{eqnarray*}
%\centering
%\includegraphics[width=0.9\textwidth]{p5Them.pdf}
\caption[]{Illustration of an application of Theorem \ref{Theorem AIS04} of \cite{Arash_Jafar_sumset}. {Note that in this figure we dropped the time index $(t)$ for convenience}. On the left is the joint entropy of  the sum (bounded density linear combination) of $N=3$ dependent random variables, $X_1(t), X_2(t),X_3(t)\in\mathcal{X}_{\max_{k\in[2]}\{\lambda_{k1}+\lambda_{k2}+\lambda_{k3}+\lambda_{k4}\}}$, $(M=4)$, which is bounded below by the joint entropy of $l_1+l_2=6$ arbitrary linear combinations, $Z_{11}, Z_{12}, Z_{13},Z_{21}, Z_{22}, Z_{23}$, comprised of power level partitions of the two random variables. In this example, $I_{11}=I_{21}=\{4\}, I_{12}=\{2,4\}, I_{22}=\{3,4\}, I_{13}=I_{23}=\{1,2,3,4\}$. Condition (\ref{condt4}) is verified as $\lambda_{11}+\lambda_{12}+\lambda_{13}\geq \mathcal{T}(Z_{12})+\mathcal{T}(Z_{13})$, $\lambda_{21}+\lambda_{22}+\lambda_{23}\geq \mathcal{T}(Z_{22})+\mathcal{T}(Z_{23})$, $\lambda_{21}\geq \mathcal{T}(Z_{23})$ and  $\lambda_{11}\geq \mathcal{T}(Z_{13})$.} \label{grant}
\end{figure}
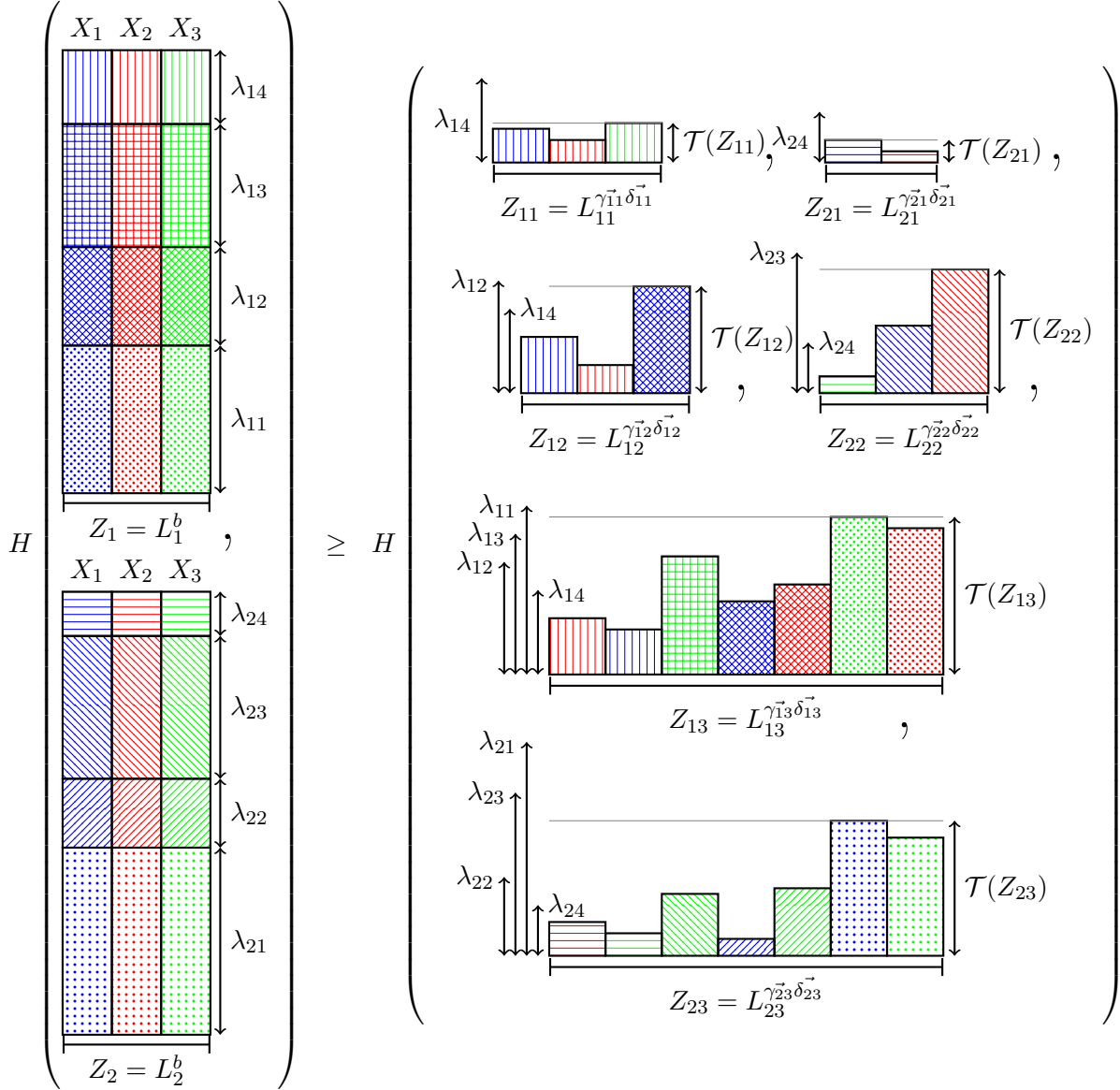

\begin{theorem}[Theorem $4$ in \cite{Arash_Jafar_sumset}] \label{Theorem AIS04}
Consider $KM$ non-negative numbers {$\{\lambda_{km}: k\in[K],m\in[M]\}$}  and random variables $X_j (t) \in \mathcal{X}_{\max_{k\in[K]}\{\lambda_{k,1}+\lambda_{k,2}+\cdots+\lambda_{k,M}\}}$, $j\in[N]$, $t\in\mathbb{N}$,  independent of $\mathcal{G}$,  and  $\forall k\in[K], K\le N$, define
\begin {eqnarray}
Z_k(t)&=&L_k^g(t)(X_1(t),X_2(t), \cdots, X_N(t))\label{mn1}\\
Z_{k,1}(t)&=& L_{k1}^{h\vec{\gamma}_{k1}\vec{\delta}_{k1}}(t)((X_{j}(t))_{\sum_{r=1}^{i-1}\lambda_{kr}}^{\sum_{r=1}^i\lambda_{kr}}, i\in I_{k,1}, j\in[N])\label{mn2}\\
Z_{k,2}(t)&=& L_{k2}^{h\vec{\gamma}_{k2}\vec{\delta}_{k2}}(t)((X_{j}(t))_{\sum_{r=1}^{i-1}\lambda_{kr}}^{\sum_{r=1}^i\lambda_{kr}}, i\in I_{k,2}, j\in[N])\label{mn3}\\
&\vdots&\notag\\
Z_{k,l_k}(t)&=&L_{kl_k}^{h\vec{\gamma}_{kl_k}\vec{\delta}_{kl_k}}(t)((X_{j}(t))_{\sum_{r=1}^{i-1}\lambda_{kr}}^{\sum_{r=1}^i\lambda_{kr}}, i\in I_{k,l_k}, j\in[N])\label{mn4}
\end{eqnarray}
The channel uses are indexed by $t\in\mathbb{N}$. $I_{kk'}\subset [M], k\in[K], k'\in[l_k],$  such that $i<j\Rightarrow m(k,i)\geq m(k,j)$, where $$m(a,b)\triangleq \min\{m: m\in I_{a,b}\}.$$ If for all $k\in[K]$ and for each $s\in\{1,2,\cdots, l_k-1\}$,
\begin {eqnarray}
\mathcal{T}(Z_{k,s+1})+\mathcal{T}(Z_{k,s+2})+\cdots+\mathcal{T}(Z_{k,l_k})&\leq& \lambda_{k,1}+\lambda_{k,2}+\cdots+\lambda_{k,(m(k,s)-1)}\label{condt4}
\end{eqnarray}
then for any acceptable random variable $W$ \footnote{ Let $\mathcal{G}(Z)\subset\mathcal{G}$ denote the set of all bounded density channel coefficients that appear in $Z_1^{[n]},\cdots,Z_K^{[n]}$, and let $W$ be a random variable such that conditioned on any $\mathcal{G}_o\subset (\mathcal{G}/\mathcal{G}(Z))\cup \{W\}$, the channel coefficients $\mathcal{G}(Z)$ satisfy the bounded density assumption. }
\begin {eqnarray}
H(Z_1^{[n]},\cdots,Z_K^{[n]}\mid W,\mathcal{G})&\geq& H(Z_{1,1}^{[n]},\cdots,Z_{K,l_K}^{[n]}\mid W)+Kn~o(\log{\bar{P}})\label{dssd4}
\end{eqnarray}
\end{theorem}

\section{System Model} {\label{sec-sys}}
In this work we will focus on the setting where all variables take only real values. Extensions to complex valued settings may be cumbersome but are expected to be conceptually straightforward as shown in  \cite{Arash_Jafar_PN}. We will focus on the two-user MIMO BC equipped with $M$ antennas at the transmitter and $N_1, N_2$ antennas at the two receivers, with the assumption throughout that
\begin{eqnarray}
N_1\leq N_2\leq M {\leq N_1+N_2},
\end{eqnarray}
since this is the only non-trivial setting where the DoF remain open. For all cases where the  condition $N_1\leq N_2\leq M$ is not true, the DoF  are already established  in \cite{Hao_Rassouli_Clerckx}.  For all cases where $M>N_1+N_2$,  it is easy to see that the DoF region is not affected if the number of transmit antennas is reduced to $N_1+N_2$, as follows. First, from the achievability side, note that the  DoF innerbound shown in \cite{Hao_Rassouli_Clerckx} remains unaffected if the number of transmit antennas is reduced to $N_1+N_2$. Then from the outer bound perspective we note that the capacity cannot be reduced if a genie informs the transmitter of the $M-N_1-N_2$ dimensional transmit signal space that is not heard by either user, which allows the transmitter to discard these $M-N_1-N_2$ transmit dimensions (antennas) without loss of generality, thereby reducing the effective number of transmit dimensions to $N_1+N_2$. Therefore in order to establish the DoF region for all cases where $M>N_1+N_2$, it suffices to show that the DoF outer bound matches the DoF inner bound for the setting $M=N_1+N_2$.
\subsection{The Channel}\label{channelmodel+}
If perfect CSIT was available, then for generic channel realizations in the two user $(M,N_1,N_2)$ MIMO broadcast channel with $N_1\leq N_2\leq M\leq N_1+N_2$, there are $M-N_1$ transmit directions available to the transmitter that are in the null-space of the channel matrix between the transmitter and the first receiver. Similarly,  there are $M-N_2$ transmit directions available to the transmitter that are in the null-space of the channel matrix between the transmitter and the second receiver. A canonical representation of the channel (obtained by applying a change of basis operation at the transmitter) makes these directions explicit by mapping them to transmit antennas, so  an $M$ dimensional input vector ${\bf X}$ is partitioned as follows.
\begin{eqnarray}
{\bf X}_{a}(t)&=&[{\bf X}(t)]_{0\rightarrow {(M-N_2)}}\\
{\bf X}_{b}(t)&=&[{\bf X}(t)]_{{(M-N_2)\rightarrow (N_1+N_2-M)}}\\
%{\bf X}_{c}(t)&=&[{\bf X}(t)]_{N_1\rightarrow (M-N_1-N_2)^+}\\
{\bf X}_{c}(t)&=&[{\bf X}(t)]_{{N_1\rightarrow (M-N_1)}}
\end{eqnarray}
Recall that the notation ${\bf X}_{m\rightarrow n}$ as used here stands for $({\bf X}_{m+1}, {\bf X}_{m+2}, \cdots, {\bf X}_{m+n})$. Thus, the partition ${\bf X}_a$ contains transmit directions that are {in the null-space of User $2$ but not User $1$, the partition ${\bf X}_c$  contains transmit directions that are in the null space of User $1$ but not  User $2$, 
and ${\bf X}_b$ contains transmit directions that are not in the null space of either user}.
Further, note that if $M=N_1+N_2$, then ${\bf X}_b$ disappears, and the partition is simply ${\bf X}={\bf X}_a\bigtriangledown{\bf X}_c$. 
If $M< N_1+N_2$ then the partition is  ${\bf X}={\bf X}_a\bigtriangledown {\bf X}_b\bigtriangledown{\bf X}_c$.
Evidently, for the first user, zero forcing is possible only along the $M-N_1$ dimensional space corresponding to ${\bf X}_c$, and for the second user, zero forcing is possible only along the $M-N_2$ dimensional space corresponding to ${\bf X}_a$.

\begin{figure}[h] 
\centering
\includegraphics[width=0.8\textwidth]{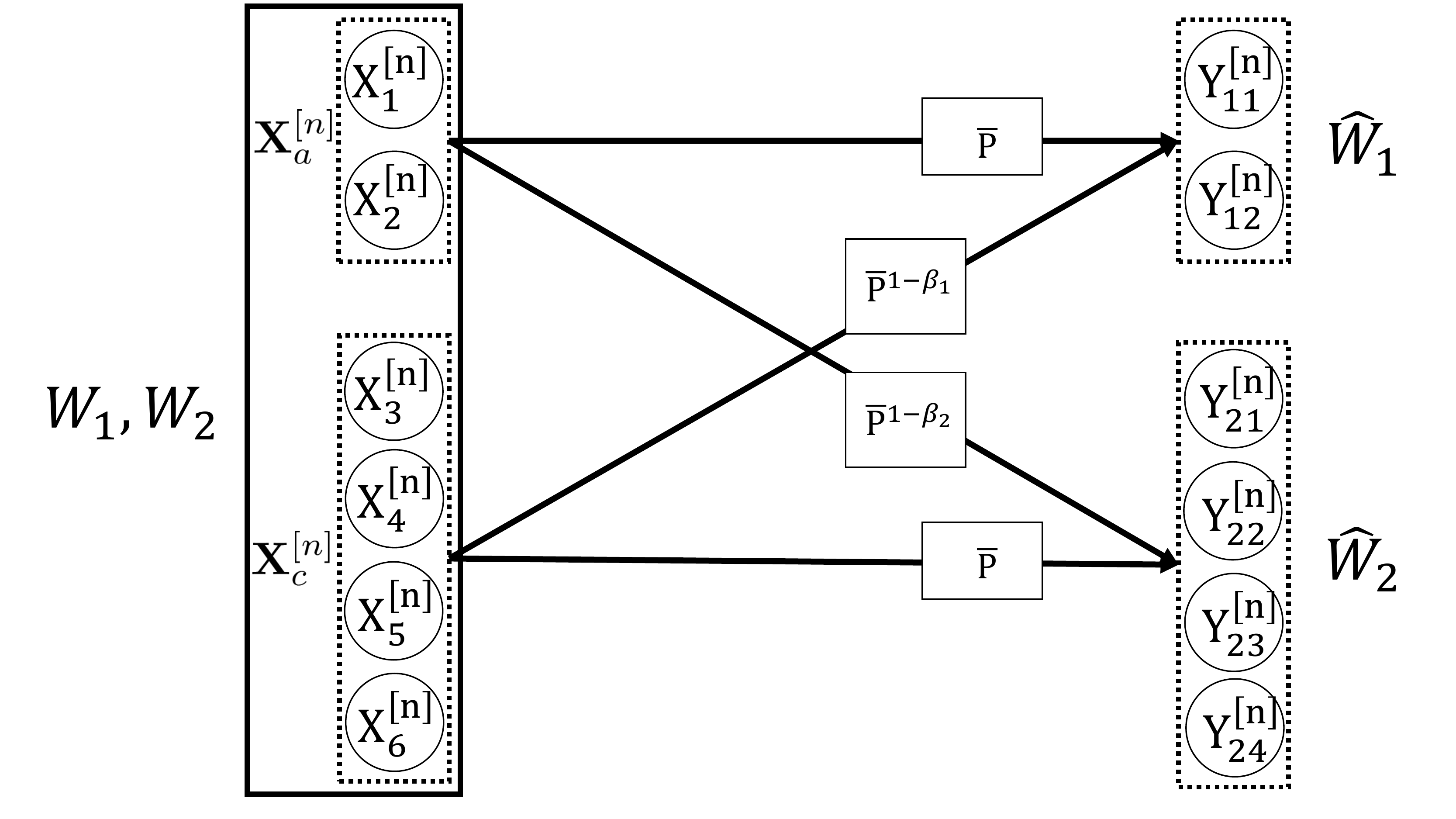}
\caption{Equivalent channel for two user $(M,N_1,N_2)=(6,2,4)$ MIMO BC.}\label{FigeqBC}
\end{figure}

With partial CSIT, only partial zero-forcing is possible based on channel estimates available to the transmitter. Therefore, the channel model for the two user $(M,N_1,N_2)$ MIMO BC with partial CSIT, is represented in its canonical form by the following input output equations. See Fig \ref{FigeqBC}. 
\begin{eqnarray}
{\bf Y}_{1}(t)&=&\sqrt{P}{\bf G}_{1ab}(t)\begin{bmatrix}
{\bf X}_{a}(t)\bigtriangledown{\bf X}_{b}(t)
\end{bmatrix}+\sqrt{P^{1-\beta_1}}{\bf G}_{1c}(t)
{\bf X}_{c}(t)
+{\bf \Gamma}_{1}(t)\label{req1}\\
{\bf Y}_{2}(t)&=&\sqrt{P}{\bf G}_{2bc}(t)\begin{bmatrix}
{\bf X}_{b}(t)\bigtriangledown
{\bf X}_{c}(t)
\end{bmatrix}+\sqrt{P^{1-\beta_2}}{\bf G}_{2a}(t)
{\bf X}_{a}(t)+{\bf \Gamma}_{2}(t)\label{req2}
\end{eqnarray}
The dimensions of these symbols are listed as follows.
\allowdisplaybreaks
\begin{align}
&{\bf Y}_{1}(t),\mathbf{\Gamma}_{1}(t): N_1\times 1,&&{\bf Y}_{2}(t),\mathbf{\Gamma}_{2}(t): N_2\times 1,\\
&{\bf G}_{1ab}(t): N_1\times N_1,&&{\bf G}_{1c}(t): N_1\times(M-N_1),\\
&{\bf G}_{2bc}(t): N_2\times N_2,&&{\bf G}_{2a}(t): N_2\times(M-N_2).
\end{align}
Here, over channel use $t\in\mathbb{N}$, the vector of symbols seen at Receiver $i$, $i\in\{1,2\}$, is ${\bf Y}_{i}(t)$, and the vector of symbols sent from the transmitter is ${\bf X}(t)={\bf X}_{a}(t)\bigtriangledown{\bf X}_{b}(t)\bigtriangledown{\bf X}_{c}(t)$. Channel matrices ${\bf G}_{1ab}(t), {\bf G}_{2bc}(t)$ correspond to directions along which no zero-forcing is possible, while ${\bf G}_{1c}, {\bf G}_{2a}$ correspond to directions that can be partially zero-forced based on channel estimates available to the transmitter. Note that due to partial CSIT, the dimensions that can be partially zero-forced have  channel strength diminished by the negative power exponents $\beta_1, \beta_2$ for users $1$ and $2$ respectively, relative to those directions along which no zero-forcing is possible. The quality of CSIT is captured by $\beta_1, \beta_2\in[0,1]$. As  the CSIT parameters $\beta_j, j\in\{1,2\}$, take values in the interval from $0$ to $1$ they cover the full range from no CSIT (i.e., no zero-forcing ability) to perfect CSIT (perfect zero-forcing ability) in the DoF sense.  $\mathbf{\Gamma}_{i}(t)$ are the zero-mean unit variance additive white Gaussian noise terms seen at outputs ${\bf Y}_{i}(t)$, independent of all inputs and channel realizations. The input vector ${\bf X}(t)$ is subject to unit power constraint. 

%Individual elements of these vector symbols are denoted through additional subscripts in parantheses. For example,  the $(k,l)^{th}$ element of the channel matrix ${\bf G}_{ij}^{rz}(t)$ is denoted as ${\bf G}_{ij(k,l)}^{rz}(t)$, and the $i^{th}$ element in the vector ${\bf X}_{1}^{z}(t)$ is represented as ${\bf X}_{1(i)}^z(t)$. 

%matrices ${\bf H}_{1}(t),{\bf H}_{2}(t)$ have determinants bounded away from zero. Mathematically, we assume that there exists $\epsilon>0$, such that,
%\begin{eqnarray}
%\left|\mbox{det}({\bf H}_{1}(t))\right|&\geq&\epsilon\label{detcon1}\\
%\left|\mbox{det}({\bf H}_{2}(t))\right|&\geq&\epsilon\label{detcon2}.
%\end{eqnarray}

All channel coefficients are distinct random variables drawn from the bounded density channel set $\mathcal{G}$, (see Definition \ref{def:bd}),   therefore all channel coefficient magnitudes are bounded above by $\Delta<\infty$. Further,  in order to avoid degenerate conditions, we assume that all channel  matrices have determinants bounded away from zero, i.e., the absolute value of the determinant of each square channel submatrix is greater than a positive constant $\epsilon$.

Perfect channel state information at the receivers (CSIR) is assumed to be available for all channels. In terms of CSIT, we assume that the  transmitter is aware of the bounded density probability density functions of all channels, but not the actual channel realizations.

%\footnote{The channels ${\bf G}_{1d}(t),{\bf G}_{2a}(t)$  exist only due to the channel estimation error terms at the transmitter that prevent perfect zero-forcing, so these channels are unknown to the transmitter. The CSIT assumption for the remaining channels,  ${\bf G}_{1ab}(t), {\bf G}_{2bd}(t)$ are in fact inconsequential because  even if these channels are assumed unknown, note that it is possible for the receivers to force arbitrary pre-determined constant realizations for channels ${\bf G}_{1ab}(t), {\bf G}_{2bd}(t)$ by  change of basis operations (e.g., by multiplying the received signal by the inverse of these channel matrices the receiver can force these channels to take the form of identity matrices), which  makes these channels perfectly known to the transmitter. Thus, the DoF result does not change whether these channels are assumed to be perfectly known to the transmitter, completely unknown, or anything in between.} Thus, the transmitted symbols ${\bf X}(t)$ must be independent of the channel realizations. Also note that the partial CSIT is already captured in the diminished channel strength parameters $\beta_1, \beta_2$ associated with partial zero-forcing.

\subsection{DoF}\label{rate}
The definitions of achievable rates $R_i(P)$ and capacity region $\mathcal{C}(P)$ are standard. The DoF region is defined as
\begin{align}
\mathcal{D}=&\{(d_1,d_2): \exists (R_1(P),R_2(P))\in\mathcal{C}(P), \mbox{ s.t. } d_k=\lim_{P\rightarrow\infty}\frac{R_k(P)}{\frac{1}{2}\log{(P)}}, \forall k\in\{1,2\}\} \label {region}
\end{align}

\section{Main Result}
%Without loss of generality let us assume $N_1\le N_2$ since the DoF region for the $N_1> N_2$ is easily obtained by switching the indices $1$ and $2$. For $M\le N_1$, the DoF region of the MIMO BC with partial CSIT is the same as the DoF region of the MIMO BC with perfect CSIT, i.e., $d_1+d_2\le M$, see \cite{Weingarten_Steinberg_Shamai}.  Thus, from now on we assume $N_1<M$. 

The following theorem characterizes the complete DoF region of the two-user  MIMO BC with arbitrary levels of partial CSIT $\beta_1, \beta_2\in[0,1]$, for arbitrary (unconstrained) choice of parameters $M, N_1, N_2$.

\begin{theorem}\label{theorem1} Without loss of generality, assume $N_1\leq N_2$. The DoF region  is expressed as follows.
\begin{enumerate}
\item For $N_2\le M$:
\begin{align}
&&\mathcal{D}=\Bigl\{(d_1,d_2)\in \mathbb{R}^{2+}&\mbox{ such that }\nonumber\\
&&&d_1\le N_1,\label{B1}\\
&&&d_2\le N_2,\label{B2}\\
&&&\frac{d_1}{N_1}+\frac{d_2}{N_2}\le1+\frac{M-N_1}{N_2}\beta_1,\label{B4}\\
&&&d_1+d_2\le N_2+(M-N_2)\beta_2,\label{B3}\\
&&&d_1+d_2\le N_2+(M-N_2)\beta_o\Bigr\}\label{B33}
\end{align}
where $\beta_o$ is defined as,
\begin{eqnarray}
\beta_o&=
&\left\{
\begin{array}{ll}  
\frac{\beta_1\beta_2(M-N_2)}{(N_2-N_1)(1-\beta_1)+(M-N_2)\beta_2}, &\beta_1+\beta_2<1\\
\frac{N_1-N_2+(N_2-N_1)\beta_2+(M-N_1)\beta_1}{M-N_1}, &\beta_1+\beta_2\ge1
\end{array}
\right.\label{B5}
\end{eqnarray}
\item For $N_2 > M$:
\begin{align}
&&\mathcal{D}=\Bigl\{(d_1,d_2)\in {\mathbb{R}^{2+}}&\mbox{ such that }\nonumber\\
&&&d_1\le N_1\label{B7}\\
&&&d_1+d_2\le M\label{B8}\\
&&&\frac{d_1}{N_1}+\frac{d_2}{M}\le1+\frac{M-N_1}{M}\beta_1\label{B9}\Bigr\}
\end{align}
\end{enumerate}
\end{theorem}

\bigskip

\begin{remark} Wang and Varanasi \cite{Yao_Varanasi_Hybrid} studied the  DoF of the two-user MIMO broadcast channel with general message set (a common message and two private messages) under hybrid CSIT models where for each user the CSIT is either  perfect (P), delayed (D), or not available (N). While the `PP', `PD', `DP', `DD', `NN' settings are fully settled,   for the `PN', `NP', `DN' and `ND' settings, only the linear DoF regions (i.e., the DoF region restricted to linear achievable schemes) are found and it is conjectured that the same regions are optimal even without restriction to linear achievable schemes. It is further explained in \cite{Yao_Varanasi_Hybrid} that the `NP' setting (where no CSIT is available for the first user and perfect CSIT is available for the second user)\footnote{Note that without loss of generality we assume that $N_1\leq N_2$ whereas \cite{Yao_Varanasi_Hybrid} assumes that $N_2\leq N_1$. Thus our user indices are switched relative to \cite{Yao_Varanasi_Hybrid}. As a consequence, what is referred to as the `PN' setting in \cite{Yao_Varanasi_Hybrid} corresponds to the `NP' setting in this paper.} is the key, i.e., if the `NP' case can be solved then the other three cases can be easily resolved. Furthermore, tight outer bounds that include a common message are found directly from the setting with only private messages by reducing the decoding requirement for the common message to only one of the receivers. Since the `NP' setting corresponds to $\beta_1=0, \beta_2=1$, evidently Theorem \ref{theorem1} settles the conjectures of Wang and Varanasi  \cite{Yao_Varanasi_Hybrid} in the affirmative.
\end{remark}

\begin{remark} Achievability of the DoF region of Theorem \ref{theorem1} is established by Hao et al. in \cite{Hao_Rassouli_Clerckx} based on a rate-splitting scheme that includes interesting `space-time' scheduling aspects. Partial converse results are also presented in \cite{Hao_Rassouli_Clerckx} based on  relatively straightforward applications of  the aligned image sets (AIS) argument \cite{Arash_Jafar_PN}. The  problem that remains open is the proof of the outer bound \eqref{B33} for $N_1\leq N_2\leq M\leq N_1+N_2$, which is the main contribution of this work. Our proof exemplifies the utility of the  `sum-set inequalities' that were recently developed from AIS arguments in \cite{Arash_Jafar_sumset}. 
\end{remark}

The proof of Theorem \ref{theorem1} (i.e., the proof of \eqref{B33} for $N_1\leq N_2\leq M\leq N_1+N_2$)  appears in Section \ref{proofthe} and is partitioned into two cases, corresponding to $\beta_1+\beta_2\geq 1$ and $\beta_1+\beta_2<1$, that are covered in Section \ref{sec:b1b2>1} and Section \ref{sec:b1b2<1}, respectively. 
For ease of exposition let us first illustrate the main ideas of the proof with two  examples --- the $(M, N_1, N_2)=(5,2,3)$ MIMO BC with  $(\beta_1,\beta_2)=(1/2, 2/3)$ as representative of the case $\beta_1+\beta_2\geq 1$, and the $(M, N_1, N_2)=(4,1,3)$ MIMO BC with $(\beta_1,\beta_2)=(1/4,1/2)$ for the case $\beta_1+\beta_2<1$.

\section{Example 1. $(M,N_1,N_2)=(5,2,3)$ with $(\beta_1,\beta_2)=(\frac{1}{2},\frac{2}{3})$}
For the two-user $(5,2,3)$ MIMO BC with  $(\frac{1}{2},\frac{2}{3})$ levels of partial CSIT, from {Theorem} \ref{theorem1} the DoF region is computed as,
\begin{align}
\mathcal{D}_1=\Bigl\{(d_1,d_2)\in\mathbb{R}^{2+}: &&d_1\le2,&&d_2\le3,&&\frac{d_1}{2}+\frac{d_2}{3}\le\frac{3}{2},&&d_1+d_2\le 3+\frac{7}{9}\Bigr\}\label{B3+}
\end{align}
and $\beta_o=\frac{7}{18}$ from \eqref{B5}. The challenge is to prove the bound  \eqref{B33}, i.e., $d_1+d_2\leq 3+7/9$.

\subsection{Deterministic Model}\label{DM_1}
The first step of the AIS approach is to transform the channel model into the deterministic setting, such that a DoF outer bound for the deterministic setting is also a DoF outer bound for the original channel.  This deterministic transformation produces a BC with input $\bar{\bf X}(t)=\bar{\bf X}_a(t)\bigtriangledown \bar{\bf X}_c(t)$, and outputs $\bar{\bf Y}_1(t), \bar{\bf Y}_2(t)$.
\begin{eqnarray}
%\bar{\bf Y}'_{1}(t)&=&\begin{bmatrix}\bar{Y}'_{11}(t)&\bar{Y}'_{12}(t)\end{bmatrix}^T\label{rre1}\\
%\bar{Y}'_{1r}(t)&=&L_{1r'(t)}^{h}\left(\bar{\mathbf{X}}_{a}(t)\right)+L_{1r'(t)}^{g}\left((\bar{\mathbf{X}}_{d}(t))^{1/2}\right),\forall r\in[2],t\in[n]\label{rre2}\\
%\bar{\bf Y}'_{2}(t)&=&\begin{bmatrix}\bar{Y}'_{21}(t)&\bar{Y}'_{22}(t)&\bar{Y}'_{23}(t)\end{bmatrix}^T\label{rre3}\\
%\bar{Y}'_{2r}(t)&=&L_{2r'(t)}^{h}\left(\bar{\mathbf{X}}_{d}(t)\right)+L_{2r'(t)}^{g}\left((\bar{\mathbf{X}}_{a}(t))^{1/3}\right),\forall r\in[3],t\in[n]\label{rre4}
\bar{\bf Y}_{1}(t)&=&\begin{bmatrix}\bar{Y}_{11}(t)&\bar{Y}_{12}(t)\end{bmatrix}^T\label{yl5}\label{xrre1BC'}\\
\bar{Y}_{1r}(t)&=&L_{1r(t)}^g\left(\bar{\mathbf{X}}_{a}(t)\bigtriangledown(\bar{\mathbf{X}}_{c}(t))^{1/2}\right),\forall r\in[2]\label{rre1BC'}
\\
\bar{\bf Y}_{2}(t)&=&\begin{bmatrix}\bar{Y}_{21}(t)&\bar{Y}_{22}(t)&\bar{Y}_{23}(t)\end{bmatrix}^T\label{yl6}\\
\bar{Y}_{2r}(t)&=&L_{2r(t)}^g\left((\bar{\mathbf{X}}_{a}(t))^{1/3}\bigtriangledown\bar{\mathbf{X}}_{c}(t)\right),\forall r\in[3]\label{rre2BC'}
\end{eqnarray}
 where $\bar{\mathbf{X}}_{a}(t)$ and $\bar{\mathbf{X}}_{c}(t)$ are defined as,
\begin{eqnarray}
\bar{\mathbf{X}}_{a}(t)&=&\begin{bmatrix}\bar{X}_{1}(t)&\bar{X}_{2}(t)\end{bmatrix}^T\\
\bar{\mathbf{X}}_{c}(t)&=&\begin{bmatrix}\bar{X}_{3}(t)&\bar{X}_{4}(t)&\bar{X}_{5}(t)\end{bmatrix}^T
\end{eqnarray}
and $\bar{X}_{m}(t)\in\{0, 1, \cdots, {\bar{P}}\}$, $\forall m\in[5]$.

\subsection{A Key Lemma} \label{essboundBC}
\begin{figure}[h] 
\centering
\includegraphics[width=\textwidth]{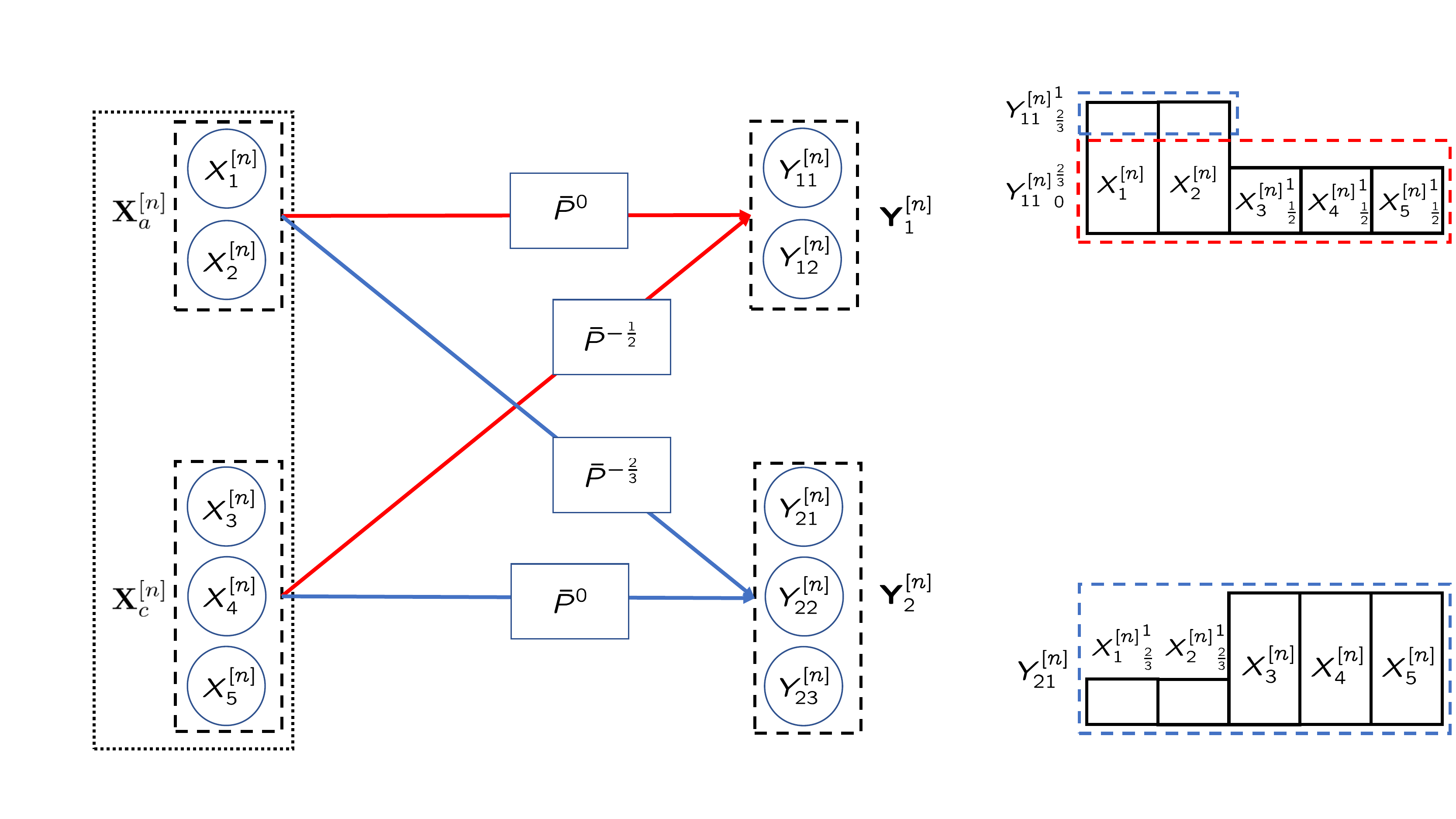}
\caption[]{Illustration corresponding to Lemma \ref{lemmaBC1}. }
\label{lemmaxx}
\end{figure}

The key to the proof of the bound, $d_1+d_2\le 3+\frac{7}{9}$, is the following lemma, which makes use of sumset inequalities from \cite{Arash_Jafar_sumset}.
\begin{lemma}\label{lemmaBC1} For the two-user  MIMO BC with $(M,N_1,N_2)=(5,2,3)$ and $(\beta_{1},\beta_2)=(\frac{1}{2},\frac{2}{3})$, 
\begin{align}
2H(\bar{\mathbf{Y}}^{[n]}_{2}\mid W_1,\mathcal{G})
&\le 3H(\bar{\mathbf{Y}}^{[n]}_{1}\mid W_1,\mathcal{G})-H((\bar{\mathbf{Y}}^{[n]}_{1})^{1/3} \mid W_1,\mathcal{G})+3n\log{\bar{P}}+n~o~(\log{\bar{P}})\label{firstlem}
\end{align}
\end{lemma}
See Figure \ref{lemmaxx} for an accompanying illustration for Lemma \ref{lemmaBC1}.
The proof of Lemma \ref{lemmaBC1} is presented in Appendix \ref{app1} .
\subsection{Proof of the Bound $d_1+d_2\le 3+\frac{7}{9}$}
\begin{enumerate}
\item{} Starting from Fano's Inequality for the first receiver and suppressing $no(\log(P))$ terms that are inconsequential for DoF, we have,
\begin{align}
3nR_1&\le 3I(\bar{\bf Y}^{[n]}_{1};W_1\mid\mathcal{G})\nonumber\\
&= 2H(\bar{\bf Y}^{[n]}_{1}\mid \mathcal{G})-3H(\bar{\bf Y}^{[n]}_{1}\mid W_1,\mathcal{G})+ H(\bar{\bf Y}^{[n]}_{1}\mid \mathcal{G})\\
%&=& 2H(\bar{\bf Y}^{[n]}_{1}\mid \mathcal{G})-2H(\bar{\bf Y}^{[n]}_{1}\mid W_1,\mathcal{G})\nonumber\\
%&&+ H((\bar{\bf Y}^{[n]}_{1})_{\frac{2}{3}},(\bar{\bf Y}^{[n]}_{1})^{1/3}\mid \mathcal{G})-H((\bar{\bf Y}^{[n]}_{1})_{\frac{2}{3}},(\bar{\bf Y}^{[n]}_{1})^{1/3}\mid W_1,\mathcal{G})\label{x1}\\
&= 2H(\bar{\bf Y}^{[n]}_{1}\mid \mathcal{G})-3H(\bar{\bf Y}^{[n]}_{1}\mid W_1,\mathcal{G})+ H((\bar{\bf Y}^{[n]}_{1})_{2/3},(\bar{\bf Y}^{[n]}_{1})^{1/3}\mid \mathcal{G})\label{x1}\\
%&=& 2H(\bar{\bf Y}^{[n]}_{1}\mid \mathcal{G})-2H(\bar{\bf Y}^{[n]}_{1}\mid W_1,\mathcal{G})\nonumber\\
%&&+ H( (\bar{\bf Y}^{[n]}_{1})^{1/3}\mid\mathcal{G})-H( (\bar{\bf Y}^{[n]}_{1})^{1/3}\mid W_1,\mathcal{G})\nonumber\\
%&&+ H((\bar{\bf Y}^{[n]}_{1})_{\frac{2}{3}}\mid (\bar{\bf Y}^{[n]}_{1})^{1/3},\mathcal{G})-H((\bar{\bf Y}^{[n]}_{1})_{\frac{2}{3}}\mid (\bar{\bf Y}^{[n]}_{1})^{1/3}, W_1,\mathcal{G})\label{x12}\\
&= 2H(\bar{\bf Y}^{[n]}_{1}\mid \mathcal{G})-3H(\bar{\bf Y}^{[n]}_{1}\mid W_1,\mathcal{G})+ H( (\bar{\bf Y}^{[n]}_{1})^{1/3}\mid\mathcal{G})+ H((\bar{\bf Y}^{[n]}_{1})_{2/3}\mid (\bar{\bf Y}^{[n]}_{1})^{1/3},\mathcal{G})\label{x12}\\
&\le \left(4+\frac{4}{3}\right)n\log{\bar{P}}-3H(\bar{\bf Y}^{[n]}_{1}\mid W_1,\mathcal{G})+ H( (\bar{\bf Y}^{[n]}_{1})^{1/3}\mid\mathcal{G})\label{x3}\\
&\le \left(7+\frac{4}{3}\right)n\log{\bar{P}}-2H(\bar{\bf Y}^{[n]}_{2}\mid W_1,\mathcal{G})+ H( (\bar{\bf Y}^{[n]}_{1})^{1/3}\mid\mathcal{G})-H( (\bar{\bf Y}^{[n]}_{1})^{1/3}\mid W_1,\mathcal{G})\label{x4}\\
&\le \left(7+\frac{4}{3}\right)n\log{\bar{P}}-2H(\bar{\bf Y}^{[n]}_{2}\mid W_1,\mathcal{G})+ H( (\bar{\bf Y}^{[n]}_{1})^{1/3}\mid W_2,\mathcal{G})\label{x5}
\end{align}
where (\ref{x1}) follows from Definition \ref{powerlevel} and   (\ref{x12})  from  the chain rule. (\ref{x3}) is implied by the fact that the entropy of a random variable is bounded by logarithm of the cardinality of its support, i.e., $2H(\bar{\bf Y}^{[n]}_{1}\mid \mathcal{G})\le 4n\log{\bar{P}}$ and $H((\bar{\bf Y}^{[n]}_{1})_{2/3}\mid (\bar{\bf Y}^{[n]}_{1})^{1/3},\mathcal{G})\le \frac{4}{3}n\log{\bar{P}}$.  (\ref{x4}) is obtained from  Lemma \ref{lemmaBC1}. \eqref{x5} follows from the property that for  independent random variables $B$ and $C$,  $I(A;B)\le I(A;B\mid C)$. As a result, we have $I( (\bar{\bf Y}^{[n]}_{1})^{1/3};W_1\mid\mathcal{G})\le I( (\bar{\bf Y}^{[n]}_{1})^{1/3};W_1\mid W_2,\mathcal{G})\le H( (\bar{\bf Y}^{[n]}_{1})^{1/3}\mid W_2,\mathcal{G})$.
\item{} Similarly, starting from  Fano's Inequality for the second receiver we have,
\begin{align}
3nR_2&\leq  I(\bar{\bf Y}^{[n]}_{2};W_2\mid\mathcal{G})+2I(\bar{\bf Y}^{[n]}_{2};W_2\mid W_1,\mathcal{G})\\
&\le H(\bar{\bf Y}^{[n]}_{2}\mid \mathcal{G})-H(\bar{\bf Y}^{[n]}_{2}\mid W_2,\mathcal{G})+2H(\bar{\bf Y}^{[n]}_{2}\mid W_1,\mathcal{G})\nonumber\\
&\le 3n\log{\bar{P}}-H(\bar{\bf Y}^{[n]}_{2}\mid W_2,\mathcal{G})+2H(\bar{\bf Y}^{[n]}_{2}\mid W_1,\mathcal{G}).\label{x8}
\end{align}
(\ref{x8}) is true for the same reason as (\ref{x3}), i.e., because the entropy of a discrete random variable is bounded by logarithm of the cardinality of its support, i.e., $H(\bar{\bf Y}^{[n]}_{2}\mid \mathcal{G})\le 3n\log{\bar{P}}$. 
\item{} Summing the inequalities \eqref{x5} and \eqref{x8} we obtain,
\begin{eqnarray}
&&3nR_1+3nR_2\nonumber\\
&\le&\left(10+\frac{4}{3}\right)n\log{\bar{P}}-H(\bar{\bf Y}^{[n]}_{2}\mid W_2,\mathcal{G})+ H( (\bar{\bf Y}^{[n]}_{1})^{1/3}\mid W_2,\mathcal{G})\\
&\le&\left(10+\frac{4}{3}\right)n\log{\bar{P}}\label{x9}
\end{eqnarray}
where \eqref{x9} is obtained by a direct application of the sumset inequalities of \cite{Arash_Jafar_sumset}, as explained below.
From \eqref{x9}, the sum DoF bound $d_1+d_2\le3+\frac{7}{9}$ follows immediately.
\item{} Finally, we explain how \eqref{x9} is implied by the sumset inequalities of \cite{Arash_Jafar_sumset}. Specifically, we need Theorem $4$ of \cite{Arash_Jafar_sumset} to prove that $H( (\bar{\bf Y}^{[n]}_{1})^{1/3}\mid W_2,\mathcal{G})\le H(\bar{\bf Y}^{[n]}_{2}\mid W_2,\mathcal{G})$.  {In order to apply the result of Theorem \ref{Theorem AIS04} (i.e., Theorem $4$ of \cite{Arash_Jafar_sumset}) to our setting, let us set  $M=1$, $K=2$, $\lambda_{1,1}=\lambda_{2,1}=1$, $l_1=l_2=1$, and $I_{11}=I_{21}=\{1\}$. Then, the inequality \eqref{dssd4} reduces to,}
\begin {eqnarray}
H(Z_1^{[n]},Z_2^{[n]}\mid W,\mathcal{G})&\geq& H(Z_{11}^{[n]},Z_{21}^{[n]}\mid W)+n~o(\log{\bar{P}}).\label{dssd4+}
\end{eqnarray}
Let us specialize $W=W_2$ and define $Z_{1}(t)$, $Z_{2}(t)$, $Z_{11}(t)$, $Z_{21}(t)$, $t\in[n]$, as,
\begin {align}
\bar{Y}_{21}(t)&=L_{21(t)}^g((\bar{X}_1(t))^{1/3},(\bar{X}_2(t))^{1/3}, \bar{X}_3(t),\bar{X}_4(t), \bar{X}_5(t))&=Z_1(t)\\
\bar{Y}_{22}(t)&=L_{22(t)}^g((\bar{X}_1(t))^{1/3},(\bar{X}_2(t))^{1/3}, \bar{X}_3(t),\bar{X}_4(t), \bar{X}_5(t))&=Z_2(t)\\
(\bar{Y}_{11}(t))^{1/3}&\doteq L_{z11(t)}^g((\bar{X}_1(t))^{1/3},(\bar{X}_2(t))^{1/3})&=Z_{11}(t)\label{eq:doteq1}\\
(\bar{Y}_{12}(t))^{1/3}&\doteq L_{z21(t)}^g(t)((\bar{X}_1(t))^{1/3},(\bar{X}_2(t))^{1/3})&=Z_{21}(t)\label{eq:doteq2}
\end{align}
Note that $L^g$ functions can be used instead of $L^h$ functions in $Z_{11}, Z_{21}$, because it only weakens the result of Theorem \ref{Theorem AIS04}. In other words,  Theorem \ref{Theorem AIS04} makes the stronger claim that \eqref{dssd4+} holds even if channel coefficients are chosen as arbitrary constants in $Z_{11}, Z_{21}$. Since the claim is true for arbitrary constants, it is also true for randomly chosen coefficients, i.e., $L^g$ functions may be used instead of $L^h$ functions in $Z_{11}, Z_{21}$. Next, consider the `$\doteq$' in \eqref{eq:doteq1} and \eqref{eq:doteq2}. This is justified as follows. 

Let us prove \eqref{eq:doteq1}, and \eqref{eq:doteq2} is similarly implied. In order to prove \eqref{eq:doteq1} we will show that $Z_{11}(t)=(\bar{Y}_{11}(t))^{1/3}-\delta_{11}(t)$ where $H(\delta_{11}(t))$ is bounded by a constant which does not scale with $P$. Since adding or subtracting bounded entropy noise terms can only make a difference of the order of $n o(\log( P))$ which is inconsequential in the DoF    sense, the $\doteq$ in \eqref{eq:doteq1} and \eqref{eq:doteq2} is justified.
\begin {eqnarray}
&&(\bar{Y}_{11}(t))^{1/3}\nonumber\\
&=&\left \lfloor \frac{\bar{Y}_{11}(t)-\bar{P}\left \lfloor\frac{\bar{Y}_{11}(t)}{ {\bar{P}}}\right \rfloor }{{\bar{P}}^{2/3}} \right \rfloor\\
&=&\left \lfloor \frac{\bar{Y}_{11}(t)}{{\bar{P}}^{2/3}} \right \rfloor-\left \lfloor \frac{\bar{P}\left \lfloor\frac{\bar{Y}_{11}(t)}{ {\bar{P}}}\right \rfloor }{{\bar{P}}^{2/3}} \right \rfloor+\delta_a(t)\\
&=&\left \lfloor \frac{\bar{Y}_{11}(t)}{{\bar{P}}^{2/3}} \right \rfloor-\delta_b(t)+\delta_a(t)\\
&=&\left \lfloor \frac{\lfloor g_1(t) X_1(t)\rfloor+\lfloor g_2(t) X_2(t)\rfloor+\sum_{k=3}^5\lfloor g_k(t) (X_k(t))^{1/2}\rfloor}{{\bar{P}}^{2/3}} \right \rfloor-\delta_b(t)+\delta_a(t)\\
&=&\lfloor g_1(t) (X_1(t))^{1/3}\rfloor+\lfloor g_2(t) (X_2(t))^{1/3}\rfloor+\delta_c(t)-\delta_b(t)+\delta_a(t)\\
&=&Z_{11}(t)+\delta_{11}(t)
\end{eqnarray}
where $\delta_{11}(t)=\delta_a(t)-\delta_b(t)+\delta_c(t)$. Here, $\delta_a(t)$ is a random variable which can only take values from the set $\{-1,0,1\}$ as $\lfloor A\rfloor+\lfloor B\rfloor-\lfloor A+B\rfloor\in\{-1,0,1\}$ for any real numbers $A$ and $B$. Next, consider $\delta_b$, whose entropy is bounded as follows.
\begin {eqnarray}
H(\delta_b)&\le& H\left(\left \lfloor\frac{\bar{Y}_{11}(t)}{ {\bar{P}}}\right \rfloor\right)\le O(1)\label{rrr+p}
\end{eqnarray}
\eqref{rrr+p} is true as $|\bar{Y}_{11}(t)|\le 5\Delta\bar{P}$. 
Similarly, $H(\delta_c)\leq O(1)$. 
%\begin {eqnarray}
%&&H(I_2)\nonumber\\
%&\le& H(\left \lfloor\frac{\bar{Y}_{11}(t)}{ {\bar{P}}}\right \rfloor)\\
%&\le& \log (5\Delta)\label{rrr+p}
%\end{eqnarray}
%\eqref{rrr+p} is true as $\bar{Y}_{11}(t)\le 5\Delta\bar{P}$ (Comment: $I_2$ is not bounded, and it scales with $\bar{P}$. However, $H(I_2)$ is bounded and does not scale with $\bar{P}$.).\\

Thus, from \eqref{dssd4+} we have,
\begin{eqnarray}
H(Z_{11}^{[n]},Z_{21}^{[n]}\mid W_2,\mathcal{G})&\leq& H(Z_{1}^{[n]},Z_{2}^{[n]}\mid W_2,\mathcal{G})+n~o(\log{\bar{P}})\\
\Rightarrow H( (\bar{\bf Y}^{[n]}_{1})^{1/3}\mid W_2,\mathcal{G})&\le& H(\bar{ Y}^{[n]}_{21},\bar{ Y}^{[n]}_{22}\mid W_2,\mathcal{G})+n~o(\log{\bar{P}})\\
&\le& H(\bar{\bf Y}^{[n]}_{2}\mid W_2,\mathcal{G})+n~o(\log{\bar{P}}).
\end{eqnarray}

\end{enumerate}

\section{Example 2. $(M,N_1,N_2)=(4,1,3)$ with $(\beta_1,\beta_2)=(\frac{1}{4},\frac{1}{2})$}
For the two-user$(M,N_1,N_2)=(4,1,3)$ with $(\beta_1,\beta_2)=(\frac{1}{4},\frac{1}{2})$ levels of partial CSIT, from {Theorem} \ref{theorem1} the DoF region is computed as,
\begin{align}
\mathcal{D}_1=\Bigl\{(d_1,d_2)\in\mathbb{R}^{2+}: &&d_1\le1,&&d_2\le3,&&{d_1}{}+\frac{d_2}{3}\le\frac{5}{4},&&d_1+d_2\le 3+\frac{1}{16}\Bigr\}\label{B3+}
\end{align}
and $\beta_o=\frac{1}{16}$ from \eqref{B5}. The challenge is to prove the bound  \eqref{B33}, i.e., $d_1+d_2\leq 3+\frac{1}{16}$.

\subsection{Deterministic Model}\label{DM_2}
Similar to Section \ref{DM_1}, the deterministic transformation produces a BC with input $\bar{\bf X}(t)=\bar{\bf X}_a(t)\bigtriangledown\bar{\bf X}_c(t)$, and outputs $\bar{\bf Y}_1(t), \bar{\bf Y}_2(t)$.
\begin{eqnarray}
\bar{Y}_{1}(t)&=&L_{1(t)}^g\left(\bar{\mathbf{X}}_{a}(t)\bigtriangledown(\bar{\mathbf{X}}_{c}(t))^{3/4}\right)\label{rre1BC'...}
\\
\bar{\bf Y}_{2}(t)&=&\begin{bmatrix}\bar{Y}_{21}(t)&\bar{Y}_{22}(t)\end{bmatrix}^T\label{yl6...}\\
\bar{Y}_{2r}(t)&=&L_{2r(t)}^g\left((\bar{\mathbf{X}}_{a}(t))^{1/2}\bigtriangledown\bar{\mathbf{X}}_{c}(t)\right),\forall r\in[2]\label{rre2BC'...}
\end{eqnarray}
 where $\bar{\mathbf{X}}_{a}(t)$ and $\bar{\mathbf{X}}_{c}(t)$ are defined as,
\begin{eqnarray}
\bar{\mathbf{X}}_{a}(t)&=&\bar{X}_{1}(t)\\
\bar{\mathbf{X}}_{c}(t)&=&\begin{bmatrix}\bar{X}_{2}(t)&\bar{X}_{2}(t)&\bar{X}_{3}(t)\end{bmatrix}^T
\end{eqnarray}
and $\bar{X}_{m}(t)\in\{0, 1, \cdots, {\bar{P}}\}$, $\forall m\in[4]$. 

\subsection{A Key Lemma} \label{essboundBC}
\begin{figure}[h] 
\centering
\includegraphics[width=\textwidth]{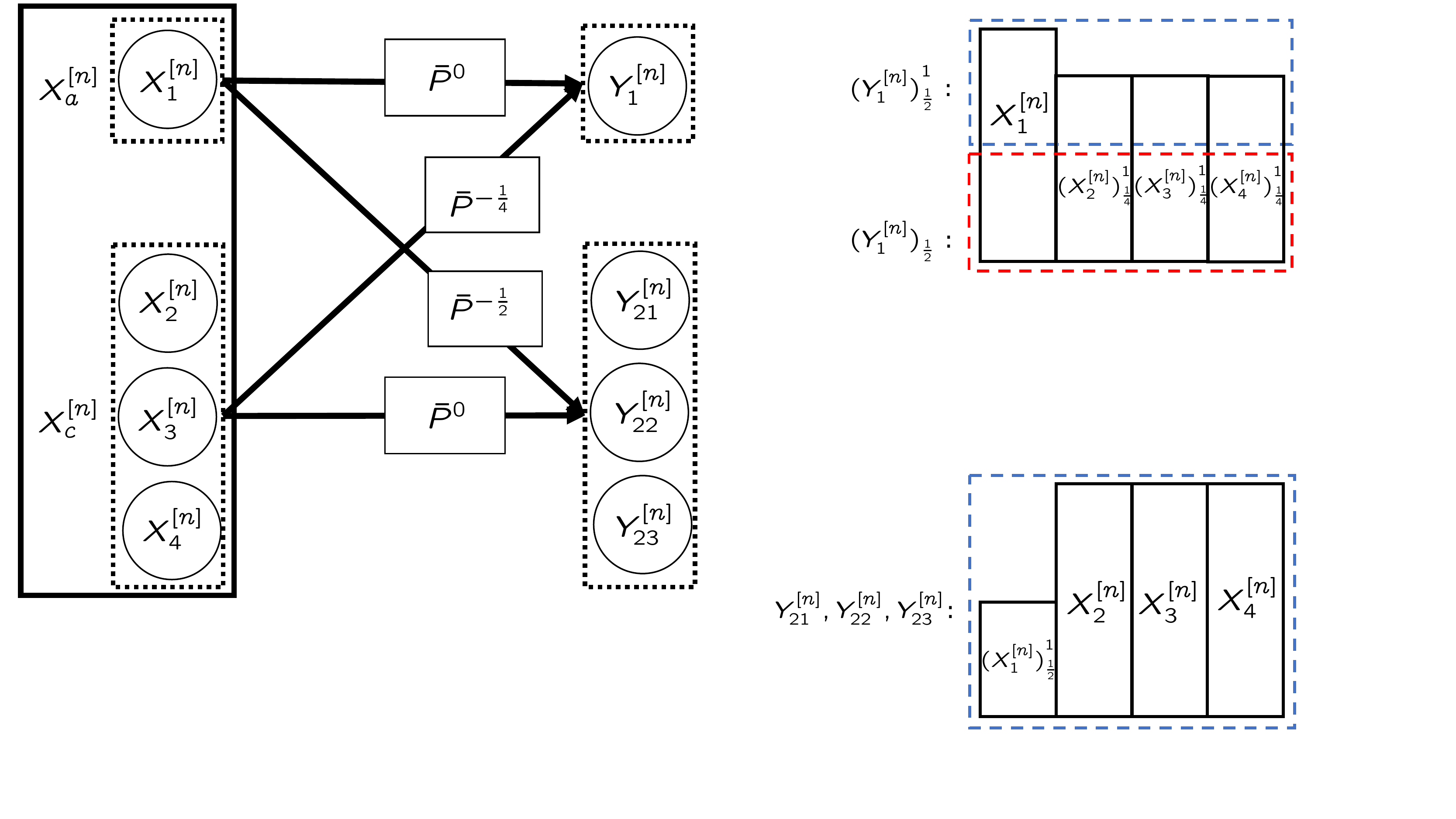}
\caption[]{Illustration corresponding to Lemma \ref{lemmaBC1...}. }
\label{lemmaxx...}
\end{figure}

To prove the bound $d_1+d_2\le 3+\frac{1}{16}$ we need the following lemma.
\begin{lemma}\label{lemmaBC1...} For the two-user  MIMO BC with $(M,N_1,N_2)=(4,1,3)$ and $(\beta_{1},\beta_2)=(\frac{1}{4},\frac{1}{2})$, 
\begin{align}
H(\bar{\mathbf{Y}}^{[n]}_{2}\mid W_1,\mathcal{G})
&\le 4H(\bar{{Y}}^{[n]}_{1}\mid W_1,\mathcal{G})-3H((\bar{{Y}}^{[n]}_{1})^{1/2} \mid W_1,\mathcal{G})+\frac{3}{4}n\log{\bar{P}}+n~o~(\log{\bar{P}})\label{firstlem...}
\end{align}
\end{lemma}
See Figure \ref{lemmaxx...} for an accompanying illustration for Lemma \ref{lemmaBC1...}.
The proof of Lemma \ref{lemmaBC1...} is presented in Appendix \ref{appendix1BC.}.
\subsection{Proof of the Bound $d_1+d_2\le3+\frac{1}{16}$}

\begin{enumerate}
\item{} Starting with Fano's Inequality for the first receiver, we have,
\begin{eqnarray}
&&4nR_1\nonumber\\
&\le& H(\bar{  Y}^{[n]}_{1}\mid \mathcal{G})-H(\bar{  Y}^{[n]}_{1}\mid W_1,\mathcal{G})+ 3H(\bar{  Y}^{[n]}_{1}\mid \mathcal{G})-3H(\bar{  Y}^{[n]}_{1}\mid W_1,\mathcal{G})\\
&=& H(\bar{  Y}^{[n]}_{1}\mid \mathcal{G})-H(\bar{  Y}^{[n]}_{1}\mid W_1,\mathcal{G})\nonumber\\
&&+ 3H((\bar{  Y}^{[n]}_{1})_{1/2},(\bar{  Y}^{[n]}_{1})^{1/2}\mid \mathcal{G})-3H((\bar{  Y}^{[n]}_{1})_{1/2},(\bar{  Y}^{[n]}_{1})^{1/2}\mid W_1,\mathcal{G})\label{BCr1.}\\
&=& H(\bar{  Y}^{[n]}_{1}\mid \mathcal{G})-H(\bar{  Y}^{[n]}_{1}\mid W_1,\mathcal{G})\nonumber\\
&&+ 3H( (\bar{  Y}^{[n]}_{1})^{1/2}\mid\mathcal{G})-3H( (\bar{  Y}^{[n]}_{1})^{1/2}\mid W_1,\mathcal{G})\nonumber\\
&&+ 3H((\bar{  Y}^{[n]}_{1})_{1/2}\mid (\bar{  Y}^{[n]}_{1})^{1/2},\mathcal{G})-3H((\bar{  Y}^{[n]}_{1})_{1/2}\mid (\bar{  Y}^{[n]}_{1})^{1/2}, W_1,\mathcal{G})\label{BCr2.}\\
&\le& \frac{5}{2}n\log{\bar{P}}-H(\bar{  Y}^{[n]}_{1}\mid W_1,\mathcal{G})-3H((\bar{  Y}^{[n]}_{1})_{1/2}\mid (\bar{  Y}^{[n]}_{1})^{1/2}, W_1,\mathcal{G})\nonumber\\
&&+ 3H( (\bar{  Y}^{[n]}_{1})^{1/2}\mid\mathcal{G})-3H( (\bar{  Y}^{[n]}_{1})^{1/2}\mid W_1,\mathcal{G})\label{BCr3.}\\
&\le& \frac{13}{4}n\log{\bar{P}}-H(\bar{  Y}^{[n]}_{21},\bar{  Y}^{[n]}_{22},\bar{  Y}^{[n]}_{23}\mid W_1,\mathcal{G})+ 3H( (\bar{  Y}^{[n]}_{1})^{1/2}\mid\mathcal{G})-3H( (\bar{  Y}^{[n]}_{1})^{1/2}\mid W_1,\mathcal{G})\nonumber\\
&&+n~o(\log{\bar{P}})\label{BCr4.}\\
&\le& \frac{13}{4}n\log{\bar{P}}-H(\bar{  Y}^{[n]}_{21},\bar{  Y}^{[n]}_{22},\bar{  Y}^{[n]}_{23}\mid W_1,\mathcal{G})+ 3H( (\bar{  Y}^{[n]}_{1})^{1/2}\mid W_2,\mathcal{G})+n~o(\log{\bar{P}})\label{rereeBC1-.}
\end{eqnarray}
where (\ref{BCr1}) follows from Definition \ref{powerlevel} and   (\ref{BCr2.})  is true from  the chain rule. (\ref{BCr3.}) is concluded as the entropy of a random variable is bounded by the logarithm of the cardinality of its support, i.e., $3H((\bar{  Y}^{[n]}_{1})_{1/2}\mid (\bar{  Y}^{[n]}_{1})^{1/2},\mathcal{G})\le \frac{3}{2}n\log{\bar{P}}$, $H(\bar{  Y}^{[n]}_{1}\mid \mathcal{G})\le n\log{\bar{P}}$.  (\ref{BCr4.}) is obtained as from Lemma \ref{lemmaBC1...} and the chain rule we have 
\begin{eqnarray}
&&H(\bar{  Y}^{[n]}_{21},\bar{  Y}^{[n]}_{22},\bar{  Y}^{[n]}_{23}\mid W_1,\mathcal{G})\nonumber\\
&\le&4H(\bar{{Y}}^{[n]}_{1}\mid W_1,\mathcal{G})-3H((\bar{{Y}}^{[n]}_{1})^{1/2} \mid W_1,\mathcal{G})+\frac{3}{4}n\log{\bar{P}}+n~o~(\log{\bar{P}})\nonumber\\
&=&H(\bar{{Y}}^{[n]}_{1}\mid W_1,\mathcal{G})+3H((\bar{{Y}}^{[n]}_{1})_{1/2} \mid (\bar{{Y}}^{[n]}_{1})^{1/2},W_1,\mathcal{G})+\frac{3}{4}n\log{\bar{P}}+n~o~(\log{\bar{P}})
\end{eqnarray}
 \eqref{rereeBC1-.} is true because $I(A;B)\le I(A;B\mid C)$ when $B$ is independent of $C$. As a result, we have $I( (\bar{  Y}^{[n]}_{1})^{1/2};W_1\mid\mathcal{G})\le H( (\bar{  Y}^{[n]}_{1})^{1/2}\mid W_2,\mathcal{G})$.
\item{} Similarly, writing Fano's Inequality for the second receiver we have,
\begin{eqnarray}
nR_2&\le& I(\bar{  Y}^{[n]}_{21},\bar{  Y}^{[n]}_{22},\bar{  Y}^{[n]}_{23};W_2\mid\mathcal{G})\nonumber\\
&=&H(\bar{  Y}^{[n]}_{21},\bar{  Y}^{[n]}_{22},\bar{  Y}^{[n]}_{23}\mid \mathcal{G})-H(\bar{  Y}^{[n]}_{21},\bar{  Y}^{[n]}_{22},\bar{  Y}^{[n]}_{23}\mid W_2,\mathcal{G})\label{rereeBC2.}\\
nR_2&\le& I(\bar{  Y}^{[n]}_{21},\bar{  Y}^{[n]}_{22},\bar{  Y}^{[n]}_{23};W_2\mid W_1,\mathcal{G})\nonumber\\
&=&H(\bar{  Y}^{[n]}_{21},\bar{  Y}^{[n]}_{22},\bar{  Y}^{[n]}_{23}\mid W_1,\mathcal{G})\label{rereeBC3.}
\end{eqnarray}
Scaling \eqref{rereeBC2.} and \eqref{rereeBC3.} by $3$ and $1$ respectively, and summing them together we have,
\begin{eqnarray}
&&4nR_2\nonumber\\
&\le&3H(\bar{  Y}^{[n]}_{21},\bar{  Y}^{[n]}_{22},\bar{  Y}^{[n]}_{23}\mid \mathcal{G})-3H(\bar{  Y}^{[n]}_{21},\bar{  Y}^{[n]}_{22},\bar{  Y}^{[n]}_{23}\mid W_2,\mathcal{G})+H(\bar{  Y}^{[n]}_{21},\bar{  Y}^{[n]}_{22},\bar{  Y}^{[n]}_{23}\mid W_1,\mathcal{G})\nonumber\\
&\le&9n\log{\bar{P}}-3H(\bar{  Y}^{[n]}_{21},\bar{  Y}^{[n]}_{22},\bar{  Y}^{[n]}_{23}\mid W_2,\mathcal{G})+H(\bar{  Y}^{[n]}_{21},\bar{  Y}^{[n]}_{22},\bar{  Y}^{[n]}_{23}\mid W_1,\mathcal{G})\label{rereeBC4.}
\end{eqnarray}
(\ref{rereeBC4.}) is concluded similar to (\ref{BCr3.}) as the entropy of a random variable is bounded by logarithm of the cardinality of its support, i.e., $H(\bar{  Y}^{[n]}_{21},\bar{  Y}^{[n]}_{22},\bar{  Y}^{[n]}_{23}\mid \mathcal{G})\le 2n\log{\bar{P}}$. 
\item{} Summing the inequalities \eqref{rereeBC1-.} and \eqref{rereeBC4.} results in,
\begin{eqnarray}
&&4nR_1+4nR_2\nonumber\\
&\le&\frac{49}{4}n\log{\bar{P}}-3H(\bar{  Y}^{[n]}_{21},\bar{  Y}^{[n]}_{22},\bar{  Y}^{[n]}_{23}\mid W_2,\mathcal{G})+ 3H( (\bar{  Y}^{[n]}_{1})^{1/2}\mid W_2,\mathcal{G})+n~o(\log{\bar{P}})\\
&\le&\frac{49}{4}n\log{\bar{P}}+n~o(\log{\bar{P}})\label{rereeBC5.}
\end{eqnarray}
Dividing (\ref{rereeBC5.}) by $4\log{\bar{P}}$, $d_1+d_2\le3+\frac{1}{16}$ is obtained.
\item{} The justification for (\ref{rereeBC5.}) is similar to \eqref{x9} as $(\bar{  Y}_{1}(t))^{1/2}$ is a bounded density linear combination of random variables  $(\bar{{X}}_{1}(t))^{1/2},(\bar{{X}}_{2}(t))^1_{\frac{3}{4}},(\bar{{X}}_{3}(t))^1_{\frac{3}{4}},(\bar{{X}}_{4}(t))^1_{\frac{3}{4}}$ while  $\bar{  Y}_{21}(t)$ is a bounded density linear combination of random variables  $(\bar{{X}}_{1}(t))^{1/2},\bar{{X}}_{2}(t),\bar{{X}}_{3}(t),\bar{{X}}_{4}(t)$. Thus, we have
\begin{eqnarray}
&&H( (\bar{  Y}^{[n]}_{1})^{1/2}\mid W_2,\mathcal{G})\nonumber\\
&\le&H(\bar{  Y}^{[n]}_{21}\mid W_2,\mathcal{G})+n~o(\log{\bar{P}})\\
&\le&H(\bar{  Y}^{[n]}_{21},\bar{  Y}^{[n]}_{22},\bar{  Y}^{[n]}_{23}\mid W_2,\mathcal{G})+n~o(\log{\bar{P}}).
\end{eqnarray}

 \end{enumerate}

\section{Proof of Theorem \ref{theorem1}}\label{proofthe}
To prove Theorem \ref{theorem1} we only need to prove the outer bound \eqref{B33}. The proof for the general setting follows closely along the lines of the examples presented above. We start, as before, with the corresponding deterministic model.
\subsection{Deterministic Channel Model}
For all $t\in[n]$, the channel outputs in the deterministic model are $\bar{\bf Y}_{1}(t)$ and $\bar{\bf Y}_{2}(t)$, which are defined as follows.
\begin{eqnarray}
\bar{\bf Y}_{1}(t)&=&\begin{bmatrix}\bar{Y}_{11}(t)&\bar{Y}_{12}(t)&\cdots&\bar{Y}_{1N_1}(t)\end{bmatrix}^T\label{yl1}\\
\bar{Y}_{1r}(t)&=&L_{1r}^g(t)\left(\bar{\mathbf{X}}_{a}(t)\bigtriangledown\bar{\mathbf{X}}_{b}(t)\bigtriangledown(\bar{\mathbf{X}}_{c}(t))^{1}_{\beta_1}\right),\forall r\in[N_1]\label{yl2}
\\
\bar{\bf Y}_{2}(t)&=&\begin{bmatrix}\bar{Y}_{21}(t)&\bar{Y}_{22}(t)&\cdots&\bar{Y}_{2N_2}(t)\end{bmatrix}^T\label{yl3}\\
\bar{Y}_{2r}(t)&=&L_{2r}^g(t)\left((\bar{\mathbf{X}}_{a}(t))^{1-\beta_2}\bigtriangledown\bar{\mathbf{X}}_{b}(t)\bigtriangledown\bar{\mathbf{X}}_{c}(t)\right),\forall r\in[N_2]\label{yl4}
%{\bar{\bf Y}_{2e}(t)}&=&{\begin{bmatrix}\bar{Y}_{21}(t)&\bar{Y}_{22}(t)&\cdots&\bar{Y}_{2N_1}(t)\end{bmatrix}^T}
\end{eqnarray}
%{(What is  $Y_{2e}$ and why is it suddenly defined here, when it is neither introduced nor otherwise mentioned in this section? Don't confuse the reader. Explain what, why.)\\}

\noindent and $\bar{\bf X}_{a}(t),\bar{\bf X}_{b}(t),{\bar{\bf X}_{c}(t)}$ are defined as
\begin{eqnarray}
\bar{\bf X}(t)&=&[\bar{X}_{1}(t),\bar{X}_{2}(t),\cdots,\bar{X}_{M}(t)]\\
\bar{\bf X}_{a}(t)&=&[\bar{\bf X}(t)]_{0\rightarrow M-N_2}\\
\bar{\bf X}_{b}(t)&=&[\bar{\bf X}(t)]_{M-N_2\rightarrow N_1+N_2-M}\\
{\bar{\bf X}_{c}(t)}&=&[\bar{\bf X}(t)]_{N_1\rightarrow M-N_1}
\end{eqnarray}
and the random variables $\bar{X}_{m}(t)$ take values from the set $\{0, 1, \cdots, {\bar{P}}\}$, i.e.,
\begin{eqnarray}
\bar{X}_{m}(t)\in\{0, 1, \cdots, {\bar{P}}\}, \forall m\in[M],t\in[n].\label{0top}
\end{eqnarray}

\subsection{ Useful  Lemma} 
 The following lemma from \cite{Arash_Jafar_KMIMOIC} will be useful, and is reproduced here for the sake of completeness.
 \begin{lemma}\label{lemmaMIMOx}[$N_1\le N_2$ \cite{Arash_Jafar_KMIMOIC}] Define the two random variables $\bar{\bf U}_1$ and  $\bar{\bf U}_2$ as,
\begin{eqnarray}
\bar{\bf U}_1&=&\left({U}_{11}^{[n]},{U}_{12}^{[n]},\cdots,{U}_{1N_1}^{[n]}\right)\label{lemmamimox1}\\
\bar{\bf U}_2&=&\left({U}_{21}^{[n]},{U}_{22}^{[n]},\cdots,{U}_{2N_2}^{[n]}\right)\label{lemmamimox2}
\end{eqnarray}
where for any $t\in[n]$, $U_{1j}(t)$ and $U_{2j}(t)$ are defined as,
\begin{eqnarray}
U_{1j}(t)&=&L_{1j}^g(t)\left((\bar{\mathbf{V}}_1(t))^{\eta}_{\eta-\lambda_{11}}\bigtriangledown(\bar{\mathbf{V}}_2(t))^{\eta}_{\eta-\lambda_{12}}\bigtriangledown\cdots\bigtriangledown(\bar{\mathbf{V}}_l(t))^{\eta}_{\eta-\lambda_{1l}}\right), \forall j\in[N_1]\label{lemmamimox3}\\
U_{2j}(t)&=&L_{2j}^g(t)\left((\bar{\mathbf{V}}_1(t))^{\eta}_{\eta-\lambda_{21}}\bigtriangledown(\bar{\mathbf{V}}_2(t))^{\eta}_{\eta-\lambda_{22}}\bigtriangledown\cdots\bigtriangledown(\bar{\mathbf{V}}_l(t))^{\eta}_{\eta-\lambda_{2l}}\right), \forall j\in[N_2]\label{lemmamimox4}
\end{eqnarray}
where $\bar{\mathbf{V}}_i(t)=\begin{bmatrix}\bar{V}_{i1}(t)&\cdots&\bar{V}_{iM_i}(t)\end{bmatrix}^T$, $\bar{V}_{im}(t)\in\mathcal{X}_{\eta}$  are all independent of $\mathcal{G}$, and $0\le\lambda_{1i},\lambda_{2i}\le\eta$ for all $i\in[l]$.  Without loss of generality,  $(\lambda_{1i}-\lambda_{2i})^+$ are sorted in descending order, i.e., $(\lambda_{1i}-\lambda_{2i})^+\ge(\lambda_{1i'}-\lambda_{2i'})^+$ if $1\le i< i'\le l$.  Then, for any acceptable\footnote{Let $\mathcal{G}(U)\subset\mathcal{G}$ denote the set of all bounded density channel coefficients that appear in $\bar{\bf U}_1,\bar{\bf U}_2$.  $W$ is acceptable if  conditioned on any $\mathcal{G}_o\subset (\mathcal{G}/\mathcal{G}(U))\cup \{W\}$, the channel coefficients $\mathcal{G}(U)$ satisfy the bounded density assumption. For instance, any random variable $W$ independent of $\mathcal{G}$ can be utilized in  Lemma \ref{lemmaMIMOx}.} random variable ${W}$, if $N_1\le \min(N_2, \sum_{i=1}^lM_i)$, then we have,
\begin{eqnarray}
&&H({\bar{\bf U}}_1\mid {W},\mathcal{G})-H({\bar{\bf U}}_2\mid {W},\mathcal{G})\nonumber\\
&\le&n\big((N_1-\sum_{i=1}^sM_i)(\lambda_{1,s+1}-\lambda_{2,s+1})^++\sum_{i=1}^sM_i(\lambda_{1i}-\lambda_{2i})^+\big)\log{\bar{P}}+n~o~(\log{\bar{P}})\label{lemmamimox5}
\end{eqnarray}
where $s$ must satisfy the condition  $\sum_{i=1}^{s}M_i\le N_1< \sum_{i=1}^{s+1}M_i$.  
\end{lemma} 
%For proof of Lemma \ref{lemmaMIMOx} see \cite{Arash_Jafar_KMIMOIC}.

%\subsection{Application of Lemma \ref{lemmaMIMOx}}
%To gain some insights into the application of Lemma \ref{lemmaMIMOx}, let us derive an outer bound for the term $H(\bar{\bf Y}^{[n]}_{2e}\mid W_1,\mathcal{G})-H(\bar{\bf Y}^{[n]}_{1}\mid W_1,\mathcal{G})$ as follows. To apply Lemma \ref{lemmaMIMOx}, the random variables  $\bar{\bf U}_1$, $\bar{\bf U}_2$, $\bar{\bf V}_1^{[n]}$, $\bar{\bf V}_2^{[n]}$, $\bar{\bf V}_3^{[n]}$, $\bar{\bf V}_4^{[n]}$ and $W$ are interpreted as $\bar{\bf Y}_{2e}^{[n]}$, $\bar{\bf Y}_1^{[n]}$, $\bar{\bf X}_d^{[n]}$, $\bar{\bf X}_c^{[n]}$, $\bar{\bf X}_a^{[n]}$, $\bar{\bf X}_b^{[n]}$ and $W_1$, respectively. $\bar{\bf Y}_{2e}(t)$ receives the top $1$ power levels of $\bar{\bf X}_d^{[n]}$ while the first receiver sees  the top $1-\beta_1$ power levels of $\bar{\bf X}_d^{[n]}$.  So we have $\eta=1, \lambda_{11}=1, \lambda_{21}=1-\beta_1$. Therefore, $(\lambda_{11}-\lambda_{21})^+=\beta_1$. From Lemma \ref{lemmaMIMOx}, \eqref{yl2} and \eqref{yl4}, we conclude that
%\begin{eqnarray}
%H(\bar{\bf Y}_{2e}^{[n]}\mid W_1,\mathcal{G})-H(\bar{\bf Y}_1^{[n]}\mid W_1,\mathcal{G})
%&\le&nN_1\beta_1\log{\bar{P}}+n~o~(\log{\bar{P}})\label{app++1}
%\end{eqnarray}}

\subsection{Split $\bar{\bf Y}_1$ into $\bar{\bf Y}_{1a}, \bar{\bf Y}_{1\tilde{a}}$}
Since $\bar{\mathbf{X}}_{a}$ is a vector random variable of size $M-N_2$, using a change of basis operation at the receiver, the $N_1$ dimensional vector $\bar{\bf Y}_1$ can be partitioned into $\bar{\bf Y}_{1\tilde{a}}$, which is its projection into the $N_1+N_2-M$ dimensional space that does not contain $\bar{\mathbf{X}}_{a}(t)$, and a projection $\bar{\bf Y}_{1{a}}$ into the $M-N_2$ dimensional space that contains $\bar{\mathbf{X}}_{a}(t)$.
\begin{eqnarray}
\bar{\bf Y}_{1\tilde a}(t)&=&\begin{bmatrix}\bar{Y}_{11\tilde a}(t)&\bar{Y}_{12\tilde a}(t)&\cdots&\bar{Y}_{1N_1+N_2-M\tilde a}(t)\end{bmatrix}^T\label{e1}\\
\bar{Y}_{1r\tilde a}(t)&=&L_{1r\tilde a}^g(t)\left(\bar{\mathbf{X}}_{b}(t)\bigtriangledown(\bar{\mathbf{X}}_{c}(t))^{1}_{\beta_1}\right),\forall r\in[N_1+N_2-M]\label{e2}
\\
\bar{\bf Y}_{1a}(t)&=&\begin{bmatrix}\bar{Y}_{11a}(t)&\bar{Y}_{12a}(t)&\cdots&\bar{Y}_{1M-N_2a}(t)\end{bmatrix}^T\label{e3}\\
\bar{Y}_{1ra}(t)&=&L_{1ra}^g(t)\left(\bar{\mathbf{X}}_{a}(t)\bigtriangledown\bar{\mathbf{X}}_{b}(t)\bigtriangledown(\bar{\mathbf{X}}_{c}(t))^{1}_{\beta_1}\right),\forall r\in[M-N_2].\label{e4}
\end{eqnarray}

\subsubsection{Proof of bound \eqref{B33} when $\beta_1+\beta_2\ge1$}\label{sec:b1b2>1}
When $\beta_1+\beta_2\ge1$, the bound \eqref{B33} reduces to,
\begin{eqnarray}
d_1+d_2&\le& N_2+(M-N_2)\beta_o\label{B33++}
\end{eqnarray}
where $\beta_o$ is equal to,
\begin{eqnarray}
\beta_o&=
&\frac{N_1-N_2+(N_2-N_1)\beta_2+(M-N_1)\beta_1}{M-N_1}\label{betao}
\end{eqnarray}
Corresponding to Lemma \ref{lemmaBC1}, in this general setting we need the following lemma which is the key to the proof of the outer bound.
\begin{lemma}\label{lemmabeta>1} For the two-user $(M,N_1,N_2)$ MIMO BC with partial CSIT where $\beta_{1}+\beta_{2}\ge1$, we have, 
\begin{eqnarray}
&&\hat{N}_1H(\bar{\bf Y}^{[n]}_{2}\mid W_1,\mathcal{G})\nonumber\\
%&{\le}&\hat{N}_1H(\bar{\bf Y}^{n}_{1}\mid W_1,\mathcal{G})+\hat{N}_2H((\bar{\bf Y}^{n}_{1a})_{\beta_2}\mid\bar{\bf Y}_{1\tilde a}^{n},(\bar{\bf Y}^{n}_{1a})^{1-\beta_2},W_1,\mathcal{G})+n\hat{N}_0\log{\bar{P}}+n~o~(\log{\bar{P}})\nonumber\\
&{\le}&(\hat{N}_1+\hat{N}_2)H(\bar{\bf Y}^{n}_{1}\mid W_1,\mathcal{G})-\hat{N}_2H(\bar{\bf Y}_{1\tilde a}^{n},(\bar{\bf Y}^{n}_{1a})^{1-\beta_2}\mid W_1,\mathcal{G})+n\hat{N}_0\log{\bar{P}}+n~o~(\log{\bar{P}})\nonumber\\
&&\label{lem>1}
\end{eqnarray}
where the numbers $\hat{N}_0$, $\hat{N}_1$ and $\hat{N}_2$ are defined as,
\begin{eqnarray}
\hat{N}_0&=&\hat{N}_1(\hat{N}_2+\hat{N}_1)\beta_1\\
\hat{N}_1&=&M-N_2\\
\hat{N}_2&=&N_2-N_1
\end{eqnarray} 
\end{lemma}
See Appendix \ref{appbeta>1} for proof of Lemma \ref{lemmabeta>1}. Now, let us prove the bound \eqref{B33++}.
\begin{enumerate}
\item{} Starting with Fano's Inequality for the first receiver and suppressing $n o(\log(P))$ terms that are inconsequential for DoF, we have,
\begin{align}
&\hspace{-0.36in}n(\hat{N}_1+\hat{N}_2)R_1\nonumber\\
&\hspace{-0.36in}\le (\hat{N}_1+\hat{N}_2) I(W_1; \bar{ \bf Y}^{[n]}_{1}\mid \mathcal{G})\\
&\hspace{-0.36in}= \hat{N}_1H(\bar{ \bf Y}^{[n]}_{1}\mid \mathcal{G})-(\hat{N}_1+\hat{N}_2)H(\bar{ \bf Y}^{[n]}_{1}\mid W_1,\mathcal{G})+ \hat{N}_2H(\bar{ \bf Y}^{[n]}_{1}\mid \mathcal{G})\\
%--
&\hspace{-0.36in}= \hat{N}_1H(\bar{ \bf Y}^{[n]}_{1}\mid \mathcal{G})-(\hat{N}_1+\hat{N}_2)H(\bar{ \bf Y}^{[n]}_{1}\mid W_1,\mathcal{G})+ \hat{N}_2H((\bar{ \bf Y}^{[n]}_{1a})_{\beta_2},(\bar{ \bf Y}^{[n]}_{1a})^{1-\beta_2},\bar{ \bf Y}^{[n]}_{1\tilde a}\mid \mathcal{G})\label{BCr1}\\
%--
&\hspace{-0.36in}\le n\hat{N}_1N_1\log(\bar{P})-(\hat{N}_1+\hat{N}_2)H(\bar{ \bf Y}^{[n]}_{1}\mid W_1,\mathcal{G})+ \hat{N}_2H( \bar{ \bf Y}^{[n]}_{1\tilde a},(\bar{ \bf Y}^{[n]}_{1a})^{1-\beta_2}\mid\mathcal{G})\nonumber\\
&+ \hat{N}_2H((\bar{ \bf Y}^{[n]}_{1a})_{\beta_2}\mid \bar{ \bf Y}^{[n]}_{1\tilde a},(\bar{ \bf Y}^{[n]}_{1a})^{1-\beta_2},\mathcal{G})\label{BCr2}\\
%--
&\hspace{-0.36in}\le n(\hat{N}_1N_1+\hat{N}_1\hat{N_2}\beta_2)\log(\bar{P})-{(\hat{N}_1+\hat{N}_2)H(\bar{ \bf Y}^{[n]}_{1}\mid W_1,\mathcal{G})}+ \hat{N}_2H( \bar{ \bf Y}^{[n]}_{1\tilde a},(\bar{ \bf Y}^{[n]}_{1a})^{1-\beta_2}\mid\mathcal{G})\label{BCr3}\\
%--
&\hspace{-0.36in}\le  n(\hat{N}_1N_1+\hat{N}_1\hat{N_2}\beta_2)\log{\bar{P}}+{n\hat{N}_o\log(\bar{P})-\hat{N}_1H(\bar{ \bf Y}^{[n]}_{2}\mid W_1,\mathcal{G})-\hat{N}_2H( \bar{ \bf Y}^{[n]}_{1\tilde a},(\bar{ \bf Y}^{[n]}_{1a})^{1-\beta_2}\mid W_1,\mathcal{G})}\nonumber\\
&+ \hat{N}_2H( \bar{ \bf Y}^{[n]}_{1\tilde a},(\bar{ \bf Y}^{[n]}_{1a})^{1-\beta_2}\mid\mathcal{G})~~\label{BCr4}\\
%--
&\hspace{-0.36in}= n(\hat{N}_0+\hat{N}_1N_1+\hat{N}_1\hat{N_2}\beta_2)\log{\bar{P}}-\hat{N}_1H(\bar{ \bf Y}^{[n]}_{2}\mid W_1,\mathcal{G})+  \hat{N}_2I(W_1; \bar{ \bf Y}^{[n]}_{1\tilde a},(\bar{ \bf Y}^{[n]}_{1a})^{1-\beta_2}\mid \mathcal{G})\\
&\hspace{-0.36in}\le n(\hat{N}_0+\hat{N}_1N_1+\hat{N}_1\hat{N_2}\beta_2)\log{\bar{P}}-\hat{N}_1H(\bar{ \bf Y}^{[n]}_{2}\mid W_1,\mathcal{G})+  \hat{N}_2H( \bar{ \bf Y}^{[n]}_{1\tilde a},(\bar{ \bf Y}^{[n]}_{1a})^{1-\beta_2}\mid W_2,\mathcal{G})\label{rereeBC1-}
\end{align}
Here (\ref{BCr1}) follows from Definition \ref{powerlevel}. The chain rule of entropy, and the fact that since User $1$ has only $N_1$ antennas, the entropy of ${\bar{\bf Y}_1^{[n]}}$ cannot be more than $nN_1\log(\bar{P})+no(\log(P))$, justifies \eqref{BCr2}. Similarly, (\ref{BCr3}) is obtained because the entropy of a random variable is bounded by logarithm of the cardinality of its support, i.e., $\hat{N}_2H((\bar{ \bf Y}^{[n]}_{1b})_{\beta_2}\mid \bar{ \bf Y}^{[n]}_{1a}, (\bar{ \bf Y}^{[n]}_{1b})^{1-\beta_2},\mathcal{G})\le n\hat{N}_2\hat{N}_1\beta_2\log{\bar{P}} + n o(\log({P}))$.  Applying Lemma \ref{lemmabeta>1} to \eqref{BCr3} produces \eqref{BCr4}. Finally, \eqref{rereeBC1-} is true because $I(A;B)\leq I(A;B|C)\leq H(B\mid C)$ whenever $A$ is independent of $C$.
\item{} Similarly, starting with Fano's Inequality for the second receiver we have,
\begin{eqnarray}
&&n(\hat{N}_1+\hat{N}_2)R_2\nonumber\\
&\le &\hat{N}_1I(\bar{ \bf Y}^{[n]}_{2};W_2\mid W_1,\mathcal{G})+ \hat{N}_2I(\bar{ \bf Y}^{[n]}_{2};W_2\mid\mathcal{G})\\
&\le&\hat{N}_1H(\bar{ \bf Y}^{[n]}_{2}\mid W_1,\mathcal{G})+\hat{N}_2H(\bar{ \bf Y}^{[n]}_{2}\mid \mathcal{G})-\hat{N}_2H(\bar{ \bf Y}^{[n]}_{2}\mid W_2,\mathcal{G})\nonumber\\
&\le&\hat{N}_1H(\bar{ \bf Y}^{[n]}_{2}\mid W_1,\mathcal{G})+n\hat{N}_2N_2\log{\bar{P}}-\hat{N}_2H(\bar{ \bf Y}^{[n]}_{2}\mid W_2,\mathcal{G})\label{rereeBC4}
\end{eqnarray}
(\ref{rereeBC4}) is justified similarly as (\ref{BCr3}), as the entropy of a random variable is bounded by logarithm of the cardinality of its support, i.e., $H(\bar{ \bf Y}^{[n]}_{2}\mid \mathcal{G})\le nN_2\log{\bar{P}}+no\log(P)$. 
\item{} Summing the inequalities \eqref{rereeBC1-} and \eqref{rereeBC4} results in,
\begin{eqnarray}
&&n(\hat{N}_1+\hat{N}_2)R_1+n(\hat{N}_1+\hat{N}_2)R_2\nonumber\\
&\le&n(\hat{N}_0+\hat{N}_1N_1+\hat{N}_1\hat{N}_2\beta_2+\hat{N}_2{N}_2)\log{\bar{P}}-\hat{N}_2H(\bar{ \bf Y}^{[n]}_{2}\mid W_2,\mathcal{G})\nonumber\\
&&+ \hat{N}_2H(  \bar{ \bf Y}^{[n]}_{1\tilde a},(\bar{ \bf Y}^{[n]}_{1a})^{1-\beta_2}\mid W_2,\mathcal{G})\\
&\le&n(\hat{N}_0+\hat{N}_1N_1+\hat{N}_1\hat{N_2}\beta_2+\hat{N}_2{N}_2)\log{\bar{P}}\label{rereeBC5}
\end{eqnarray}
(\ref{rereeBC5}) is immediately obtained from Lemma \ref{lemmaMIMOx} by defining ${\bar{\bf U}}_1= (\bar{ \bf Y}^{[n]}_{1\tilde a},(\bar{ \bf Y}^{[n]}_{1a})^{1-\beta_2})$ and ${\bar{\bf U}}_2=\bar{ \bf Y}^{[n]}_{2}$, we have,
\begin{eqnarray}
H(  \bar{ \bf Y}^{[n]}_{1\tilde a},(\bar{ \bf Y}^{[n]}_{1a})^{1-\beta_2}\mid W_2,\mathcal{G})-H(\bar{ \bf Y}^{[n]}_{2}\mid W_2,\mathcal{G})&\le&n~o~(\log{\bar{P}})\label{ffgt}
\end{eqnarray}
where \eqref{ffgt} is true as all the parameters $(\lambda_{1i}-\lambda_{2i})^+$ in Lemma \ref{lemmaMIMOx} are equal to zero. 
Finally, applying the DoF limit in (\ref{rereeBC5}), the bound $d_1+d_2\le N_2+(M-N_2)\beta_o$ is concluded. 
\end{enumerate}

\subsubsection{Proof of bound \eqref{B33} when $\beta_1+\beta_2<1$}\label{sec:b1b2<1}
\begin{figure}[h] 
\centering
\includegraphics[width=0.8\textwidth]{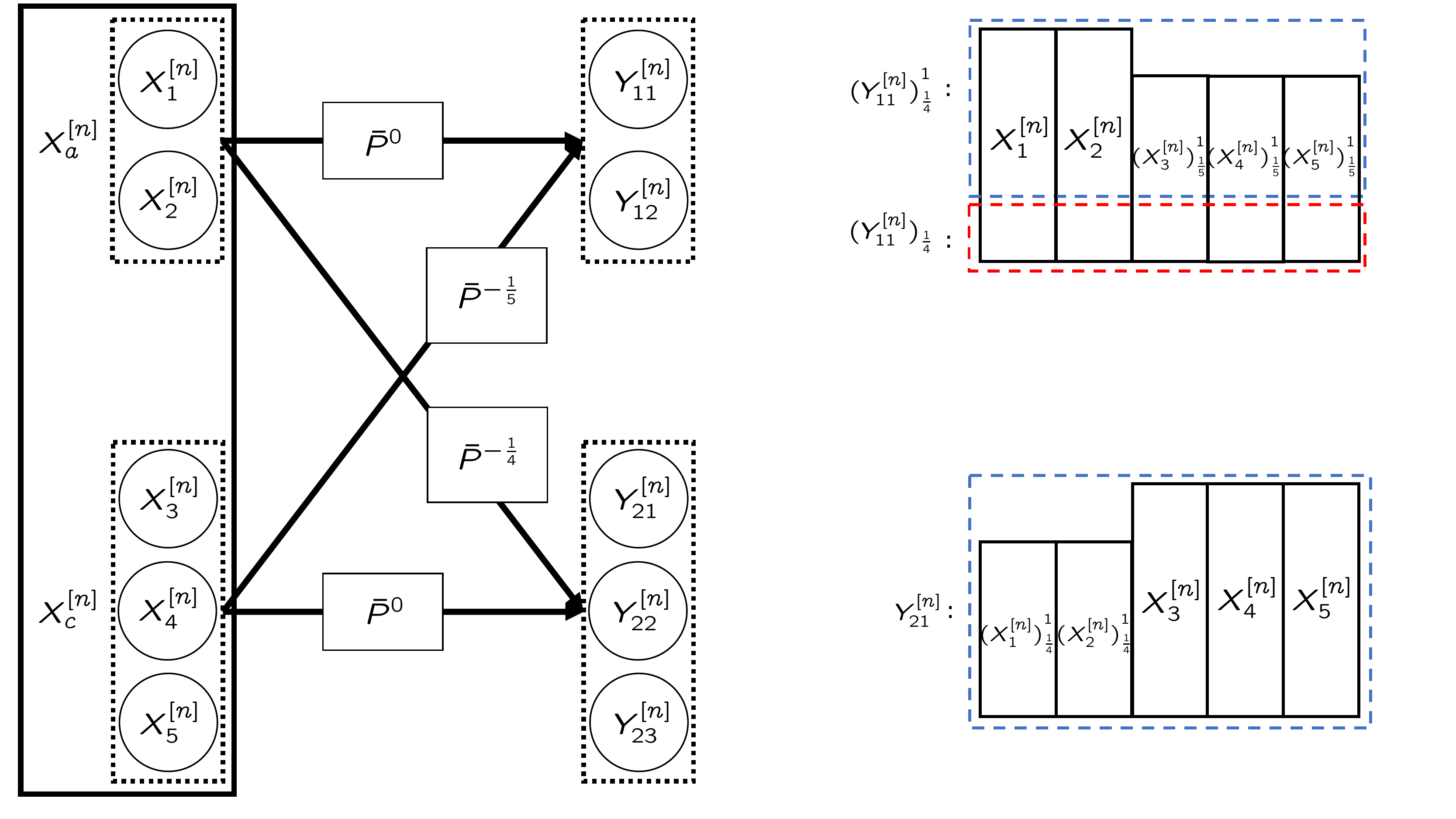}
\caption[]{Illustration for Lemma \ref{lemma1BC}}\label{figxxxBC}
\end{figure}

 In this section we prove the bound \eqref{B33} for  general $(M,N_1,N_2)$ when $\beta_1+\beta_2<1$. The bound \eqref{B33} simplifies to,
\begin{eqnarray}
d_1+d_2&\le& N_2+(M-N_2)\beta_o\label{B33+}
\end{eqnarray}
where $\beta_o$ is equal to,
\begin{eqnarray}
\beta_o&=
&\frac{\beta_1\beta_2(M-N_2)}{(N_2-N_1)(1-\beta_1)+(M-N_2)\beta_2}
\end{eqnarray}
The  proof relies on the following lemma.
\begin{lemma}\label{lemma1BC} For the two-user $(M,N_1,N_2)$ MIMO BC with partial CSIT where $\beta_{1}+\beta_{2}<1$, we have, 
\begin{eqnarray}
&&\breve{N}_1H(\bar{\bf Y}^{[n]}_{2}\mid W_1,\mathcal{G})\nonumber\\
&{\le}&(\breve{N}_1+\breve{N}_2)H(\bar{\bf Y}^{n}_{1}\mid W_1,\mathcal{G})-\breve{N}_2H(\bar{\bf Y}_{1a}^{n},(\bar{\bf Y}^{n}_{1b})^{1-\beta_2}\mid W_1,\mathcal{G})+n\breve{N}_0\log{\bar{P}}+n~o~(\log{\bar{P}})\nonumber\\
\label{firstlemBC+}
\end{eqnarray}
where the numbers $\breve{N}_0$, $\breve{N}_1$ and $\breve{N}_2$ are defined as,
\begin{eqnarray}
\breve{N}_0&=&\breve{N}_1(M-N_1)\beta_1\\
\breve{N}_1&=&(M-N_2)\beta_2\\
\breve{N}_2&=&(N_2-N_1)(1-\beta_1)
\end{eqnarray}
See Figure \ref{figxxxBC} for the comparison of the two sides of (\ref{firstlemBC+}). 
\end{lemma}
See Appendix \ref{appbeta<1} for proof of Lemma \ref{lemma1BC}. With the aid of Lemma \ref{lemma1BC}, the proof of the bound \eqref{B33+} proceeds along the lines of the corresponding proof for $\beta_1+\beta_2\geq 1$ that was presented in Section \ref{sec:b1b2>1}.
\begin{enumerate}
\item{} Starting from Fano's Inequality for the first receiver and proceeding through the same set of steps as (\eqref{BCr1}-\eqref{rereeBC1-}) we have,
\begin{eqnarray}
&&n(\breve{N}_1+\breve{N}_2)R_1\nonumber\\
&\le& n(\breve{N}_0+\breve{N}_1N_1+\breve{N}_1\breve{N_2})\log{\bar{P}}-\breve{N}_1H(\bar{ \bf Y}^{[n]}_{2}\mid W_1,\mathcal{G})+ \breve{N}_2H( \bar{ \bf Y}^{[n]}_{1a},(\bar{ \bf Y}^{[n]}_{1b})^{1-\beta_2}\mid W_2,\mathcal{G})\nonumber\\
&&+n~o~(\log{\bar{P}})\label{rereeBC1--}
\end{eqnarray}
where the difference between \eqref{rereeBC1-} and \eqref{rereeBC1--} is due to the difference in the definitions of $(\hat{N}_1,\hat{N}_2)$ versus $(\breve{N}_1,\breve{N}_2)$.
\item{} Similarly, for the second receiver we have,
\begin{eqnarray}
&&n(\breve{N}_1+\breve{N}_2)R_2\nonumber\\
&\le&\breve{N}_2H(\bar{ \bf Y}^{[n]}_{2}\mid \mathcal{G})-\breve{N}_2H(\bar{ \bf Y}^{[n]}_{2}\mid W_2,\mathcal{G})+\breve{N}_1H(\bar{ \bf Y}^{[n]}_{2}\mid W_1,\mathcal{G})\nonumber\\
&\le&n\breve{N}_2N_2\log{\bar{P}}-\breve{N}_2H(\bar{ \bf Y}^{[n]}_{2}\mid W_2,\mathcal{G})+\breve{N}_1H(\bar{ \bf Y}^{[n]}_{2}\mid W_1,\mathcal{G})\label{rereeBC4+}
\end{eqnarray}
\item{} Summing the inequalities \eqref{rereeBC1--} and \eqref{rereeBC4+} results in,
\begin{eqnarray}
&&n(\breve{N}_1+\breve{N}_2)R_1+n(\breve{N}_1+\breve{N}_2)R_2\nonumber\\
&\le&n(\breve{N}_0+\breve{N}_1N_1+\breve{N}_1\breve{N_2}+\breve{N}_2{N}_2)\log{\bar{P}}-\breve{N}_2H(\bar{ \bf Y}^{[n]}_{2}\mid W_2,\mathcal{G})\nonumber\\
&&+ \breve{N}_2H(  \bar{ \bf Y}^{[n]}_{1a},(\bar{ \bf Y}^{[n]}_{1b})^{1-\beta_2}\mid W_2,\mathcal{G})+n~o~(\log{\bar{P}})\\
&\le&n(\breve{N}_0+\breve{N}_1N_1+\breve{N}_1\breve{N_2}+\breve{N}_2{N}_2)\log{\bar{P}}+n~o~(\log{\bar{P}})\label{rereeBC5+}
\end{eqnarray}
(\ref{rereeBC5+}) follows from Lemma \ref{lemmaMIMOx} similar to (\ref{rereeBC5}). Finally, applying the DoF limit in (\ref{rereeBC5+}) we obtain the bound, $d_1+d_2\le N_2+(M-N_2)\beta_o$. 
\end{enumerate}

\section{Conclusion}
The DoF region of the the  two-user MIMO BC with arbitrary levels of partial CSIT was characterized as a function of the number of antennas and the levels of CSIT while perfect CSIR is assumed. The main challenge was  deriving  an outer bound that captures the difference of entropies caused by asymmetric number of antennas and asymmetric levels of partial CSIT which was accomplished with the aid of sum-set inequalities and AIS approach.

\appendix

\section{Proof of Lemma \ref{lemmaBC1}}\label{app1}
\begin{figure}[h] 
\centering
\includegraphics[width=0.6\textwidth]{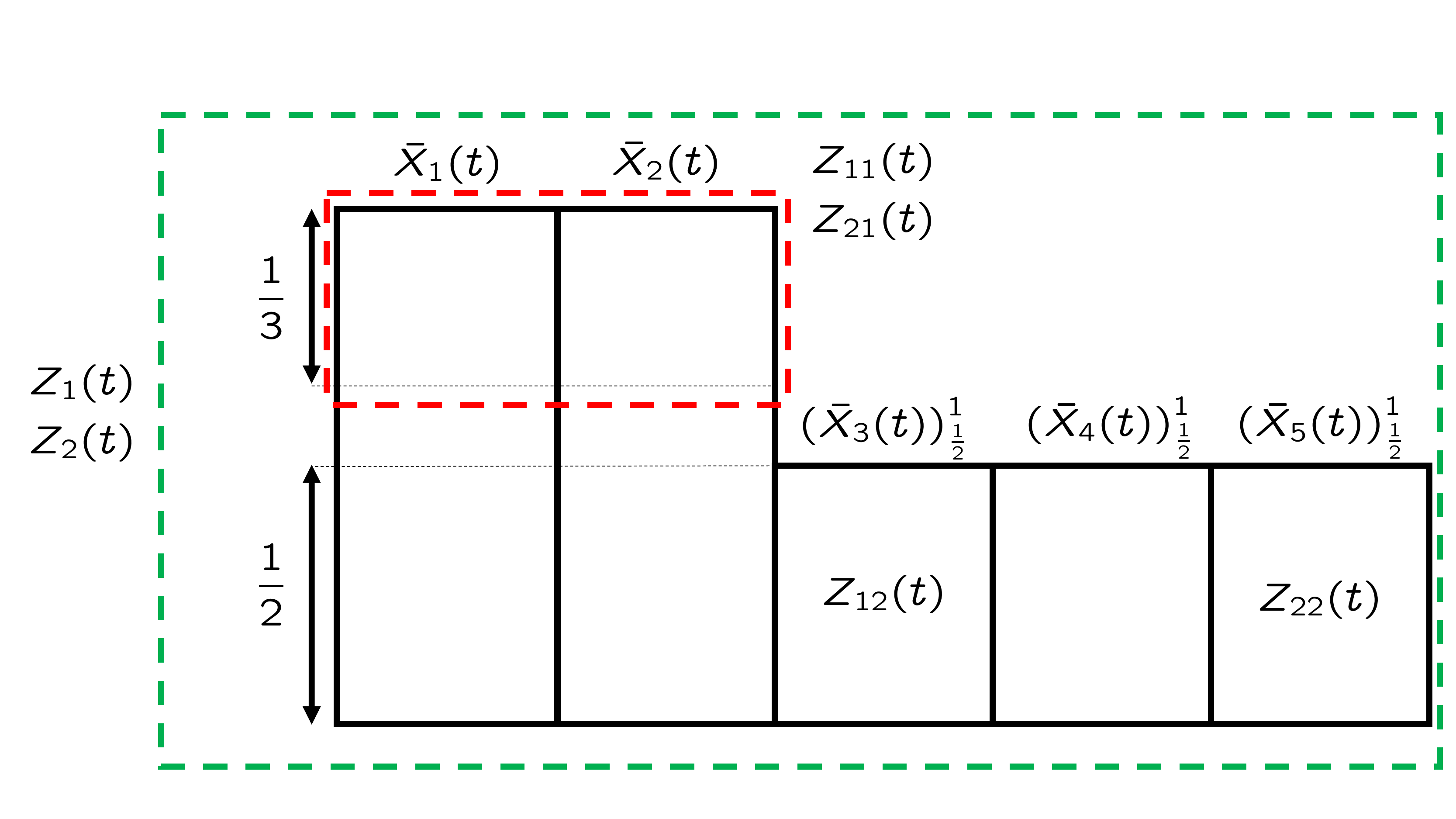}
\caption[]{The random variables in (\eqref{uio00}-\eqref{uio0}) are illustrated.}
\label{lem1}
\end{figure}

Suppressing $n o(\log(P))$ that are inconsequential for DoF, we proceed as follows.
\begin{align}
&2H(\bar{\bf{Y}}^{[n]}_{2}\mid W_1,\mathcal{G})+H( (\bar{\mathbf{Y}}^{[n]}_{1})^{1/3}\mid W_1,\mathcal{G})\nonumber\\
&=2H((\bar{\bf{Y}}^{[n]}_{2}))^{1/2}, (\bar{\bf{Y}}^{[n]}_{2})_{1/2}\mid W_1,\mathcal{G})+H( (\bar{\mathbf{Y}}^{[n]}_{1})^{1/3}\mid W_1,\mathcal{G})\label{qqe1}\\
&\le 2H((\bar{\bf{Y}}^{[n]}_{2}))^{1/2}\mid W_1,\mathcal{G})+H( (\bar{\mathbf{Y}}^{[n]}_{1})^{1/3}\mid W_1,\mathcal{G})+3n\log{\bar{P}}\label{qqe2}\\
&\leq 2H((\bar{{X}}^{[n]}_{3}))^{1/2},(\bar{{X}}^{[n]}_{4}))^{1/2}, (\bar{X}^{[n]}_{5}))^{1/2}\mid W_1,\mathcal{G})+H( (\bar{\mathbf{Y}}^{[n]}_{1})^{1/3}\mid W_1,\mathcal{G})+3n\log{\bar{P}}\label{qqe3}\\
%&\le&2H((\bar{{X}}^{[n]}_{1})^{1}_{\frac{2}{3}},(\bar{{X}}^{[n]}_{2})^{1}_{\frac{2}{3}},(\bar{{X}}^{[n]}_{3})^{1}_{1/2},(\bar{{X}}^{[n]}_{4})^{1}_{1/2},(\bar{X}^{[n]}_{5})^{1}_{1/2}\mid W_1,\mathcal{G})+H( (\bar{\mathbf{Y}}^{[n]}_{1})^{1}_{\frac{2}{3}}\mid W_1,\mathcal{G})+3n\log{\bar{P}}\nonumber\\\label{qqe4}\\
&\leq 2H((\bar{\bf Y}^{[n]}_{1})^{1/3},(\bar{{X}}^{[n]}_{3}))^{1/2},(\bar{{X}}^{[n]}_{4}))^{1/2},(\bar{X}^{[n]}_{5}))^{1/2}\mid W_1,\mathcal{G})+H( (\bar{\mathbf{Y}}^{[n]}_{1})^{1/3}\mid W_1,\mathcal{G})+3n\log{\bar{P}}\label{qqe5}\\
&= 2H((\bar{{X}}^{[n]}_{3}))^{1/2},(\bar{{X}}^{[n]}_{4}))^{1/2},(\bar{X}^{[n]}_{5}))^{1/2}\mid (\bar{\bf Y}^{[n]}_{1})^{1/3},W_1,\mathcal{G})+3H( (\bar{\mathbf{Y}}^{[n]}_{1})^{1/3}\mid W_1,\mathcal{G})+3n\log{\bar{P}}\label{qqe6}\\
%--
&\le H((\bar{{X}}^{[n]}_{3}))^{1/2},(\bar{{X}}^{[n]}_{4}))^{1/2}\mid (\bar{\bf Y}^{[n]}_{1})^{1/3},W_1,\mathcal{G})+H((\bar{{X}}^{[n]}_{3}))^{1/2},(\bar{{X}}^{[n]}_{5}))^{1/2}\mid (\bar{\bf Y}^{[n]}_{1})^{1/3},W_1,\mathcal{G})\nonumber\\
&~~~~+H((\bar{{X}}^{[n]}_{4}))^{1/2},(\bar{{X}}^{[n]}_{5}))^{1/2}\mid (\bar{\bf Y}^{[n]}_{1})^{1/3},W_1,\mathcal{G})+3H( (\bar{\mathbf{Y}}^{[n]}_{1})^{1/3}\mid W_1,\mathcal{G})+3n\log{\bar{P}}\label{qqe7}\\
&=H((\bar{{X}}^{[n]}_{3}))^{1/2},(\bar{{X}}^{[n]}_{4}))^{1/2}, (\bar{\bf Y}^{[n]}_{1})^{1/3}\mid W_1,\mathcal{G})+H((\bar{{X}}^{[n]}_{3}))^{1/2}, (\bar{{X}}^{[n]}_{5}))^{1/2}, (\bar{\bf Y}^{[n]}_{1})^{1/3}\mid W_1,\mathcal{G})\nonumber\\
&~~~~~+H((\bar{{X}}^{[n]}_{4}))^{1/2},(\bar{{X}}^{[n]}_{5}))^{1/2}, (\bar{\bf Y}^{[n]}_{1})^{1/3}\mid W_1,\mathcal{G})+3n\log{\bar{P}}\label{qqe8}\\
&\le 3H(\bar{\bf Y}^{[n]}_{1}\mid W_1,\mathcal{G})+3n\log{\bar{P}}\label{qqe9}
\end{align}
(\ref{qqe1}) follows from Definition \ref{powerlevel}, and (\ref{qqe2}) is true because ${\bf Y}_2$ has $3$ antennas and the entropy of the bottom half of the signal power levels on each of them can altogether contribute at most  $\frac{3}{2}n\log{\bar{P}}$. Next, (\ref{qqe3}) is true because $({\bf Y}_2)^{1/2}$ is a function of $(\bar{{X}}^{[n]}_{3}))^{1/2},(\bar{{X}}^{[n]}_{4}))^{1/2}, (\bar{X}^{[n]}_{5}))^{1/2}$, and (\ref{qqe5})  simply uses the property that $H(A|C)\leq H(A,B|C)$. The chain rule of entropy produces (\ref{qqe6}) and (\ref{qqe8}), and (\ref{qqe7}) follows from the sub-modularity property of the entropy function, i.e., for any three random variables $A$, $B$ and $C$,
\begin{eqnarray}
2H(A, B,C)&\le&H(A,B)+H(A,C)+H(B,C)
\end{eqnarray}
Finally, \eqref{qqe9}  is where we use the sumset inequality from Theorem \ref{Theorem AIS04}. For instance, 
\begin{eqnarray}
H((\bar{{X}}^{[n]}_{3}))^{1/2},(\bar{{X}}^{[n]}_{5}))^{1/2}, (\bar{\bf Y}^{[n]}_{1})^{1/3}\mid W_1,\mathcal{G})&\le& H(\bar{\bf Y}^{[n]}_{1}\mid W_1,\mathcal{G})+n~o(\log{\bar{P}})
\end{eqnarray}
 is obtained from Theorem \ref{Theorem AIS04} by setting $\lambda_1=\lambda_2=\frac{1}{2}$, $I_{11}=I_{21}=2$, $I_{12}=I_{22}=1$  and defining $Z_{11}(t)$, $Z_{12}(t)$, $Z_{21}(t)$, $Z_{22}(t)$, $Z_{1}(t)$, $Z_{2}(t)$ as,
\begin {eqnarray}
Z_{11}(t)&=&(\bar{Y}_{11}(t))^{1/3}\doteq L_{11(t)}^h((\bar{X}_1(t))^{1/3},(\bar{X}_2(t))^{1/3})\label{uio00}\\
Z_{12}(t)&=&L_{12(t)}^h((\bar{X}_1(t))_{1/2},(\bar{X}_2(t))_{1/2},(\bar{X}_3(t))^{1/2},(\bar{X}_4(t))^{1/2},(\bar{X}_5(t))^{1/2})=(\bar{X}_3(t))^{1/2}\\
Z_{21}(t)&=&(\bar{Y}_{12}(t))^{1/3}\doteq L_{21(t)}^h((\bar{X}_1(t))^{1/3},(\bar{X}_2(t))^{1/3})\\
Z_{22}(t)&=&L_{12(t)}^h((\bar{X}_1(t))_{1/2},(\bar{X}_2(t))_{1/2},(\bar{X}_3(t))^{1/2},(\bar{X}_4(t))^{1/2},(\bar{X}_5(t))^{1/2})=(\bar{X}_5(t))^{1/2}\\
Z_{1}(t)&=&\bar{Y}_{11}(t)=L_{y11(t)}^g(\bar{X}_1(t),\bar{X}_2(t),(\bar{X}_3(t))^{1/2},(\bar{X}_4(t))^{1/2},(\bar{X}_5(t))^{1/2})\\
Z_{2}(t)&=&\bar{Y}_{12}(t)=L_{y12(t)}^g(\bar{X}_1(t),\bar{X}_2(t),(\bar{X}_3(t))^{1/2},(\bar{X}_4(t))^{1/2},(\bar{X}_5(t))^{1/2})\label{uio0}
\end{eqnarray}
Note that for any $i,j$ the linear combination $L_{ij}^h(t)$ is arbitrary linear combination of random variables satisfying the condition in Definition \ref{deflc}. Thus, some of the coefficients in $L_{ij}^h(t)$ can be chosen to be zero, e.g., we choose $L_{12}^h(t)((\bar{X}_1(t))_{1/2},(\bar{X}_2(t))_{1/2},(\bar{X}_3(t))^{1/2},(\bar{X}_4(t))^{1/2},(\bar{X}_5(t))^{1/2})$ to be $(\bar{X}_3(t))^{1/2}$. Thus, from Theorem \ref{Theorem AIS04} we have,
\begin{eqnarray}
H(Z_{11}^{[n]},Z_{12}^{[n]},Z_{21}^{[n]},Z_{22}^{[n]}\mid W_1,\mathcal{G})&\leq& H(Z_{1}^{[n]},Z_{2}^{[n]}\mid W_1,\mathcal{G})+n~o(\log{\bar{P}})\label{uio1}\\
\Rightarrow H((\bar{{X}}^{[n]}_{3}))^{1/2},(\bar{{X}}^{[n]}_{5}))^{1/2}, (\bar{\bf Y}^{[n]}_{1})^{1/3}\mid W_1,\mathcal{G})&\le& H(\bar{\bf Y}^{[n]}_{1}\mid W_1,\mathcal{G})+n~o(\log{\bar{P}})\label{uio2}
\end{eqnarray}
See Figure \ref{lem1} and \ref{uio2++}. 
\begin{figure}[!h] 
\begin{eqnarray*}
&&H\left(\begin{tikzpicture}[scale=0.7, baseline=(current bounding box.center)]
\draw[ thick, pattern= dots, pattern color=orange] (-1,0) rectangle (0,3);
\draw[ thick, pattern=grid, pattern color=orange] (-1,3) rectangle (0,4);
\draw[ thick, pattern=vertical lines, pattern color=orange] (-1,4) rectangle (0,6);
\draw[ thick, pattern=crosshatch, pattern color=blue] (0,0) rectangle (1,3);
\draw[ thick, pattern=grid, pattern color=blue] (0,3) rectangle (1,4);
\draw[ thick, pattern=vertical lines, pattern color=blue] (0,4) rectangle (1,6);
\draw[ thick, pattern=vertical lines, pattern color=red] (1,2-2) rectangle (2,3);
\draw[ thick, pattern=vertical lines, pattern color=brown] (2,2-2) rectangle (3,3);
\draw[ thick, pattern=vertical lines, pattern color=green] (3,2-2) rectangle (4,3);
\draw[thick, <->] (4.2,0)--(4.2,3) node[midway, right]{$\frac{1}{2}$};
\draw[thick, <->] (4.2,3)--(4.2,4) node[midway, right]{$\frac{1}{6}$};
\draw[thick, <->] (4.2,4)--(4.2,6) node[midway, right]{$\frac{1}{3}$};
\draw[thick, |-|] (-1,-0.2)--(4,-0.2) node[midway, below]{$Z_1(t)$};
\draw (-0.5, 6.5) node[above]{$X_1$};
\draw (0.5, 6.5) node[above]{$X_2$};
\draw (1.5, 6.5) node[above][scale=0.7]{${(X_3)}_{1/2}^1$};
\draw (2.5, 6.5) node[above][scale=0.7]{${(X_4)}_{1/2}^1$};
\draw (3.5, 6.5) node[above][scale=0.7]{${(X_5)}_{1/2}^1$};
\end{tikzpicture}\begin{tikzpicture}[scale=0.7, baseline=(current bounding box.center)]
\draw[ thick, pattern= dots, pattern color=orange] (-1,0) rectangle (0,3);
\draw[ thick, pattern=grid, pattern color=orange] (-1,3) rectangle (0,4);
\draw[ thick, pattern=vertical lines, pattern color=orange] (-1,4) rectangle (0,6);
\draw[ thick, pattern=crosshatch, pattern color=blue] (0,0) rectangle (1,3);
\draw[ thick, pattern=grid, pattern color=blue] (0,3) rectangle (1,4);
\draw[ thick, pattern=vertical lines, pattern color=blue] (0,4) rectangle (1,6);
\draw[ thick, pattern=vertical lines, pattern color=red] (1,2-2) rectangle (2,3);
\draw[ thick, pattern=vertical lines, pattern color=brown] (2,2-2) rectangle (3,3);
\draw[ thick, pattern=vertical lines, pattern color=green] (3,2-2) rectangle (4,3);
\draw[thick, <->] (4.2,0)--(4.2,3) node[midway, right]{$\frac{1}{2}$};
\draw[thick, <->] (4.2,3)--(4.2,4) node[midway, right]{$\frac{1}{6}$};
\draw[thick, <->] (4.2,4)--(4.2,6) node[midway, right]{$\frac{1}{3}$};
\draw[thick, |-|] (-1,-0.2)--(4,-0.2) node[midway, below]{$Z_2(t)$};
\draw (-0.5, 6.5) node[above]{$X_1$};
\draw (0.5, 6.5) node[above]{$X_2$};
\draw (1.5, 6.5) node[above][scale=0.7]{${(X_3)}_{1/2}^1$};
\draw (2.5, 6.5) node[above][scale=0.7]{${(X_4)}_{1/2}^1$};
\draw (3.5, 6.5) node[above][scale=0.7]{${(X_5)}_{1/2}^1$};
\end{tikzpicture}
\right)\nonumber\\
&\geq&
H\left(
\begin{tikzpicture}[scale=0.7, baseline=(current bounding box.center)]
\draw[ thick, pattern=vertical lines, pattern color=orange] (-1,9) rectangle (0,11);
\draw[ thick, pattern=vertical lines, pattern color=blue] (0,9) rectangle (1,11);

\draw[thick,|-|](-1,8.8)--(1,8.8) node[midway, below]{ $Z_{11}(t)$};
\draw[thick,<->](-1.2,9)--(-1.2,11) node[midway, left]{$\frac{1}{3}$};
\draw[help lines](-1,11.1)--(1,11.1);
\draw[thick,<->](1.2,9)--(1.2,11) node[midway, right]{ $\mathcal{T}(Z_{11})$};
\path (1.5, 9) node[right]{\Huge ,};
\end{tikzpicture}
\begin{tikzpicture}[scale=0.7, baseline=(current bounding box.center)]
\draw[ thick, pattern=vertical lines, pattern color=orange] (-1,9) rectangle (0,11);
\draw[ thick, pattern=vertical lines, pattern color=blue] (0,9) rectangle (1,11);

\draw[thick,|-|](-1,8.8)--(1,8.8) node[midway, below]{ $Z_{21}(t)$};
\draw[thick,<->](-1.2,9)--(-1.2,11) node[midway, left]{$\frac{1}{3}$};
\draw[help lines](-1,11.1)--(1,11.1);
\draw[thick,<->](1.2,9)--(1.2,11) node[midway, right]{ $\mathcal{T}(Z_{21})$};
\path (1.5, 9) node[right]{\Huge ,};
\end{tikzpicture}
\begin{tikzpicture}[scale=0.7, baseline=(current bounding box.center)]
\draw[ thick, pattern=vertical lines, pattern color=red] (-1,9) rectangle (0,12);

\draw[thick,|-|](-1,8.8)--(0,8.8) node[midway, below]{ $Z_{12}(t)$};
\draw[thick,<->](-1.2,9)--(-1.2,12) node[midway, left]{$\frac{1}{2}$};
\draw[help lines](-1,12.1)--(0,12.1);
\draw[thick,<->](0.2,9)--(0.2,12) node[midway, right]{ $\mathcal{T}(Z_{12})$};
\path (0.5, 9) node[right]{\Huge ,};
\end{tikzpicture}
\begin{tikzpicture}[scale=0.7, baseline=(current bounding box.center)]
\draw[ thick, pattern=vertical lines, pattern color=green] (-1,9) rectangle (0,12);

\draw[thick,|-|](-1,8.8)--(0,8.8) node[midway, below]{ $Z_{22}(t)$};
\draw[thick,<->](-1.2,9)--(-1.2,12) node[midway, left]{$\frac{1}{2}$};
\draw[help lines](-1,12.1)--(0,12.1);
\draw[thick,<->](0.2,9)--(0.2,12) node[midway, right]{ $\mathcal{T}(Z_{22})$};
\end{tikzpicture}\right)
\end{eqnarray*}
\caption{ Illustration of \eqref{uio2}.}\label{uio2++}
\end{figure}

\section{Proof of Lemma \ref{lemmaBC1...}} \label{appendix1BC.}
Note that $2H((\bar{{Y}}^{[n]}_{21})_{1/4},(\bar{{Y}}^{[n]}_{22})_{1/4}\mid W_1,\mathcal{G})\le\log{\bar{P}}$. Therefore, it is sufficient to prove,
\begin{eqnarray}
&&2H((\bar{{Y}}^{[n]}_{21})^{3/4},(\bar{{Y}}^{[n]}_{22})^{3/4}\mid W_1,\mathcal{G})+3H((\bar{ Y}^{n}_{1})^{1/2}\mid W_1,\mathcal{G})\nonumber\\
&{\le}&5H(\bar{Y}^{n}_{1}\mid W_1,\mathcal{G})+n~o~(\log{\bar{P}})\label{ko0}
\end{eqnarray}
For any $i\in[7]$, define ${C}_i(t)$ as
\begin{eqnarray}
{C}_i(t)&=&\left\{\begin{matrix}
(\bar{{X}}_{5}(t))_{\frac{3}{4}}^{1}, i=1\\ 
(\bar{{X}}_{5}(t))_{1/2}^{\frac{3}{4}}, i=2\\ 
L_{3}^g((\bar{{X}}_{1}(t))_{\frac{3}{4}}^{1},(\bar{{X}}_{2}(t))_{\frac{3}{4}}^{1},(\bar{{X}}_{3}(t))_{\frac{3}{4}}^{1},(\bar{{X}}_{4}(t))_{1/4}^{1/2},(\bar{{X}}_{5}(t))_{1/4}^{1/2}), i=3\\ 
(\bar{{X}}_{4}(t))_{\frac{3}{4}}^{1}, i=4\\ 
(\bar{{X}}_{4}(t))_{1/2}^{\frac{3}{4}}, i=5\\ 
L_{6}^g((\bar{{X}}_{1}(t))_{\frac{3}{4}}^{1},(\bar{{X}}_{2}(t))_{\frac{3}{4}}^{1},(\bar{{X}}_{3}(t))_{\frac{3}{4}}^{1},(\bar{{X}}_{4}(t))_{1/4}^{1/2},(\bar{{X}}_{5}(t))_{1/4}^{1/2}), i=6\\
{C}_1(t), i=7
\end{matrix}\right.\label{ko00}
\end{eqnarray}
\begin{figure}[h] 
\centering
\includegraphics[width=0.8\textwidth]{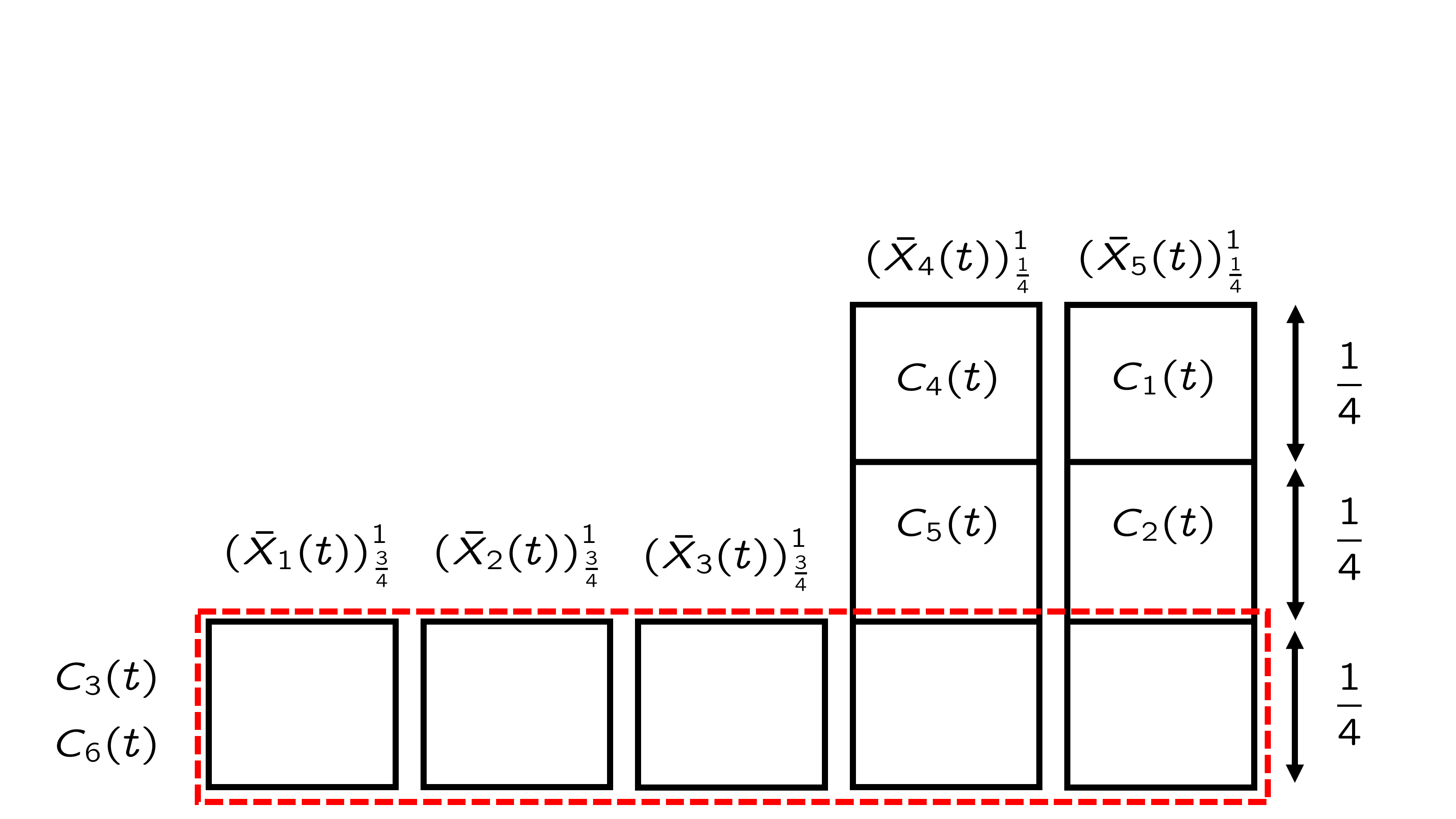}
\caption[]{The random variables $(\bar{X}_1(t))^1_{\frac{3}{4}},(\bar{X}_2(t))^1_{\frac{3}{4}},(\bar{X}_3(t))^1_{\frac{3}{4}},(\bar{X}_4(t))^{3/4},(\bar{X}_5(t))^{3/4}$ and their partitions ${C}_{i}(t)$ are specified.}
\label{figko}
\end{figure}
Starting from the left side of  \eqref{ko0}, we have
\begin{eqnarray}
&&2H((\bar{{Y}}^{[n]}_{21})^{3/4},(\bar{{Y}}^{[n]}_{22})^{3/4}\mid W_1,\mathcal{G})+3H((\bar{ Y}^{n}_{1})^{1/2}\mid W_1,\mathcal{G})\nonumber\\
&=&2H({{C}}_1^{[n]},{{C}}_2^{[n]},{{C}}_{3}^{[n]},{{C}}_4^{[n]},{{C}}_5^{[n]},{{C}}_{6}^{[n]}\mid W_1,\mathcal{G})+3H((\bar{ Y}^{n}_{1})^{1/2}\mid W_1,\mathcal{G})\label{ko1}\\
&\le&2H((\bar{{Y}}_1^{[n]})^{1/2},{{C}}_2^{[n]},{{C}}_{3}^{[n]},{{C}}_4^{[n]},{{C}}_5^{[n]},{{C}}_{6}^{[n]}\mid W_1,\mathcal{G})+3H((\bar{ Y}^{n}_{1})^{1/2}\mid W_1,\mathcal{G})+n~o~(\log{\bar{P}})\label{ko2}\\
&=&2H({{C}}_2^{[n]},{{C}}_{3}^{[n]},{{C}}_4^{[n]},{{C}}_5^{[n]},{{C}}_{6}^{[n]}\mid (\bar{{Y}}_1^{[n]})^{1/2},W_1,\mathcal{G})+5H((\bar{ Y}^{n}_{1})^{1/2}\mid W_1,\mathcal{G})+n~o~(\log{\bar{P}})\label{ko3}\\
&\le&\sum_{i=2}^6H({{C}}_i^{[n]},{{C}}_{i+1}^{[n]}\mid (\bar{{Y}}_1^{[n]})^{1/2},W_1,\mathcal{G})+5H((\bar{ Y}^{n}_{1})^{1/2}\mid W_1,\mathcal{G})+n~o~(\log{\bar{P}})\label{ko4}\\
&=&\sum_{i=2}^6H((\bar{{Y}}_1^{[n]})^{1/2},{{C}}_i^{[n]},{{C}}_{i+1}^{[n]}\mid W_1,\mathcal{G})+n~o~(\log{\bar{P}})\label{ko5}\\
&\le&5H(\bar{ Y}^{n}_{1}\mid W_1,\mathcal{G})+n~o~(\log{\bar{P}})\label{ko6}
\end{eqnarray}
\eqref{ko1} follows from Definition \ref{powerlevel} and definition of ${C}_i(t)$ in \eqref{ko00}. \eqref{ko3} is true from the chain rule and \eqref{ko4} is concluded from sub-modularity properties of entropy function, i.e., for any $m$ random variables $\{X_1,X_2,\cdots,X_m\}$ where we define $X_{k+m}$ as $X_k$ for positive numbers  $k$ we have, 
\begin{eqnarray}
nH(X_1,\cdots,X_m\mid W)\le \sum_{i=1}^mH(X_i,X_{i+1},\cdots,X_{i+n}\mid W)
\end{eqnarray}
for any random variable $W$ if $n\le m$. \eqref{ko2} follows from Theorem \ref{Theorem AIS04} to our setting. Let us set  $M=1$, $N=5$, $K=1$, $\lambda_{1,1}=1$, $l_1=1$, and $I_{1,1}=\{1\}$. Then, the inequality \eqref{dssd4} reduces to,
\begin {eqnarray}
H(Z_1^{[n]}\mid W,\mathcal{G})&\geq& H(Z_{11}^{[n]}\mid W)+n~o(\log{\bar{P}}).\label{dssd4+}
\end{eqnarray}
where
\begin {eqnarray}
Z_1(t)&=&(\bar{ Y}_{1}(t))^{1/2}\\
Z_{11}(t)&=&C_1(t)\\
W&=&{{C}}_2^{[n]},{{C}}_{3}^{[n]},{{C}}_4^{[n]},{{C}}_5^{[n]},{{C}}_{6}^{[n]},W_1
\end{eqnarray}
\eqref{ko6} follows from Theorem \ref{Theorem AIS04} to our setting similar to \eqref{ko2}. Let us set $N=5$, $M=3$, $K=1$, $\lambda_{1,3}=\frac{1}{2},\lambda_{1,1}=\lambda_{1,2}=\frac{1}{4}$, $I_{1,1}=3,I_{1,2}=\{2,3\}$, and $I_{1,2,3}=\{1,2,3\}$. Then, the inequality \eqref{dssd4} reduces to,
\begin {eqnarray}
H(Z_1^{[n]}\mid W,\mathcal{G})&\geq& H(Z_{11}^{[n]},Z_{12}^{[n]},Z_{13}^{[n]}\mid W)+n~o(\log{\bar{P}}).\label{dssd4+}
\end{eqnarray}
where
\begin {eqnarray}
Z_1(t)&=&\bar{ Y}_{1}(t)\\
Z_{11}(t)&=&(\bar{ Y}_{1}(t))^{1/2}\nonumber\\
&\doteq& L^g((\bar{ X}_{1}(t))^{1/2},(\bar{ X}_{2}(t))^{1/4},(\bar{ X}_{3}(t))^{1/4},(\bar{ X}_{4}(t))^{1/4},(\bar{ X}_{5}(t))^{1/4})\label{tmtm1}\\
Z_{12}(t)&=&C_i(t)\\
Z_{13}(t)&=&C_{i+1}(t)\\
W&=&W_1
\end{eqnarray}
\eqref{tmtm1} follows similar to \eqref{eq:doteq1}. Note that the condition \eqref{condt4} is satisfied as
\begin{eqnarray}
\mathcal{T}(Z_{1,2})+\mathcal{T}(Z_{1,3})&\leq&\lambda_{1,1}+\lambda_{1,2}\\
\mathcal{T}(Z_{1,3})&\leq&\lambda_{1,1}
\end{eqnarray}
Note that $\mathcal{T}(C_i(t))=\frac{1}{4}$ from Figure \ref{figko}. For instance, illustration of 
\begin{eqnarray}
H((\bar{{Y}}_1^{[n]})^{1/2},{{C}}_2^{[n]},{{C}}_{3}^{[n]}\mid W_1,\mathcal{G})\le H(\bar{ Y}^{n}_{1}\mid W_1,\mathcal{G})\label{ko++}
\end{eqnarray}
 is shown in Figure \ref{figko+}.
\begin{figure}[!h] 
\begin{eqnarray*}
&&H\left(\begin{tikzpicture}[scale=0.7, baseline=(current bounding box.center)]
\draw[ thick, pattern= dots, pattern color=orange] (-1,0) rectangle (0,2);
\draw[ thick, pattern=crosshatch, pattern color=orange] (-1,2) rectangle (0,4);
\draw[ thick, pattern=grid, pattern color=orange] (-1,4) rectangle (0,6);
\draw[ thick, pattern=vertical lines, pattern color=orange] (-1,6) rectangle (0,8);
\draw[ thick, pattern=crosshatch, pattern color=blue] (0,2-2) rectangle (1,4-2);
\draw[ thick, pattern=grid, pattern color=blue] (0,4-2) rectangle (1,6-2);
\draw[ thick, pattern=vertical lines, pattern color=blue] (0,6-2) rectangle (1,8-2);
\draw[ thick, pattern=crosshatch, pattern color=red] (1,2-2) rectangle (2,4-2);
\draw[ thick, pattern=grid, pattern color=red] (1,4-2) rectangle (2,6-2);
\draw[ thick, pattern=vertical lines, pattern color=red] (1,6-2) rectangle (2,8-2);
\draw[ thick, pattern=crosshatch, pattern color=green] (2,2-2) rectangle (3,4-2);
\draw[ thick, pattern=grid, pattern color=green] (2,4-2) rectangle (3,6-2);
\draw[ thick, pattern=vertical lines, pattern color=green] (2,6-2) rectangle (3,8-2);
\draw[ thick, pattern=crosshatch, pattern color=brown] (3,2-2) rectangle (4,4-2);
\draw[ thick, pattern=grid, pattern color=brown] (3,4-2) rectangle (4,6-2);
\draw[ thick, pattern=vertical lines, pattern color=brown] (3,6-2) rectangle (4,8-2);
\draw[thick, <->] (4.2,0)--(4.2,2) node[midway, right]{$\frac{1}{4}$};
\draw[thick, <->] (4.2,2)--(4.2,4) node[midway, right]{$\frac{1}{4}$};
\draw[thick, <->] (4.2,4)--(4.2,8) node[midway, right]{$\frac{1}{2}$};
\draw[thick, |-|] (-1,-0.2)--(4,-0.2) node[midway, below]{$Z_1(t)=\bar{Y}_1(t)$};
\draw (-0.5, 9) node[above]{$X_1$};
\draw (0.5, 9) node[above]{$X_2$};
\draw (1.5, 9) node[above]{$X_3$};
\draw (2.5, 9) node[above]{$X_4$};
\draw (3.5, 9) node[above]{$X_5$};
\end{tikzpicture}
\right)\nonumber\\
&\geq&
H\left(
\begin{tikzpicture}[scale=0.7, baseline=(current bounding box.center)]
\draw[ thick, pattern=grid, pattern color=orange] (-1,9) rectangle (0,11);
\draw[ thick, pattern=vertical lines, pattern color=orange] (-1,11) rectangle (0,13);
\draw[ thick, pattern=vertical lines, pattern color=blue] (0,9) rectangle (1,11);
\draw[ thick, pattern=vertical lines, pattern color=red] (1,9) rectangle (2,11);
\draw[ thick, pattern=vertical lines, pattern color=green] (2,9) rectangle (3,11);
\draw[ thick, pattern=vertical lines, pattern color=brown] (3,9) rectangle (4,11);
\draw[thick,|-|](-1,8.8)--(4,8.8) node[midway, below]{ $Z_{11}(t)=(\bar{Y}_1(t))_{1/2}^1$};
\draw[thick,<->](-1.2,9)--(-1.2,13) node[midway, left]{$\frac{1}{2}$};
\draw[help lines](-1,13.1)--(4,13.1);
\draw[thick,<->](4.2,9)--(4.2,13) node[midway, right]{ $\mathcal{T}(Z_1)$};
\path (5, 9) node[right]{\Huge ,};

\draw[ thick, pattern=grid, pattern color=brown] (2,5) rectangle (3,7);
\draw[thick,<->](1.8,5)--(1.8,7) node[midway, left]{$\frac{1}{4}$};
\draw[thick,<->](3.2,5)--(3.2,7) node[midway, right]{ $\mathcal{T}(Z_2)$};
\draw[help lines](1.8,7.2)--(3.25,7.2);
\path (5, 5) node[right]{\Huge ,};
\draw[thick,|-|](2,4.8)--(3,4.8) node[midway, below]{ $Z_{12}(t)=C_2(t)$};

\draw[ thick, pattern=vertical lines, pattern color=orange] (0,0) rectangle (1,2);
\draw[ thick, pattern=vertical lines, pattern color=blue] (1,0) rectangle (2,2);
\draw[ thick, pattern=vertical lines, pattern color=red] (2,0) rectangle (3,2);
\draw[ thick, pattern=grid, pattern color=green] (3,0) rectangle (4,2);
\draw[ thick, pattern=grid, pattern color=brown] (4,0) rectangle (5,2);
\draw[thick,<->](-0.4,0)--(-0.4,2) node[midway,left]{$\frac{1}{4}$};
\draw[thick,<->](5.2,0)--(5.2,2) node[midway, right]{ $\mathcal{T}(Z_3)$};
\draw[help lines](0,2.2)--(5,2.2);
\draw[thick,|-|](0,-0.2)--(5,-0.2) node[midway, below]{ $Z_3(t)=C_3(t)$};
\end{tikzpicture}
\right)
\end{eqnarray*}
\caption{ Illustration of \eqref{ko++}.}\label{figko+}
\end{figure}

\section{Proof of Lemma \ref{lemmabeta>1}}\label{appbeta>1}

\begin{enumerate}
\item{} Define the random variables $\bar{\bf Y}_{1c}(t)$ and $\bar{\bf Y}_{1d}(t)$ from the random variables $\bar{\bf Y}_{1a}(t)$ and $\bar{\bf Y}_{1b}(t)$ in (\eqref{e1}-\eqref{e4}) as,
\begin{eqnarray}
\bar{\bf Y}_{1c}(t)&=&\begin{bmatrix}\bar{y}_{11c}(t)&\bar{y}_{12c}(t)&\cdots&\bar{y}_{1,N_1+N_2-M,c}(t)\end{bmatrix}^T\label{e1+}\\
\bar{y}_{1rc}(t)&=&L_{3r}^g(t)\left((\bar{\mathbf{X}}_{a}(t))^{1-\beta_2}\bigtriangledown\bar{\mathbf{X}}_{b}(t)\bigtriangledown(\bar{\mathbf{X}}_{c}(t))^{1-\beta_1}\right),\forall r\in[N_1+N_2-M]\nonumber\\\label{e2+}
\\
\bar{\bf Y}_{1d}(t)&=&\begin{bmatrix}\bar{y}_{11d}(t)&\bar{y}_{12d}(t)&\cdots&\bar{y}_{1,M-N_2,d}(t)\end{bmatrix}^T\label{e3+}\\
\bar{y}_{1rd}(t)&=&L_{3r}^g(t)\left(\bar{\mathbf{X}}_{a}(t)\bigtriangledown(\bar{\mathbf{X}}_{c}(t))^{1}_{\beta_1}\right),\forall r\in[M-N_2]\label{e4+}
\end{eqnarray}
As $[\bar{\bf Y}_{1c}(t)\bigtriangledown(\bar{\bf Y}_{1d}(t))^{1-\beta_2}]={\bf A}'(t)[\bar{\bf Y}_{1a}(t)\bigtriangledown(\bar{\bf Y}_{1b}(t))^{1-\beta_2}]$ for some invertible matrix ${\bf A}'(t)$, the terms $\bar{\bf Y}_{1a}(t)$ and $\bar{\bf Y}_{1b}(t)$ are replaced with $\bar{\bf Y}_{1c}(t)$ and $\bar{\bf Y}_{1d}(t)$, respectively.\footnote{Note that $\bar{\bf Y}_{1c}(t)$ and $\bar{\bf X}_{c}(t)$ do not exist simultaneously as $\bar{\bf Y}_{1c}(t)$ and $\bar{\bf X}_{c}(t)$ are vector random variables of size $(N_1+N_2-M)\times1$ and $(M-N_1-N_2)\times1$, respectively.}
\item{} As $\bar{\mathbf{X}}_{b}(t)$ and $\bar{\mathbf{X}}_{c}(t)$ are vector random variables of size $N_1+N_2-M$ and $M-N_1$, the random variables $\bar{\bf Y}_{2a}(t)$ and $\bar{\bf Y}_{2b}(t)$ are defined from the random variable $\bar{\bf Y}_{2}(t)$ in \eqref{yl3} and \eqref{yl4} as,
\begin{eqnarray}
\bar{\bf Y}_{2a}(t)&=&\begin{bmatrix}\bar{y}_{21a}(t)&\bar{y}_{22a}(t)&\cdots&\bar{y}_{2,N_1+N_2-M,a}(t)\end{bmatrix}^T\label{frreqqq1}\\
\bar{y}_{2ra}(t)&=&L_{4r}^g(t)\left((\bar{\mathbf{X}}_{a}(t))^{1-\beta_2}\bigtriangledown\bar{\mathbf{X}}_{b}(t)\right),\forall r\in[N_1+N_2-M]\label{yl2qq1}
\\
\bar{\bf Y}_{2b}(t)&=&\begin{bmatrix}\bar{y}_{21b}(t)&\bar{y}_{22b}(t)&\cdots&\bar{y}_{2,M-N_2,b}(t)\end{bmatrix}^T\label{frreqqq1}\\
\bar{y}_{2rb}(t)&=&L_{4r}^g(t)\left((\bar{\mathbf{X}}_{a}(t))^{1-\beta_2}\bigtriangledown\bar{\mathbf{X}}_{c}(t)\right),\forall r\in[M-N_1]\label{rre1BqqCq2}
\end{eqnarray} 
Similarly, as $[\bar{\bf Y}_{2a}(t)\bigtriangledown(\bar{\bf Y}_{2b}(t))^{1-\beta_2}]={\bf A}(t)[\bar{\bf Y}_{2}(t)]$ for some invertible matrix ${\bf A}(t)$, the term $\bar{\bf Y}_{2}(t)$ is replaced with $\bar{\bf Y}_{2a}(t)\bigtriangledown\bar{\bf Y}_{2b}(t)$.
\item{} The entropy of a discrete random variable is bounded by logarithm of the cardinality of it, i.e., $H((\bar{\bf Y}^{[n]}_{2b})_{\beta_1}\mid W_1,\mathcal{G})\le n\hat{N}_2\beta_1\log{\bar{P}} $. Thus, from the chain rule we have,
\begin{eqnarray}
&&\hat{N}_1H(\bar{\bf Y}^{[n]}_{2a},\bar{\bf Y}^{[n]}_{2b}\mid W_1,\mathcal{G})-n\hat{N}_0\log{\bar{P}}\nonumber\\
&=&\hat{N}_1H(\bar{\bf Y}^{[n]}_{2a},\bar{\bf Y}^{[n]}_{2b}\mid W_1,\mathcal{G})-n\hat{N}_1\hat{N}_2\beta_1\log{\bar{P}}\\
&\le&\hat{N}_1H(\bar{\bf Y}^{[n]}_{2a},\bar{\bf Y}^{[n]}_{2b}\mid W_1,\mathcal{G})-\hat{N}_1H((\bar{\bf Y}^{[n]}_{2b})_{\beta_1}\mid W_1,\mathcal{G})\\
&{\le}&\hat{N}_1H(\bar{\bf Y}^{[n]}_{2a},(\bar{\bf Y}^{[n]}_{2b})^{1}_{\beta_1}\mid W_1,\mathcal{G})\label{fr2q}
\end{eqnarray}
where  \eqref{fr2q} is true from the chain rule. 
\item{} Note that,
\begin{eqnarray}
H((\bar{\bf Y}^{n}_{1d})_{\beta_2}\mid\bar{\bf Y}_{1c}^{n},(\bar{\bf Y}^{n}_{1d})^{1-\beta_2},W_1,\mathcal{G})&{=}&H(\bar{\bf Y}^{n}_{1}\mid W_1,\mathcal{G})-H(\bar{\bf Y}_{1c}^{n},(\bar{\bf Y}^{n}_{1d})^{1-\beta_2}\mid W_1,\mathcal{G})\nonumber\\
\label{fr1}
\end{eqnarray}
where (\ref{fr1}) follows from the chain rule.
\end{enumerate}
From the above observations in order to prove \eqref{B33++}, it is sufficient to demonstrate the following inequality,
\begin{eqnarray}
&&\hat{N}_1H(\bar{\bf Y}^{[n]}_{2a},(\bar{\bf Y}^{[n]}_{2b})^{1-\beta_1}\mid W_1,\mathcal{G})+\hat{N}_2H(\bar{\bf Y}_{1c}^{n},(\bar{\bf Y}^{n}_{1d})^{1-\beta_2}\mid W_1,\mathcal{G})\nonumber\\
&{\le}&(\hat{N}_1+\hat{N}_2)H(\bar{\bf Y}^{n}_{1}\mid W_1,\mathcal{G})+n~o~(\log{\bar{P}})\label{firstlemBC-1}
\end{eqnarray}
Before proceeding to proof of \eqref{firstlemBC-1} let us define the random variables $\bar{C}_{i}(t)$ as the components of $\bar{\bf X}_c(t)$, i.e.,
\begin{eqnarray}
\bar{C}_{i}(t)&=&\left\{\begin{matrix}
\bar{X}_{i-\hat{N}_1-\hat{N}_2+M}(t) &0<i\le \hat{N}_1+\hat{N}_2 \\ 
\bar{C}_{i-\hat{N}_1-\hat{N}_2}(t) & i>\hat{N}_1+\hat{N}_2
\end{matrix}\right.\label{defcn1}
\end{eqnarray}
Starting from the left side of  \eqref{firstlemBC-1}, we have
\begin{eqnarray}
&&\hat{N}_1H(\bar{\bf Y}^{[n]}_{2a},(\bar{\bf Y}^{[n]}_{2b})^{1}_{\beta_1}\mid W_1,\mathcal{G})+\hat{N}_2H(\bar{\bf Y}_{1c}^{n},(\bar{\bf Y}^{n}_{1d})^{1-\beta_2}\mid W_1,\mathcal{G})\nonumber\\
&\le&\hat{N}_1H(\bar{\bf Y}^{[n]}_{1c},(\bar{\bf Y}^{[n]}_{2b})^{1}_{\beta_1},(\bar{\bf Y}^{[n]}_{1d})^{1-\beta_2}\mid W_1,\mathcal{G})+\hat{N}_2H(\bar{\bf Y}_{1c}^{n},(\bar{\bf Y}^{n}_{1d})^{1-\beta_2}\mid W_1,\mathcal{G})\nonumber\\
&&+n~o~(\log{\bar{P}})\label{eed1}\\
&=&\hat{N}_1H((\bar{\bf Y}^{[n]}_{2b})^{1}_{\beta_1}\mid \bar{\bf Y}^{[n]}_{1c},(\bar{\bf Y}^{[n]}_{1d})^{1-\beta_2},W_1,\mathcal{G})+(\hat{N}_1+\hat{N}_2)H(\bar{\bf Y}_{1c}^{n},(\bar{\bf Y}^{n}_{1d})^{1-\beta_2}\mid W_1,\mathcal{G})\nonumber\\
&&+n~o~(\log{\bar{P}})\label{eed2}\\
&=&\hat{N}_1H((\bar{\bf X}^{[n]}_{c})^{1}_{\beta_1}\mid \bar{\bf Y}^{[n]}_{1c},(\bar{\bf Y}^{[n]}_{1d})^{1-\beta_2},W_1,\mathcal{G})+(\hat{N}_1+\hat{N}_2)H(\bar{\bf Y}_{1c}^{n},(\bar{\bf Y}^{n}_{1d})^{1-\beta_2}\mid W_1,\mathcal{G})\nonumber\\
&&+n~o~(\log{\bar{P}})\label{eed3}\\
&=&\hat{N}_1H(\bar{C}^{[n]}_{1},\bar{C}^{[n]}_{2},\cdots,\bar{C}^{[n]}_{\hat{N}_1+\hat{N}_2}\mid \bar{\bf Y}^{[n]}_{1c},(\bar{\bf Y}^{[n]}_{1d})^{1-\beta_2},W_1,\mathcal{G})\nonumber\\
&&+(\hat{N}_1+\hat{N}_2)H(\bar{\bf Y}_{1c}^{n},(\bar{\bf Y}^{n}_{1d})^{1-\beta_2}\mid W_1,\mathcal{G})+n~o~(\log{\bar{P}})\label{eed4}\\
&\le&\sum_{j=1}^{\hat{N}_1+\hat{N}_2}H(\bar{C}^{[n]}_{i_1},\bar{C}^{[n]}_{i_2},\cdots,\bar{C}^{[n]}_{i_{\hat{N}_1}}\mid \bar{\bf Y}^{[n]}_{1c},(\bar{\bf Y}^{[n]}_{1d})^{1-\beta_2},W_1,\mathcal{G})\nonumber\\
&&+(\hat{N}_1+\hat{N}_2)H(\bar{\bf Y}_{1c}^{n},(\bar{\bf Y}^{n}_{1d})^{1-\beta_2}\mid W_1,\mathcal{G})+n~o~(\log{\bar{P}})\label{eed5}\\
&=&\sum_{j=1}^{\hat{N}_1+\hat{N}_2}H(\bar{C}^{[n]}_{i_1},\bar{C}^{[n]}_{i_2},\cdots,\bar{C}^{[n]}_{i_{\hat{N}_1}}, \bar{\bf Y}^{[n]}_{1c},(\bar{\bf Y}^{[n]}_{1d})^{1-\beta_2}\mid W_1,\mathcal{G})+n~o~(\log{\bar{P}})\label{eed6}\\
&\le&(\hat{N}_1+\hat{N}_2)H(\bar{\bf Y}^{[n]}_{1}\mid W_1,\mathcal{G})+n~o~(\log{\bar{P}})\label{eed7}
\end{eqnarray}
Let us explain how \eqref{eed1}  follows from Lemma \ref{lemmaMIMOx}. Set $M_1=M_2=2$, and define $\bar{\bf U}_1$ and  $\bar{\bf U}_2$ as,
\begin{eqnarray}
\bar{\bf U}_1&=&\left(\bar{\bf Y}^{[n]}_{2a}\right)\label{lemmamimox1c}\\
\bar{\bf U}_2&=&\left(\bar{\bf Y}^{[n]}_{1c}\right)\label{lemmamimox2c}\\
W&=&\left((\bar{\bf Y}^{[n]}_{2b})_{\beta_1}^{1},(\bar{\bf Y}^{[n]}_{1d})^{1-\beta_2}\right)
\end{eqnarray}
From \eqref{lemmamimox5}, \eqref{eed1} is concluded as all the $(\lambda_{1i}-\lambda_{2i})^+$ are zero in the right side of \eqref{lemmamimox5}. \eqref{eed2} is true from the chain rule, \eqref{eed3} is concluded as $\beta_1+\beta_2\ge1$,  \eqref{eed4} is obtained from the definition of $\bar{C}^{[n]}_{i}$ in \eqref{defcn1} and (\ref{eed5}) follows from sub-modularity properties of the entropy function, i.e., for any $m$ random variables $\{X_1,X_2,\cdots,X_m\}$ where we define $X_{k+m}$ as $X_k$ for positive numbers  $k$ we have, 
\begin{eqnarray}
nH(X_1,\cdots,X_m)\le \sum_{i=1}^mH(X_i,X_{i+1},\cdots,X_{i+n}) \label{submod}
\end{eqnarray}
if $n\le m$.  \eqref{eed6} is true from the chain rule. \eqref{eed7}  is concluded from Theorem \ref{Theorem AIS04}. From \eqref{e4+}, define the random variables $Z_r(t)$, $Z_{r1}(t)$ and $Z_{r2}(t)$ for all $ r\in[\hat{N}_1]$ and $t\in[n]$ as
\begin {eqnarray}
Z_r(t)&=&\bar{y}_{1rd}(t)\label{etrt3}\\
Z_{r1}(t)&=&(\bar{y}_{1rd}(t))^{1}_{\beta_2}\nonumber\\
&\doteq& L_{3r}^g(t)\left((\bar{\mathbf{X}}_{a}(t))^{1}_{\beta_2}\right)\label{etrt4}\\
Z_{r2}(t)&=&\bar{C}_{i_{r}}(t)
\end{eqnarray}
where  $\lambda_{ri}$ is derived for any $i\in\{1,2\}$ as
\begin {eqnarray}
\lambda_{ri}&=&\left\{\begin{matrix}
\beta_2 &i=1 \\ 
1-\beta_2& i=2
\end{matrix}\right.
\end{eqnarray}

\section{Proof of Lemma \ref{lemma1BC}}\label{appbeta<1}
Define the set $s_\beta$ as the set $\{(\beta_1,\beta_2)\in \mathbb{R}^{2+};\beta_1,\beta_2\le1\}$. We claim that for any positive numbers  $Q_1$ and $Q_2$ the real-valued function $S(\beta_{1},\beta_{2})$ defined as,
\begin{eqnarray}
&&S(\beta_{1},\beta_{2})\nonumber\\
\overset{\underset{\mathrm{def}}{}}{=}&&\sum_{r=1}^{Q_1}f_r(\beta_{1},\beta_{2})\nonumber\\
&&\times H(L_{rs}^g({({\bf X}_{a}^n)}^{g_{rsa}(\beta_{1},\beta_{2})}\bigtriangledown{({\bf X}_{b}^n)}^{g_{rsb}(\beta_{1},\beta_{2})}\bigtriangledown{({\bf X}_{c}^n)}^{g_{rsc}(\beta_{1},\beta_{2})}),\forall s\in[Q_2])\nonumber\\
\end{eqnarray}
is a continuous function on the set $s_{\beta}$ under the conditions specified in the following lemma .  
\begin{lemma}\label{lemmacef}
$S(\beta_{1},\beta_{2})$ is a continuous function on the set $s_{\beta}$ if $f_r(\beta_{1},\beta_{2})$, $g_{rsl}(\beta_{1},\beta_{2})$ are bounded continuous functions on $s_{\beta}$ for any $r\in[Q_1],s\in[Q_2],l\in\{a,b,c,d\}$.
\end{lemma}
Proof of Lemma \ref{lemmacef} is relegated to Section \ref{lemmacefp}. Therefore, it is sufficient to prove Lemma \ref{lemma1BC} for the non-negative rational numbers $(\beta_{1},\beta_{2})=(\frac{m}{q},\frac{e}{q})$ where $m+e\le q$ \footnote{Note that for any real-valued continuous function $f(x_1,x_2,\cdots,x_n)$ on $\mathbb{R}^{n}$
\begin{eqnarray}
&\mbox{If}&f(x_1,x_2,\cdots,x_n)\le A, \forall (x_1,x_2,\cdots,x_n)\in \mathbb{Q}^{n} \nonumber\\
&\mbox{then}&f(x_1,x_2,\cdots,x_n)\le A, \forall (x_1,x_2,\cdots,x_n)\in \mathbb{R}^{n}\label{qert}
\end{eqnarray}
where \eqref{qert} follows from continuity of function $f(x_1,x_2,\cdots,x_n)$ on $\mathbb{R}^{n}$. Therefore, proving that a function $f(x_1,x_2,\cdots,x_n)$ on $\mathbb{R}^{n}$ is bounded by some number $A$ is equivalent to proving that the function is bounded by $A$ for  $(x_1,x_2,\cdots,x_n)\in \mathbb{Q}^{n}$. Thus, proving $\eqref{firstlemBC+}$ for the set $\{(\beta_1,\beta_2)\in \mathbb{R}^{2+};\beta_1+\beta_2\le1\}$ is equivalent to proving $\eqref{firstlemBC+}$ for the set $\{(\beta_1,\beta_2)\in \mathbb{Q}^{2+};\beta_1+\beta_2\le1\}$, i.e.,
\begin{eqnarray}
&&\breve{N}_1H(\bar{\bf Y}^{[n]}_{2}\mid W_1,\mathcal{G})\nonumber\\
&{\le}&\breve{N}_1H(\bar{\bf Y}^{n}_{1}\mid W_1,\mathcal{G})+\breve{N}_2H((\bar{\bf Y}^{n}_{1b})_{\beta_2}\mid\bar{\bf Y}_{1a}^{n},(\bar{\bf Y}^{n}_{1b})^{1-\beta_2},W_1,\mathcal{G})+n\breve{N}_0\log{\bar{P}}+n~o~(\log{\bar{P}})\nonumber\\
&&, (\beta_1,\beta_2)\in \mathbb{Q}^{2+};\beta_1+\beta_2\le1
\end{eqnarray}

}.
From the definition of $\bar{\bf Y}_{1c}(t)$, $\bar{\bf Y}_{1d}(t)$, $\bar{\bf Y}_{2a}(t)$, $\bar{\bf Y}_{2c}(t)$ and $\bar{\bf Y}_{2d}(t)$ in (\eqref{e2+}-\eqref{rre1BqqCq2}) we have,
\begin{eqnarray}
&&\breve{N}_1H(\bar{\bf Y}^{[n]}_{2a},\bar{\bf Y}^{[n]}_{2c},\bar{\bf Y}^{[n]}_{2d}\mid W_1,\mathcal{G})-n\breve{N}_0\log{\bar{P}}\nonumber\\
&=&\breve{N}_1H(\bar{\bf Y}^{[n]}_{2a},\bar{\bf Y}^{[n]}_{2c},\bar{\bf Y}^{[n]}_{2d}\mid W_1,\mathcal{G})-n\breve{N}_1(M-N_1)\beta_1\log{\bar{P}}\\
&\le&\breve{N}_1H(\bar{\bf Y}^{[n]}_{2a},\bar{\bf Y}^{[n]}_{2c},\bar{\bf Y}^{[n]}_{2d}\mid W_1,\mathcal{G})-\breve{N}_1H((\bar{\bf Y}^{[n]}_{2c})_{\beta_1},(\bar{\bf Y}^{[n]}_{2d})_{\beta_1}\mid W_1,\mathcal{G})\\
&{\le}&\breve{N}_1H(\bar{\bf Y}^{[n]}_{2a},(\bar{\bf Y}^{[n]}_{2c})^{1}_{\beta_1},(\bar{\bf Y}^{[n]}_{2d})^{1}_{\beta_1}\mid W_1,\mathcal{G})\label{fr2}
\end{eqnarray}
where $\bar{\bf Y}_{2c}(t)$ and $\bar{\bf Y}_{2d}(t)$ are defined from the random variable $\bar{\bf Y}_{2b}(t)$ in \eqref{frreqqq1} and \eqref{rre1BqqCq2} as,
\begin{eqnarray}
\bar{\bf Y}_{2c}(t)&=&[\bar{\bf Y}_{2b}(t)]_{0\rightarrow(M-N_2)}\\
\bar{\bf Y}_{2d}(t)&=&[\bar{\bf Y}_{2b}(t)]_{(M-N_2)\rightarrow N_2-N_1}
\end{eqnarray}
\eqref{fr2} is true from the chain rule. Similar to \eqref{firstlemBC-1}, from \eqref{fr1} it is sufficient to demonstrate the following inequality,
\begin{eqnarray}
&&\breve{N}_1H(\bar{\bf Y}^{[n]}_{2a},(\bar{\bf Y}^{[n]}_{2c})^{1}_{\frac{m}{q}},(\bar{\bf Y}^{[n]}_{2d})^{1}_{\frac{m}{q}}\mid W_1,\mathcal{G})+\breve{N}_2H(\bar{\bf Y}_{1c}^{n},(\bar{\bf Y}^{n}_{1d})^{1}_{\frac{e}{q}}\mid W_1,\mathcal{G})\nonumber\\
&{\le}&(\breve{N}_1+\breve{N}_2)H(\bar{\bf Y}^{n}_{1}\mid W_1,\mathcal{G})+n~o~(\log{\bar{P}})\label{firstlemBC-p}
\end{eqnarray}
where the numbers $\breve{N}_1$ and $\breve{N}_2$ can be rewritten as,
\begin{eqnarray}
\breve{N}_1&=&(M-N_2)\frac{e}{q}\\
\breve{N}_2&=&(N_2-N_1)\frac{q-m}{q}
\end{eqnarray}
Before preceding to proof of \eqref{firstlemBC-p} let us define the random variables $\bar{C}^{[n]}_{i}$ as the distinct $\frac{1}{q}$ power levels of $(\bar{\bf Y}^{[n]}_{2c})_{\frac{m}{q}}^{\frac{m+e}{q}}$ and $(\bar{\bf Y}^{[n]}_{2d})^{1}_{\frac{m}{q}}$, i.e.,
\begin{eqnarray}
\bar{C}_{i}(t)&=&\left\{\begin{matrix}
(\bar{Y}_{2c}^{\lfloor\frac{i-1}{e}\rfloor+1}(t))_{\frac{m+i-1-e\lfloor\frac{i-1}{e}\rfloor}{q}}^{\frac{m+i-e\lfloor\frac{i-1}{e}\rfloor}{q}}&1\le i\le q\breve{N}_1\\ 
(\bar{Y}_{2d}^{\lfloor\frac{\bar{i}}{q-m}\rfloor+1}(t))_{\frac{\bar{i}-(q-m)\lfloor\frac{\bar{i}}{q-m}\rfloor}{q}}^{\frac{\bar{i}+1-(q-m)\lfloor\frac{\bar{i}}{q-m}\rfloor}{q}}&q\breve{N}_1< {i}\le q(\breve{N}_1+\breve{N}_2)\\
\bar{C}_{i-q(\breve{N}_1+\breve{N}_2)}(t)&q(\breve{N}_1+\breve{N}_2)< i
\end{matrix}\right.\label{defcn1}
\end{eqnarray}
where $\bar{i}=i-e(M-N_2)-1$. For instance, $C_{1}(t)$ is defined as $(\bar{Y}_{2c}^{1}(t))_{\frac{m}{q}}^{\frac{m+1}{q}}$, i.e., the bottom $\frac{1}{q}$ power level of $\bar{Y}_{2c}^{1}(t)$. Starting from the left side of  \eqref{firstlemBC-p}, we have
\begin{eqnarray}
&&\breve{N}_1H(\bar{\bf Y}^{[n]}_{2a},(\bar{\bf Y}^{[n]}_{2c})^{1}_{\frac{m}{q}},(\bar{\bf Y}^{[n]}_{2d})^{1}_{\frac{m}{q}}\mid W_1,\mathcal{G})+\breve{N}_2H(\bar{\bf Y}_{1c}^{n},(\bar{\bf Y}^{n}_{1d})^{1}_{\frac{e}{q}}\mid W_1,\mathcal{G})\nonumber\\
&=&\breve{N}_1H(\bar{\bf Y}^{[n]}_{2a},(\bar{\bf Y}^{[n]}_{2c})_{\frac{m}{q}}^{\frac{m+e}{q}},(\bar{\bf Y}^{[n]}_{2c})^{1}_{\frac{m+e}{q}},(\bar{\bf Y}^{[n]}_{2d})^{1}_{\frac{m}{q}}\mid W_1,\mathcal{G})+\breve{N}_2H(\bar{\bf Y}_{1c}^{n},(\bar{\bf Y}^{n}_{1d})^{1}_{\frac{e}{q}}\mid W_1,\mathcal{G})\label{qqee1}\\
&\le&\breve{N}_1H(\bar{\bf Y}^{[n]}_{1c},(\bar{\bf Y}^{[n]}_{2c})_{\frac{m}{q}}^{\frac{m+e}{q}},(\bar{\bf Y}^{[n]}_{1d})^{1}_{\frac{e}{q}},(\bar{\bf Y}^{[n]}_{2d})^{1}_{\frac{m}{q}}\mid W_1,\mathcal{G})+\breve{N}_2H(\bar{\bf Y}_{1c}^{n},(\bar{\bf Y}^{n}_{1d})^{1}_{\frac{e}{q}}\mid W_1,\mathcal{G})\nonumber\\
&&+n~o~(\log{\bar{P}})\label{qqee2}\\
&=&\breve{N}_1H((\bar{\bf Y}^{[n]}_{2c})_{\frac{m}{q}}^{\frac{m+e}{q}},(\bar{\bf Y}^{[n]}_{2d})^{1}_{\frac{m}{q}}\mid \bar{\bf Y}^{[n]}_{1c},(\bar{\bf Y}^{[n]}_{1d})^{1}_{\frac{e}{q}},W_1,\mathcal{G})+(\breve{N}_1+\breve{N}_2)H(\bar{\bf Y}_{1c}^{n},(\bar{\bf Y}^{n}_{1d})^{1}_{\frac{e}{q}}\mid W_1,\mathcal{G})\nonumber\\
&&+n~o~(\log{\bar{P}})\label{qqee3}\\
&=&(M-N_2)\frac{e}{q}H(\bar{C}^{[n]}_{1},\bar{C}^{[n]}_{2},\cdots,\bar{C}^{[n]}_{(M-N_2)e+(N_2-N_1)(q-m)}\mid \bar{\bf Y}^{[n]}_{1c},(\bar{\bf Y}^{[n]}_{1d})^{1}_{\frac{e}{q}},W_1,\mathcal{G})\nonumber\\
&&+((M-N_2)\frac{e}{q}+(N_2-N_1)\frac{q-m}{q})H(\bar{\bf Y}_{1c}^{n},(\bar{\bf Y}^{n}_{1d})^{1}_{\frac{e}{q}}\mid W_1,\mathcal{G})+n~o~(\log{\bar{P}})\label{qqee4}\\
&\le&\frac{1}{q}\sum_{j=1}^{(M-N_2)e+(N_2-N_1)(q-m)}H(\bar{C}^{[n]}_{i_1},\bar{C}^{[n]}_{i_2},\cdots,\bar{C}^{[n]}_{i_{(M-N_2)e}}\mid \bar{\bf Y}^{[n]}_{1c},(\bar{\bf Y}^{[n]}_{1d})^{1}_{\frac{e}{q}},W_1,\mathcal{G})\nonumber\\
&&+((M-N_2)\frac{e}{q}+(N_2-N_1)\frac{q-m}{q})H(\bar{\bf Y}_{1c}^{n},(\bar{\bf Y}^{n}_{1d})^{1}_{\frac{e}{q}}\mid W_1,\mathcal{G})+n~o~(\log{\bar{P}})\label{qqee5}\\
&=&\frac{1}{q}\sum_{j=1}^{(M-N_2)e+(N_2-N_1)(q-m)}H(\bar{C}^{[n]}_{i_j},\bar{C}^{[n]}_{i_{j+1}},\cdots,\bar{C}^{[n]}_{i_{j-1+(M-N_2)e}},\bar{\bf Y}^{[n]}_{1c},(\bar{\bf Y}^{[n]}_{1d})^{1}_{\frac{e}{q}}\mid W_1,\mathcal{G})\nonumber\\
&&+n~o~(\log{\bar{P}})\label{qqee6}\\
&\le&\frac{(M-N_2)e+(N_2-N_1)(q-m)}{q}H(\bar{\bf Y}^{[n]}_{1}\mid W_1,\mathcal{G})+n~o~(\log{\bar{P}})\label{qqee7}\\
&=&(\breve{N}_1+\breve{N}_2)H(\bar{\bf Y}^{[n]}_{1}\mid W_1,\mathcal{G})+n~o~(\log{\bar{P}})\label{qqee8}
\end{eqnarray}
\eqref{qqee1} follows from Definition \ref{powerlevel}. Let us explain how \eqref{qqee2}  follows from Lemma \ref{lemmaMIMOx}. Set $M_1=M_2=2$, and define $\bar{\bf U}_1$ and  $\bar{\bf U}_2$ as,
\begin{eqnarray}
\bar{\bf U}_1&=&\left(\bar{\bf Y}^{[n]}_{2a},(\bar{\bf Y}^{[n]}_{2c})^{1}_{\frac{m+e}{q}}\right)\label{lemmamimox1c}\\
\bar{\bf U}_2&=&\left(\bar{\bf Y}^{[n]}_{1c},(\bar{\bf Y}^{[n]}_{1d})^{1}_{\frac{e}{q}}\right)\label{lemmamimox2c}\\
W&=&\left((\bar{\bf Y}^{[n]}_{2c})_{\frac{m}{q}}^{\frac{m+e}{q}},(\bar{\bf Y}^{[n]}_{2d})^{1}_{\frac{m}{q}}\right)
\end{eqnarray}
From \eqref{lemmamimox5}, \eqref{qqee2} is concluded as all the $(\lambda_{1i}-\lambda_{2i})^+$ are zero in the right side of \eqref{lemmamimox5}. \eqref{qqee3} is true from the chain rule, \eqref{qqee4} is obtained from the definition of $\bar{C}^{[n]}_{i}$ in \eqref{defcn1} and (\ref{qqee5}) follows from sub-modularity properties of the entropy function, see \eqref{submod}. \eqref{qqee6} and \eqref{qqee8} are true from the chain rule. Let us clarify how \eqref{qqee7} is concluded from Theorem \ref{Theorem AIS04} [Theorem $4$ in \cite{Arash_Jafar_sumset}], i.e.,
\begin {eqnarray}
&&H(\bar{C}^{[n]}_{i_j},\bar{C}^{[n]}_{i_{j+1}},\cdots,\bar{C}^{[n]}_{i_{j-1+(M-N_2)e}},(\bar{\bf Y}^{[n]}_{1d})^{1}_{\frac{e}{q}}\mid \bar{\bf Y}^{[n]}_{1c},W_1,\mathcal{G})\\
&\le&H(\bar{\bf Y}^{[n]}_{1d}\mid \bar{\bf Y}^{[n]}_{1c},W_1,\mathcal{G})+n~o~(\log{\bar{P}})\\
\end {eqnarray}
 From \eqref{e4+}, define the random variables $Z_r(t)$, and $Z_{ri}(t)$ for all $ r\in[(M-N_2)]$ and $t\in[n]$ as
\begin {eqnarray}
Z_r(t)&=&\bar{y}_{1rd}(t)\label{etrt3}\\
Z_{r1}(t)&=&(\bar{y}_{1rd}(t))^{1}_{\frac{e}{q}}\label{etrt4}\\
Z_{r,m+1}(t)&=&\bar{C}_{i_{j+(r-1)e+m-1}}(t),\forall m\in[e]\label{etrt5}
\end{eqnarray}
where  $\lambda_{ri}$ is derived for any $i\in[e+1]$ as
\begin {eqnarray}
\lambda_{ri}&=&\left\{\begin{matrix}
\frac{1}{q} &0<i\le e \\ 
\frac{q-e}{q} & i=e+1
\end{matrix}\right.
\end{eqnarray}
Let us prove \eqref{qqee7} in detail.

\section{Proof of \eqref{qqee7}}
We wish to prove that for the random variables $Z_r(t)$, and $Z_{ri}(t)$ defined in (\eqref{etrt3}-\eqref{etrt5}),
\begin {eqnarray}
&&H(Z_{11}^{[n]},\cdots,Z_{(M-N_2),(e+1)}^{[n]}\mid W)-H(Z_1^{[n]},\cdots,Z_{(M-N_2)}^{[n]}\mid W,\mathcal{G}) \nonumber\\
&\leq&+n~o(\log{\bar{P}})\label{pr1}
\end{eqnarray}

\subsection{Main idea of the proof}
First of all, let us go over a toy example in order to better grasp the main idea of the proof. Assume $n=1,M-N_2=1$. Moreover, define
\begin {eqnarray}
Z&=&L_1^g(X_1,X_2)\\
Z'&=&L_2^g(X_1,X_2)\\
Z_a&=&(Z)_{0.8}^1\\
Z_b&=&(Z')_{0.8}
\end{eqnarray}
First of all, note that we can assume that $Z_a=(Z)_{0.8}^1\doteq L_1^g({(X_1)}_{0.8}^1,{(X_2)}_{0.8}^1)$ which is concluded similar to \eqref{eq:doteq1}. Let us prove that,
\begin {eqnarray}
H(Z_a,Z_b)-H(Z\mid \mathcal{G})
&\leq&o(\log{\bar{P}})\label{5t}
\end{eqnarray}
Intuitively, \eqref{5t} is true as treating $Z'$ the same as $Z$ we have,
\begin {eqnarray}
H((Z)_{0.8},(Z)_{0.8}^1)-H(Z\mid \mathcal{G})
&\leq&o(\log{\bar{P}})\label{5t//}
\end{eqnarray}
which is true from Definition \ref{powerlevel}. Let us prove \eqref{5t} as follows. Define $U'$ and $U$ as $(Z_a,Z_b)$ and $Z$, respectively.  Then, our goal is to prove that
\begin{eqnarray}
H(U')-H( U\mid  \mathcal{G})&\le& o~(\log{\bar{P}})\label{lemmaxq}
\end{eqnarray}
 Similar to the AIS approach in \cite{Arash_Jafar_sumset}, we can assume that $ U$ is a function of $U', \mathcal{G}$. See section \ref{sec_AIS+} for details. Moreover, we have
\begin{eqnarray}
H(U')-H( U\mid  \mathcal{G})
&\le&\log\left\{\mbox{E}_{\mathcal{G}}\left|\mathcal{S}_{\nu}(\mathcal{G})\right| \right\}
\end{eqnarray}  
where $\mathcal{S}_{\nu}(\mathcal{G})$ is the set of all $ U'$ which result in the same $ U$ as $\nu$. 
\begin{eqnarray}
\mbox{E}_{\mathcal{G}}\{|S_{\nu}(\mathcal{G})|\}=\sum_{{\lambda}}\mathbb{P}({\lambda}) 
\end{eqnarray}
where $\mathbb{P}({\lambda}) $ is defined as the probability that ${\lambda}$ and $\nu$ correspond to the same $ U$. 
\subsubsection{ Bounding the Probability of Image Alignment}
Given $\mathcal{G}$, consider two distinct instances of $ U'$ ~denoted as $\lambda=(\lambda_1,\lambda_2)$ and $\nu=(\nu_1,\nu_2)$ produced by corresponding  realizations of codewords $(X_1,X_2)$ denoted by $(E_1,E_2)$ and $(F_1,F_2)$, respectively.
\begin{eqnarray}
\lambda_1&=&L_1^g({(E_1)}_{0.8}^1,{(E_2)}_{0.8}^1)\\
\lambda_2&=&{(L_2^g(E_1,E_2))}_{0.8}\\
\nu_1&=&L_1^g({(F_1)}_{0.8}^1,{(F_2)}_{0.8}^1)\\
\nu_2&=&{(L_2^g(F_1,F_2))}_{0.8}
\end{eqnarray}
Now, let us bound $\mathbb{P}(\lambda\in \mathcal{S}_{\nu})$ from above. We wish to bound the probability that the images of these two codewords align, or in other words $ U(\lambda,\mathcal{G})=U(\nu,\mathcal{G})$. Thus, we have
\begin{eqnarray}
L_{1}^g(E_1,E_2)&=&L_{1}^g(F_1,F_2)\\
\rightarrow \lfloor g_{1}E_1\rfloor+\lfloor g_{2}E_2\rfloor&=&\lfloor g_{1}F_1\rfloor+\lfloor g_{2}F_2\rfloor
\end{eqnarray}
For fixed value of $g_{1}$ the random variable $g_2(E_2-F_2)$ must take values within an interval of length no more than $4$. Thus, the probability of which is no more than $\frac{4f_{\max}}{|E_2-F_2|}$ if $E_2\neq F_2$.  The probability of alignment is  bounded by $\frac{4f_{\max}}{\max(|E_1-F_1|,|E_2-F_2|)}$ if either $E_1-F_1\neq0$ or $E_2-F_2\neq0$.
\subsubsection{\bf Bounding the Average Size of Aligned Image Sets}
Our goal is to prove that,
\begin{eqnarray}
\mbox{E}_{\mathcal{G}}\left|\mathcal{S}_{\nu}(\mathcal{G})\right|=\sum_{|\lambda_{1}-\nu_{1}|\in {\mathcal{Z}_{1}},|\lambda_{2}-\nu_{2}|\in {\mathcal{Z}_{2}}}\frac{4f_{\max}}{\max(|E_1-F_1|,|E_2-F_2|)}&\le& c{\log{\bar{P}}}
\end{eqnarray}
for some positive constant $c$  not depending on $P$ which results in $H(U')-H( U\mid  \mathcal{G})
\le\log(c\log P)$. ${\mathcal{Z}_{i}}$ is defined as the support of  the random variable $|\lambda_{i}-\nu_{i}|$ \footnote{From Definition \ref{def:length}, we know that $\mathcal{Z}_{1}=\{a:a\in\mathbb{Z}, |a| \le 2\Delta\bar{P}^{0.2}\}$, $\mathcal{Z}_{2}=\{a:a\in\mathbb{Z}, |a| \le 2\Delta\bar{P}^{0.8}\}$.}. Define $\Delta_j$ as follows
 \begin{eqnarray}
{\Delta}_{j}&\triangleq&{(E_j)}_{0.8}^1-{(F_j)}_{0.8}^1, \forall j\in\{1,2\}
\end{eqnarray}
Consider the following two cases of $\max_{j\in \{1,2\}}|{\Delta}_{j}|\ge2$ and $\max_{j\in \{1,2\}}|{\Delta}_{j}|\le1$.
\begin{enumerate}
\item{$\max_{j\in \{1,2\}}|{\Delta}_{j}|\ge2$}\\
Any number $X\in\mathcal{X}_\delta$ can be written as $(X)_{a}^{\delta}\bar{P}^a+X_a$ for any non-negative number $a$ less than $\delta$. Thus, we have
\begin{eqnarray}
&&|E_j-F_j|\nonumber\\
&=&|{\Delta}_{j}\bar{P}^{0.8}+(E_j)_{0.8}-(F_j)_{0.8}|\\
&\ge&|{\Delta}_{j}\bar{P}^{0.8}|-|(E_j)_{0.8}-(F_j)_{0.8}|\\
&\ge&|{\Delta}_{j}\bar{P}^{0.8}|-\bar{P}^{0.8}\\
&=&(|{\Delta}_{j}|-1)\bar{P}^{0.8}
\end{eqnarray}
Therefore, $\mbox{E}_{\mathcal{G}}\left|\mathcal{S}_{\nu}(\mathcal{G})\right|$ is bounded as follows.
\begin{eqnarray}
&&\mbox{E}_{\mathcal{G}}\left|\mathcal{S}_{\nu}(\mathcal{G})\right|\nonumber\\
&=&\sum_{|\lambda_{1}-\nu_{1}|\in {\mathcal{Z}_{1}},|\lambda_{2}-\nu_{2}|\in {\mathcal{Z}_{2}}}\frac{4f_{\max}}{\max(|E_1-F_1|,|E_2-F_2|)}\\
&\le& \sum_{|\lambda_{1}-\nu_{1}|\in {\mathcal{Z}_{1}},|\lambda_{2}-\nu_{2}|\in {\mathcal{Z}_{2}}, 2\le\max_{j\in \{1,2\}}|{\Delta}_{j}|}\frac{4f_{\max}}{\bar{P}^{0.8}\max(|{\Delta}_{1}|-1,|{\Delta}_{2}|-1)}\\
&\le& \sum_{|\lambda_{1}-\nu_{1}|\in {\mathcal{Z}_{1}},|\lambda_{2}-\nu_{2}|\in {\mathcal{Z}_{2}},|\lambda_{1}-\nu_{1}|<2\Delta+5}\frac{4f_{\max}}{\bar{P}^{0.8}}\label{f5+}\\
&&+ \sum_{|\lambda_{1}-\nu_{1}|\in {\mathcal{Z}_{1}},|\lambda_{2}-\nu_{2}|\in {\mathcal{Z}_{2}},2\Delta+5\le|\lambda_{1}-\nu_{1}|}\frac{4f_{\max}}{\bar{P}^{0.8}(\frac{|\lambda_{1}-\nu_{1}|-4}{2\Delta}-1)}\label{f5+}\\
&\le&\frac{8f_{\max}|\mathcal{Z}_{2}|(2\Delta+5)}{\bar{P}^{0.8}}+\frac{4f_{\max}|\mathcal{Z}_{2}|}{\bar{P}^{0.8}} \sum_{|\lambda_{1}-\nu_{1}|\in {\mathcal{Z}_{1}},2\Delta+5\le|\lambda_{1}-\nu_{1}|}\frac{1}{(|\lambda_{1}-\nu_{1}|-2\Delta-5)}\label{f6+}\\
&\le&{8f_{\max}\Delta(2\Delta+5)}+8 f_{\max}\Delta\log(2\Delta\bar{P}^{0.2})\label{f7+}
\end{eqnarray}
where \eqref{f5+} is true as,
\begin{eqnarray}
|\lambda_{1}-\nu_{1}|&=&|L_1^g({(E_1)}_{0.8}^1,{(E_2)}_{0.8}^1)-L_1^g({(F_1)}_{0.8}^1,{(F_2)}_{0.8}^1)|\\
&\le&4+2\Delta\max(|{(E_1)}_{0.8}^1-{(F_1)}_{0.8}^1,{(E_2)}_{0.8}^1-{(F_2)}_{0.8}^1|)\label{mm+1}
\end{eqnarray}
\eqref{f6+} is concluded as for any summation $\sum_{a\in S_a,b\in S_b}f(a,b)$ and the real-valued function $f(a,b)$ we have,
\begin{eqnarray}
\sum_{a\in S_a,b\in S_b}f(a,b)\le |S_a|\sum_{b\in S_b}\max_{a\in S_a}f(a,b)\le|S_a||S_b|\max_{a\in S_a,b\in S_b}f(a,b)\label{summation}
\end{eqnarray}
\eqref{f7+} ) is true as the partial sum of harmonic series can be bounded above by logarithmic function, i.e., $\sum_{i=1}^n\frac{1}{i}\le\log n$.

\item{$\max_{j\in \{1,2\}}|{\Delta}_{j}|\le1$.}\\
In this case, from \eqref{mm+1} the random variable $|\lambda_{1}-\nu_{1}|$ can only takes values from the set $\{a:a\in\mathbb{Z}, |a| \le 4+2\Delta\}$. Therefore, we bound the expected value of $\mbox{E}_{\mathcal{G}}$ as follows
\begin{eqnarray}
&&\sum_{|\lambda_{1}-\nu_{1}|\in {\mathcal{Z}_{1}},|\lambda_{2}-\nu_{2}|\in {\mathcal{Z}_{2}}}\frac{4f_{\max}}{\max(|E_1-F_1|,|E_2-F_2|)}\\
&\le&\sum_{|\lambda_{2}-\nu_{2}|\in {\mathcal{Z}_{2}}}\frac{8f_{\max}(2+\Delta)}{\max(|E_1-F_1|,|E_2-F_2|)}\label{mm+9}
\end{eqnarray}
Let us first compute the term $|\lambda_{2}-\nu_{2}|$.
\begin{eqnarray}
&&|\lambda_{2}-\nu_{2}|\nonumber\\
&=&|{(L_2^g(E_1,E_2))}_{0.8}-{(L_2^g(F_1,F_2))}_{0.8}|\\
&=&|L_2^g(E_1,E_2)-\bar{P}^{0.8}\lfloor\frac{L_2^g(E_1,E_2)}{\bar{P}^{0.8}} \rfloor-L_2^g(F_1,F_2)+\bar{P}^{0.8}\lfloor\frac{L_2^g(F_1,F_2)}{\bar{P}^{0.8}} \rfloor|\nonumber\\
&=&I_1|L_2^g(E_1,E_2)-L_2^g(F_1,F_2)|+I_2\bar{P}^{0.8}\left(\lfloor\frac{L_2^g(E_1,E_2)}{\bar{P}^{0.8}} \rfloor-\lfloor\frac{L_2^g(F_1,F_2)}{\bar{P}^{0.8}} \rfloor\right)\\
&=&I_1|L_2^g(E_1,E_2)-L_2^g(F_1,F_2)|+I_2\bar{P}^{0.8}I_3\label{jhgf1}
\end{eqnarray}
where $I_1,I_2\in\{-1,1\}$ and $I_3$ is  an integer-valued random variable taking numbers from the set $\{a:a\in\mathbb{Z}, |a| \le 4+2\Delta\}$. $I_1,I_2$ take numbers from the set $\{-1,1\}$ as for any real numbers $a$ and $b$, we have $|a+b|=I_1|a|+I_2b$ for  $I_1,I_2\in\{-1,1\}$. Moreover, $|I_3|\le 4+2\Delta$ follows similar to \eqref{mm+1} and the fact that  $\max_{j\in \{1,2\}}|{\Delta}_{j}|\le1$. Therefore, from \eqref{mm+9} we have
\begin{eqnarray}
&&\sum_{|\lambda_{2}-\nu_{2}|\in {\mathcal{Z}_{2}}}\frac{8f_{\max}(2+\Delta)}{\max(|E_1-F_1|,|E_2-F_2|)}\nonumber\\
&\le&\sum_{|\lambda_{2}-\nu_{2}|\in {\mathcal{Z}_{2}}}\frac{8f_{\max}(2+\Delta)}{\max\left(\frac{|L_2^g(E_1,E_2)-L_2^g(F_1,F_2)|-4}{2\Delta},1\right)}\label{kklj1}\\
&=&\sum_{|\lambda_{2}-\nu_{2}|\in {\mathcal{Z}_{2}}}\frac{16\Delta f_{\max}(2+\Delta)}{\max\left(||\lambda_{2}-\nu_{2}|-I_2I_3\bar{P}^{0.8}-4I_1|,1\right)}\label{jhgf2}\\
&\le&16\Delta f_{\max}(2+\Delta)2\Delta(4\Delta+13)\log(2\Delta\bar{P}^{0.8})\label{tt+12}\\
&=&O(\log \bar{P})
\end{eqnarray}
\eqref{kklj1} is concluded similar to \eqref{f5+} as,
\begin{eqnarray}
|L_2^g(E_1,E_2)-L_2^g(F_1,F_2)|&\le&4+2\Delta\max(|E_1-F_1|,|E_2-F_2|)
\end{eqnarray}
\eqref{jhgf2} is true from \eqref{jhgf1} and \eqref{tt+12} follows from the following inequality. For any   positive integer number $a$ and any integer-valued functions $b(n)$ and $c(n)$ whose absolute values are bounded by $M$ we have
\begin{eqnarray}
\sum_{i=1}^{an}\frac{1}{\max(|i-b(n)n-c(n)|,1)}\le a(2M+5)\log((a+M)n+M)\label{tt+13}
\end{eqnarray}
 \eqref{tt+13} is true as we count each number at most $a(2M+5)$ times, e.g., consider the term $\frac{1}{2}$. $i$ can be  $b(n)n+c(n)+2$ or $b(n)n+c(n)-2$ for any $b(n)\in\{1,2,\cdots,a\},c(n)\in\{-M,\cdots,0,\cdots,M\}$. 
\end{enumerate}

\subsection{Aligned Image Sets}\label{sec_AIS+} Define ${\bar{\bf U}}'$ and ${\bar{\bf U}}$ as $\big(Z_{11}^{[n]},\cdots,Z_{(M-N_2),(e+1)}^{[n]}\big)$ and $\big(Z_1^{[n]},\cdots,Z_{(M-N_2)}^{[n]}\big)$, respectively.  Let us prove,
\begin{eqnarray}
H({\bar{\bf U}}'\mid  {W},\mathcal{G})-H({\bar{\bf U}}\mid  {W},\mathcal{G})&\le& n~o~(\log{\bar{P}})\label{lemmaxq}
\end{eqnarray}
 We are only interested in the difference of entropies of ${\bar{\bf U}}'$ and ${\bar{\bf U}}$ conditioned on $ W$ and $\mathcal{G}$.  Similar to the AIS approach in \cite{Arash_Jafar_sumset}, we first claim that from the functional dependence, ${\bar{\bf U}}$ can be made a function of ${\bar{\bf U}}', W,\mathcal{G}$. Consider some instance of ${\bar{\bf U}}'$, e.g., $\nu^{[n]}$.  For given $ W$ and channel realization $\mathcal{G}$, define aligned image set $\mathcal{S}_{\nu^{[n]}}( W,\mathcal{G})$ as the set of all ${\bar{\bf U}}'$ which result in the same ${\bar{\bf U}}$ as $\nu^{[n]}$. Since uniform distribution maximizes the entropy,
\begin{eqnarray}
\mathcal{D}_{\Delta}&\triangleq&H({\bar{\bf U}}'\mid  W,\mathcal{G})-H({\bar{\bf U}}\mid  W,\mathcal{G})\nonumber\\
&\le&H({\bar{\bf U}}'\mid {\bar{\bf U}},W,\mathcal{G})\label{+jen00}\\
&\le&\mbox{E}_{\mathcal{G}}\left\{\log{\left|\mathcal{S}_{\nu^{[n]}}(W,\mathcal{G})\right| }\right\}\label{+jen0}\\
&=&\mbox{E}_{W}\left\{\mbox{E}_{\mathcal{G}}\left\{\log{\left|\mathcal{S}_{\nu^{[n]}}(W,\mathcal{G})\right| }\mid W\right\}\right\}\label{+jen0}\\
&\le&\max_{w\in\mathcal{W}}\mbox{E}_{\mathcal{G}}\left\{\log{\left|\mathcal{S}_{\nu^{[n]}}(W,\mathcal{G})\right| }\mid W=w\right\}\label{+jen0}\\
&\le& \max_{w\in\mathcal{W}}\log\left\{\mbox{E}_{\mathcal{G}}{\left|\mathcal{S}_{\nu^{[n]}}(W,\mathcal{G})\right| }{\mid W=w}\right\}\label{+jen}
\end{eqnarray}  
where $\mathcal{W}$ is defined as the support of the random variable $W$. We are only interested in the difference of entropies of ${\bar{\bf U}}'$ and ${\bar{\bf U}}$ conditioned on $W$ and $\mathcal{G}$, i.e., $H({\bar{\bf U}}'\mid W,\mathcal{G})-H({\bar{\bf U}}\mid W,\mathcal{G})$. Similar to the AIS approach in \cite{Arash_Jafar_sumset}, we start with functional dependence. From the functional dependence argument, without loss of generality ${\bar{\bf U}}$ can be made a function of ${\bar{\bf U}}',W,\mathcal{G}$.  So, from (\ref{+jen}), $\mbox{E}_{\mathcal{G}}\{|S_{\nu^n}(W,\mathcal{G})|\mid W=w\}$ is what needed to be calculated. Expected value of size of the cardinality of aligned image set is equal to the summation of probability of alignment over all ${\lambda}^n$, or in the other words,
\begin{eqnarray}
\mbox{E}_{\mathcal{G}}\{|S_{\nu^n}(W,\mathcal{G})|{\mid W=w}\}=\sum_{{\lambda}^n}\mathbb{P}({\lambda}^n) \label{vbcc+}
\end{eqnarray}
where $\mathbb{P}({\lambda}^n) $ is defined as the probability that ${\lambda}^n$ and $\nu^n$ correspond to the same ${\bar{\bf U}}$.

\subsection{ Bounding the Probability of Image Alignment}
Given ~$\mathcal{G}$ and $W=w$, ~consider~ two ~distinct ~instances ~of ~${\bar{\bf U}}'$ ~denoted as $\lambda^{[n]}=(\lambda_{11}^{[n]},\cdots,$ $\lambda_{(M-N_2),(e+1)}^{[n]})$ and $\nu^{[n]}=(\nu_{11}^{[n]},\cdots,\nu_{(M-N_2),(e+1)}^{[n]})$ produced by corresponding  realizations of codewords $(X_1^n,X_2^n,\cdots,X_M^n)$ denoted by $(E_1^n,E_2^n,\cdots,E_M^n)$ and $(F_1^n,F_2^n,\cdots,F_M^n)$, respectively. For any $k\in[(M-N_2)]$, $l\in [e+1]$, $t\in[n]$, the random variables $\lambda_{kl}(t)$ and $\nu_{kl}(t)$ are derived as, 
\begin{eqnarray}
\lambda_{kl}(t)&=&\left\{\begin{matrix}
L_{k1}^h(t)\left((\bar{\mathbf{E}}_{a}(t))^{1}_{\frac{e}{q}}\bigtriangledown(\bar{\mathbf{E}}_{c}(t))^{1}_{\frac{m+e}{q}}\right) &l=1 \\ 
\left(L_{kl}^g(t)((\bar{\mathbf{E}}_{a}(t))^{1}_{\frac{e}{q}}\bigtriangledown\bar{\mathbf{E}}_{c}(t))\right)_{\frac{a_{kl}-1}{q}}^{\frac{a_{kl}}{q}}& 1<l\le e+1
\end{matrix}\right.\label{+x1}\\
\nu_{kl}(t)&=&\left\{\begin{matrix}
L_{k1}^h(t)\left((\bar{\mathbf{F}}_{a}(t))^{1}_{\frac{e}{q}}\bigtriangledown(\bar{\mathbf{F}}_{c}(t))^{1}_{\frac{m+e}{q}}\right) &l=1 \\ 
\left(L_{kl}^g(t)((\bar{\mathbf{F}}_{a}(t))^{1}_{\frac{e}{q}}\bigtriangledown\bar{\mathbf{F}}_{c}(t))\right)_{\frac{a_{kl}-1}{q}}^{\frac{a_{kl}}{q}}& 1<l\le e+1
\end{matrix}\right.\label{+x2}
\end{eqnarray}
where for any $k\in[(M-N_2)]$ we assume that $a_{kl}$ are  arbitrary distinct decreasing numbers belonging to the set  $\{m+1,m+2,\cdots,q\}$, i.e., we assume that  for any $2\le l<l'\le e+1$, $a_{kl}>a_{kl'}$ and $\{a_{k2},a_{k3},\cdots,a_{k,e+1}\}\in\{m+1,m+2,\cdots,q\}$. Without loss of generality, let us assume that $a_{kl}=m+e+2-l$. The random variables ${\bf E}(t)$, ${\bf E}_{a}(t)$,  ${\bf E}_{c}(t)$, ${\bf F}(t)$, ${\bf F}_{a}(t)$ and ${\bf F}_{c}(t)$ are also defined as,
\begin{eqnarray}
{\bf E}(t)&=&\begin{bmatrix}{E}_{1}(t)&{E}_{2}(t)&\cdots&{E}_{M}(t)\end{bmatrix}^T\\
{\bf E}_{a}(t)&=&[{\bf E}(t)]_{0\rightarrow M-N_2}\\
{\bf E}_{c}(t)&=&[{\bf E}(t)]_{N_1\rightarrow M-N_1}\\
{\bf F}(t)&=&\begin{bmatrix}{F}_{1}(t)&{F}_{2}(t)&\cdots&{F}_{M}(t)\end{bmatrix}^T\\
{\bf F}_{a}(t)&=&[{\bf F}(t)]_{0\rightarrow M-N_2}\\
{\bf F}_{c}(t)&=&[{\bf F}(t)]_{N_1\rightarrow M-N_1}
\end{eqnarray}
Note that for any $m\in[M]$ and $t\in[n]$, $\bar{E}_{m}(t),\bar{F}_{m}(t)\in\{0, 1, \cdots, {\bar{P}}\}$, see \eqref{0top}. In the next step we bound $\mathbb{P}(\lambda^{[n]}\in \mathcal{S}_{\nu^{[n]}})$ from above. We wish to bound the probability that the images of these two codewords align, or in other words ${\bar{\bf U}}({\lambda}^n,W,\mathcal{G})={\bar{\bf U}}({\nu}^n,W,\mathcal{G})$. Thus, for any $k\in[K]$ and $t\in[n]$ we have,
\begin{eqnarray}
L_{k}^g(t)\left(\bar{\mathbf{E}}_{a}(t)\bigtriangledown(\bar{\mathbf{E}}_{c}(t))^{1}_{\frac{m}{q}}\right)&=&L_{k}^g(t)\left(\bar{\mathbf{F}}_{a}(t)\bigtriangledown(\bar{\mathbf{F}}_{c}(t))^{1}_{\frac{m}{q}}\right)\label{pr01}
\end{eqnarray}
From \eqref{pr01}, we have, 
\begin{eqnarray}
&&\sum_{j=1}^{M-N_2}\lfloor g_{kj}(t) E_{j}(t)\rfloor+\sum_{j=N_1+1}^{M}\lfloor g_{kj}(t) (E_{j}(t))^{1}_{\frac{m}{q}}\rfloor\nonumber\\
&=&\sum_{j=1}^{M-N_2}\lfloor g_{kj}(t) F_{j}(t)\rfloor+\sum_{j=N_1+1}^{M}\lfloor g_{kj}(t) (F_{j}(t))^{1}_{\frac{m}{q}}\rfloor\label{ret+1}\\
&\Rightarrow&|\sum_{j=1}^{M-N_2} g_{kj}(t) (F_{j}(t)-E_{j}(t))+\sum_{j=N_1+1}^{M} g_{kj}(t) ((F_{j}(t))^{1}_{\frac{m}{q}}-(E_{j}(t))^{1}_{\frac{m}{q}})|\le M\nonumber\\
\label{ret+2}
\end{eqnarray}
where  (\ref{ret+2}) follows from (\ref{ret+1}) as for any real number $x$, $|x-\lfloor x\rfloor|<1$. For any $j\in s_{M}\overset{\underset{\mathrm{def}}{}}{=}\{1,2,\cdots,(M-N_2),N_1+1,\cdots,M\}$, define $A_{j}(t)$ as,
\begin{eqnarray}
A_{j}(t)&=&\left\{\begin{matrix}F_{j}(t)-E_{j}(t) &1\le j\le(M-N_2) \\ 
(F_{j}(t))^{1}_{\frac{m}{q}}-(E_{j}(t))^{1}_{\frac{m}{q}}& N_1<j\le M
\end{matrix}\right.\label{pr6}
\end{eqnarray}
For any $k\in[K],t\in[n],l\in\{N\}$ and any fixed values of $g_{k1}(t),\cdots,g_{k(l-1)}(t),g_{k(l+1)}(t),\cdots,g_{kM}(t)$ the random variable $g_{kl}(t)A_l(t)$ must take values within an interval of length no more than $2{M}$. Therefore, for any $k\in[K],t\in[n],l\in s_M$ if $A_{l}(t)\neq 0$, then $g_{kl}(t)$ must take values in an interval of length no more than $\frac{2{M}}{|A_{l}(t)|}$, the probability of which is no more than $\frac{2{M}f_{\max}}{|A_{l}(t)|}$.  The probability of alignment is  bounded by
\begin{eqnarray}
\mathbb{P}(\lambda^{[n]}\in \mathcal{S}_{\nu^{[n]}})&\le&{\left\{\prod_{t=1,\max_{j\in s_M}|A_{j}(t)|\neq0}^n\frac{2Mf_{\max}}{\max_{j\in s_M}|A_{j}(t)|}\right\}}^K\\
&=&\prod_{k=1}^K~\prod_{t=1,\max_{j\in s_M}|A_{j}(t)|\neq0}^n\frac{2Nf_{\max}}{\max_{j\in s_M}|A_{j}(t)|}\label{pale}
\end{eqnarray}

\subsection{\bf Bounding the Average Size of Aligned Image Sets}

Let us assume $n=1$, $K=1$ as the generalization to arbitrary $n$ and $K$ follows similar to proof of Theorem $4$ in \cite{Arash_Jafar_sumset}. Without loss of generality let us drop the time index $(t)$. Thus, our goal is to prove that,
\begin{eqnarray}
&&\sum_{|\lambda_{11}-\nu_{11}|\in {\mathcal{Z}_{11}},\cdots,|\lambda_{1,e+1}-\nu_{1,e+1}|\in {\mathcal{Z}}_{1,e+1},\max_{j\in s_M}|A_{j}|\neq0}\frac{2Mf_{\max}}{\max_{j\in s_M}|A_{j}|}\nonumber\\
&&\le c_1+c_2{\log{\bar{P}}}\label{boundf++}
\end{eqnarray}
where $A_j$ is defined from $A_j(t)$ in \eqref{pr6} by dropping the time index $(t)$. Note that, \eqref{pr1} is concluded from \eqref{+jen} and \eqref{boundf++} as $\log{\log{\bar{P}}}=o~(\log{\bar{P}})$. $\mathcal{Z}_{1l}$ is  defined as the set $\{0\}\cup[M+1+\lfloor M\Delta{\bar{P}}^{1}_{\frac{e}{q}}\rfloor]$ for $l=1$ and the set $\{0\}\cup[ {\bar{P}}^{\frac{1}{q}}]$ for $l\in[e+1],l\neq1$. We define $\breve{\Delta}_{l}$ as,
\begin{eqnarray}
\breve{\Delta}_{j}&\triangleq&\left\{\begin{matrix}
(\bar{{E}}_{j})^{1}_{\frac{e}{q}}-(\bar{{F}}_{j})^{1}_{\frac{e}{q}} &1\le j\le M-N_2 \\ 
(\bar{{E}}_{j})^{1}_{\frac{m+e}{q}}-(\bar{{F}}_{j})^{1}_{\frac{m+e}{q}} &N_1<j\le M 
\end{matrix}\right.\label{pr7}
\end{eqnarray}
where \eqref{pr7} is derived from \eqref{+x1} and \eqref{+x2} dropping the time index $(t)$. Consider the following two cases of $\max_{j\in s_M}|\breve{\Delta}_{j}|\ge2$ and $\max_{j\in s_M}\breve|{\Delta}_{j}|\le1$.
\begin{enumerate}
\item{} $\max_{j\in s_M}|\breve{\Delta}_{j}|\ge2$ \\
Any number $X\in\mathcal{X}_\delta$ can be written as $(X)_{a}^{\delta}\bar{P}^a+X_a$ for any non-negative number $a$ less than $\delta$. Thus, when $1\le j\le(M-N_2)$, the term $|E_j-F_j|$  is bounded from below as,
\begin{eqnarray}
|A_j|&=&|E_j-F_j|\nonumber\\
&=&|\breve{\Delta}_{j}\bar{P}^{\frac{e}{q}}+(E_j)_{\frac{e}{q}}-(F_j)_{\frac{e}{q}}|\\
&\ge&|\breve{\Delta}_{j}\bar{P}^{\frac{e}{q}}|-|(E_j)_{\frac{e}{q}}-(F_j)_{\frac{e}{q}}|\\
&\ge&|\breve{\Delta}_{j}\bar{P}^{\frac{e}{q}}|-\bar{P}^{\frac{e}{q}}\\
&=&(|\breve{\Delta}_{j}|-1)\bar{P}^{\frac{e}{q}}\label{pr8}
\end{eqnarray}
and if $N_1< j\le M$, the term $|(E_j)^{1}_{\frac{m}{q}}-(F_j)^{1}_{\frac{m}{q}}|$  is bounded from below as,
\begin{eqnarray}
|A_j|&=&|(E_j)^{1}_{\frac{m}{q}}-(F_j)^{1}_{\frac{m}{q}}|\nonumber\\
&=&|\breve{\Delta}_{j}\bar{P}^{\frac{e}{q}}+(E_j)_{\frac{e}{q}}-(F_j)_{\frac{e}{q}}|\\
&\ge&|\breve{\Delta}_{j}\bar{P}^{\frac{e}{q}}|-|(E_j)_{\frac{e}{q}}-(F_j)_{\frac{e}{q}}|\\
&\ge&(|\breve{\Delta}_{j}|-1)\bar{P}^{\frac{e}{q}}\label{pr9}
\end{eqnarray}
Moreover, from \eqref{+x1}, \eqref{+x2}, and \eqref{pr7}, the term $|\lambda_{11}-\nu_{11}|$  is bounded from above as follows,
\begin{eqnarray}
|\lambda_{11}-\nu_{11}|&<&M+M\Delta\max_{j\in s_M}|\breve{\Delta}_{j}|\label{pr10}
\end{eqnarray}
The left side of  \eqref{boundf++} is bounded as,
\begin{eqnarray}
&&\sum_{|\lambda_{11}-\nu_{11}|\in {\mathcal{Z}_{11}},\cdots,|\lambda_{1,e+1}-\nu_{1,e+1}|\in {\mathcal{Z}}_{1,e+1},\max_{j\in s_M}|A_{j}|\neq0}\frac{2Mf_{\max}}{\max_{j\in s_M}|A_{j}|}\nonumber\\
&\le&\sum_{|\lambda_{11}-\nu_{11}|\in {\mathcal{Z}_{11}},\cdots,|\lambda_{1,e+1}-\nu_{1,e+1}|\in {\mathcal{Z}}_{1,e+1}}\frac{2Mf_{\max}}{\max_{j\in s_M}(|\breve{\Delta}_{j}|-1)\bar{P}^{\frac{e}{q}}}\label{bmn1}\\
&\le&\sum_{|\lambda_{11}-\nu_{11}|\in \{0\}\cup[M+M\Delta],|\lambda_{12}-\nu_{12}|\in {\mathcal{Z}}_{12},\cdots,|\lambda_{1,e+1}-\nu_{1,e+1}|\in {\mathcal{Z}}_{1,e+1}}\frac{2Mf_{\max}}{\bar{P}^{\frac{e}{q}}}\nonumber\\
&&+\sum_{|\lambda_{11}-\nu_{11}|\in {\mathcal{Z}}^{\star}_{11},|\lambda_{12}-\nu_{12}|\in {\mathcal{Z}}_{12},\cdots,|\lambda_{1,e+1}-\nu_{1,e+1}|\in {\mathcal{Z}}_{1,e+1}}\frac{2M^2\Delta f_{\max}}{(|\lambda_{11}-\nu_{11}|-M-M\Delta)\bar{P}^{\frac{e}{q}}}\nonumber\\
\label{bmn3}\\
&\le&\frac{2Mf_{\max}(1+M+M\Delta)\prod_{j=2}^{e+1}|{\mathcal{Z}}_{1j}|}{\bar{P}^{\frac{e}{q}}}\nonumber\\
&&+\frac{2M^2\Delta f_{\max}\prod_{j=2}^{e+1}|{\mathcal{Z}}_{1j}|}{\bar{P}^{\frac{e}{q}}}\sum_{|\lambda_{11}-\nu_{11}|\in {\mathcal{Z}}^{\star}_{11}}\frac{1}{(|\lambda_{11}-\nu_{11}|-M-M\Delta)}\label{bmn4}\\
&\le&2^{e+1}M^2f_{\max}(1+\Delta)(2+\log{(2+M+M\Delta\bar{P}^{1}_{\frac{e}{q}})})\label{bmn5}
\end{eqnarray}
where ${\mathcal{Z}}^{\star}_{11}$ is defined as the set of  $\bar{\mathcal{Z}}_{11}\cap[M+M\Delta]^c$. \eqref{bmn1} follows from \eqref{pr8} and \eqref{pr9}. Note that $\max_{j\in s_M}|A_{j}|$ is positive number as $\max_{j\in s_M}|\breve{\Delta}_{j}|\ge2$. \eqref{bmn3}  is obtained from \eqref{pr10} and \eqref{bmn4} is true as for any summation $\sum_{a\in S_a,b\in S_b}f(a,b)$ and the real-valued function $f(a,b)$ we have,
\begin{eqnarray}
\sum_{a\in S_a,b\in S_b}f(a,b)\le |S_a|\sum_{b\in S_b}\max_{a\in S_a}f(a,b)\le|S_a||S_b|\max_{a\in S_a,b\in S_b}f(a,b)\label{summation}
\end{eqnarray}
Finally, \eqref{bmn5} is concluded as the partial sum of harmonic series can be bounded above by logarithmic function i.e., $\sum_{i=1}^n\frac{1}{n}\le 1+\ln{n}$. 

\item{} $\max_{j\in s_M}|\breve{\Delta}_{j}|\le1$ \\

First of all note that from \eqref{pr10}, the term $|\lambda_{11}-\nu_{11}|$ only gets values from the set $\{0\}\cup[M+M\Delta]$. Let us  define $\hat{\Delta}_{lj}$ as,
\begin{eqnarray}
\hat{\Delta}_{lj}&\triangleq&\left\{\begin{matrix}
(\bar{{E}}_{j})^1_{\frac{e-l+1}{q}}-(\bar{{F}}_{j})^1_{\frac{e-l+1}{q}} &1\le j\le M-N_2 \\ 
(\bar{{E}}_{j})^1_{\frac{e+m-l+1}{q}}-(\bar{{F}}_{j})^1_{\frac{e+m-l+1}{q}} &N_1<j\le M 
\end{matrix}\right.\label{pr7+}
\end{eqnarray}
and define the number $l^\star$ as the smallest integer from the set $\{2,3,\cdots,e+1\}$ where
\begin{eqnarray}
\max_{j\in\{N_1+1,\cdots,M\}}|\hat{\Delta}_{l^{\star}j}|\ge2\label{conl}
\end{eqnarray}
Consider the following two cases.
\begin{enumerate}
\item{} $l^\star$ doesn't exist. \\

If there doesn't exist any $l\in\{2,\cdots,e+1\}$ and $j\in\{\max(N_1,M-N_2)+1,\cdots,M\}$ satisfying the condition \eqref{conl}, i.e., $\forall l,j, l\in\{2,\cdots,e+1\},  j\in\{\max(N_1,M-N_2)+1,\cdots,M\}, \hat{\Delta}_{lj}\in\{-1,0,1\}$, then we have,
\begin{eqnarray}
&&|L_{kl}^g(t)((\bar{\mathbf{E}}_{a}(t))^{1}_{\frac{e}{q}}\bigtriangledown\bar{\mathbf{E}}_{c}(t))-L_{kl}^g(t)((\bar{\mathbf{F}}_{a}(t))^{1}_{\frac{e}{q}}\bigtriangledown\bar{\mathbf{F}}_{c}(t))|\nonumber\\
&\le&2M\Delta\bar{P}^{\frac{e+m-l+1}{q}}\\
&\le&2M\Delta\bar{P}^{\frac{a_{1l}-1}{q}}
\end{eqnarray}
Thus, each of the variables $|\lambda_{1l}-\nu_{1l}|$ is bounded by $2M\Delta+1$ for any $l\in\{2,\cdots,e+1\}$. Therefore, \eqref{boundf++} is true as the summation in  \eqref{boundf++} is the summation of positive numbers less than $2Mf_{\max}$ over at most ${(2M\Delta+1)}^{e+1}$ numbers, see \eqref{summation}. 

\item{} $2\le l^\star\le e+1$.\\
Similar to \eqref{pr8} and \eqref{pr9}, when $1\le j\le(M-N_2)$, the term $|E_j-F_j|$  is bounded from below as,
\begin{eqnarray}
|A_j|&=&|E_j-F_j|\nonumber\\
&=&|\hat{\Delta}_{l^\star j}\bar{P}^{\frac{e-l^{\star}+1}{q}}+(E_{j})_{\frac{e-l^{\star}+1}{q}}-(F_{j})_{\frac{e-l^{\star}+1}{q}}|\\
&\ge&(|\hat{\Delta}_{l^\star j}|-1)\bar{P}^{\frac{e-l^{\star}+1}{q}}\label{jstar+}
\end{eqnarray}
and if $N_1< j\le M$, the term $|(E_j)^{1}_{\frac{m}{q}}-(F_j)^{1}_{\frac{m}{q}}|$  is bounded from below as,
\begin{eqnarray}
|A_j|&=&|(E_j)^{1}_{\frac{m}{q}}-(F_j)^{1}_{\frac{m}{q}}|\nonumber\\
&=&|\hat{\Delta}_{l^\star j}\bar{P}^{\frac{e-l^{\star}+1}{q}}+(E_{j})_{\frac{e-l^{\star}+1}{q}}-(F_{j})_{\frac{e-l^{\star}+1}{q}}|\\
&\ge&(|\hat{\Delta}_{l^\star j}|-1)\bar{P}^{\frac{e-l^{\star}+1}{q}}\label{jstar++}
\end{eqnarray}
The left side of  \eqref{boundf++} is bounded as,
\begin{eqnarray}
&&\sum_{|\lambda_{11}-\nu_{11}|\in \{0\}\cup[M+M\Delta],|\lambda_{12}-\nu_{12}|\in {\mathcal{Z}}_{12},\cdots,|\lambda_{1,e+1}-\nu_{1,e+1}|\in {\mathcal{Z}}_{1,e+1},\max_{j\in s_M}|A_{j}|\neq0}\nonumber\\
&&\frac{2Mf_{\max}}{\max_{j\in s_M}|A_{j}|}\nonumber\\
&\le&(M+1+M\Delta)(2M\Delta+1)^{l^\star-2}\frac{{(1+{\bar{P}}^{\frac{1}{q}})}^{e-l^{\star}+1}}{\bar{P}^{\frac{e-l^{\star}+1}{q}}}\nonumber\\
&&\times\sum_{|\lambda_{1l^\star}-\nu_{1l^\star}|\in {\mathcal{Z}}_{1l^\star}}\frac{2Mf_{\max}}{\max_{j\in s_M}(|\hat{\Delta}_{l^\star j}|-1)}\label{bmn1+}\\
&\le&(M+1+M\Delta)(2M\Delta+1)^{l^\star-2}\frac{{(1+{\bar{P}}^{\frac{1}{q}})}^{e-l^{\star}+1}}{\bar{P}^{\frac{e-l^{\star}+1}{q}}}\nonumber\\
&&\times\left(2M^2(1+\Delta)f_{\max}+\sum_{|\lambda_{1l^\star}-\nu_{1l^\star}|\in \bar{\mathcal{Z}}_{1l^\star}}\frac{2M^2\Delta f_{\max}}{|\lambda_{1l^\star}-\nu_{1l^\star}|-M-M\Delta}\right)\nonumber\\
\label{bmn2+}\\
&\le&2^{e+1}(M+1+M\Delta)(2M\Delta+1)^{l^\star-2}M^2f_{\max}(1+\Delta)(2+\log{\bar{P}^{\frac{1}{q}}})\label{bmn5+}\\
&\le&c_1+c_2\log{\bar{P}}
\end{eqnarray}
for some constants $c_1$ and $c_2$ not depending on $\bar{P}$. Note that, $\bar{\mathcal{Z}}_{1l^\star}$ is defined as the set ${\mathcal{Z}}_{1l^\star}\cap{[M+M\Delta]}^c$ and \eqref{bmn2+} follows similar to \eqref{bmn3}.

\end{enumerate}
\end{enumerate}

\section{Proof of Lemma \ref{lemmacef}}\label{lemmacefp}
As the functions $g_{rsl}(\beta_{1},\beta_{2})$ are bounded continuous functions on $s_{\beta}$ for any $r\in[Q_1],s\in[Q_2],l\in\{a,b,c,d\}$ for any positive real number $\epsilon$ there exists some $\delta$ that if $|\beta_{1}-\beta'_{1}|<\delta$,$|\beta_{2}-\beta'_{2}|<\delta$ then 
\begin{align}
|g_{rsl}(\beta_{1},\beta_{2})-g_{rsl}(\beta'_{1},\beta'_{2})|&\le \epsilon, \forall r\in[Q_1],s\in[Q_2],l\in\{a,b,c,d\}\label{fffd2}
\end{align}
Therefore, from Lemma \ref{lemmaMIMOx} we have,
\begin {eqnarray}
&|&H(L_{rs}^g({({\bf X}_{a}^n)}^{g_{rsa}(\beta_{1},\beta_{2})}\bigtriangledown{({\bf X}_{b}^n)}^{g_{rsb}(\beta_{1},\beta_{2})}\bigtriangledown{({\bf X}_{c}^n)}^{g_{rsc}(\beta_{1},\beta_{2})}),\forall s\in[Q_2])
\nonumber\\
&&-H(L_{rs}^g({({\bf X}_{a}^n)}^{g_{rsa}(\beta'_{1},\beta'_{2})}\bigtriangledown{({\bf X}_{b}^n)}^{g_{rsb}(\beta'_{1},\beta'_{2})}\bigtriangledown{({\bf X}_{c}^n)}^{g_{rsc}(\beta'_{1},\beta'_{2})}),\forall s\in[Q_2])
|\nonumber\\
\le&& nMs\epsilon\log{\bar{P}}+n~o~(\log{\bar{P}}) \label{rmt1}
\end {eqnarray}
Now as the sum and multiplications of bounded continuous functions are continuous, $S(\beta_{1},\beta_{2})$ is continuous function on $s_{\beta}$.


\begin{thebibliography}{10}
\providecommand{\url}[1]{#1}
\csname url@samestyle\endcsname
\providecommand{\newblock}{\relax}
\providecommand{\bibinfo}[2]{#2}
\providecommand{\BIBentrySTDinterwordspacing}{\spaceskip=0pt\relax}
\providecommand{\BIBentryALTinterwordstretchfactor}{4}
\providecommand{\BIBentryALTinterwordspacing}{\spaceskip=\fontdimen2\font plus
\BIBentryALTinterwordstretchfactor\fontdimen3\font minus
  \fontdimen4\font\relax}
\providecommand{\BIBforeignlanguage}[2]{{%
\expandafter\ifx\csname l@#1\endcsname\relax
\typeout{** WARNING: IEEEtran.bst: No hyphenation pattern has been}%
\typeout{** loaded for the language `#1'. Using the pattern for}%
\typeout{** the default language instead.}%
\else
\language=\csname l@#1\endcsname
\fi
#2}}
\providecommand{\BIBdecl}{\relax}
\BIBdecl

\bibitem{Cadambe_Jafar_int}
V.~Cadambe and S.~Jafar, ``{Interference Alignment and the Degrees of Freedom
  of the $K$ user Interference Channel},'' \emph{IEEE Transactions on
  Information Theory}, vol.~54, no.~8, pp. 3425--3441, Aug. 2008.

\bibitem{Caire_Shamai}
G.~Caire and S.~Shamai, ``On the achievable throughput of a multiantenna
  {G}aussian broadcast channel,'' \emph{IEEE Transactions on Information Theory}, vol.~49,
  no.~7, pp. 1691--1706, July 2003.

\bibitem{Lapidoth_Shamai_Wigger_BC}
A.~Lapidoth, S.~Shamai, and M.~Wigger, ``On the capacity of fading {MIMO}
  broadcast channels with imperfect transmitter side-information,'' in
  \emph{Proceedings of 43rd Annual Allerton Conference on Communications,
  Control and Computing}, Sep. 28-30, 2005.

\bibitem{Arash_Jafar_PN}
A.~G. Davoodi and S.~A. Jafar, ``Aligned image sets under channel uncertainty:
  Settling conjectures on the collapse of degrees of freedom under finite
  precision {CSIT},'' \emph{IEEE Transactions on Information Theory}, vol.~62,
  no.~10, pp. 5603--5618, 2016.




\bibitem{Arash_Jafar_sumset}
------, ``Sum-set inequalities from aligned image sets: Instruments for robust
  {GDoF} bounds,'' \emph{arXiv preprint arXiv:1703.01168}, 2017.

\bibitem{Hao_Rassouli_Clerckx}
C.~Hao, B.~Rassouli, and B.~Clerckx, "Achievable {DoF} regions of {MIMO} networks with imperfect {CSIT}," \emph{IEEE Transactions on Information Theory}, vol.~63,
  no.~10, pp. 6587--6606, 2017.


\bibitem{Weingarten_Steinberg_Shamai}
H.~Weingarten, Y.~Steinberg, and S.~Shamai, ``The capacity region of the
  {G}aussian {MIMO} broadcast channel,'' \emph{IEEE Transactions on Information Theory}, vol.~52, pp. 3936--3964, Sep. 2006.

\bibitem{Jafar_scalar}
S.~Jafar and A.~Goldsmith, ``Isotropic fading vector broadcast channels: the
  scalar upperbound and loss in degrees of freedom,'' \emph{IEEE Transactions on Information Theory}, vol.~51, no.~3, pp. 848--857, March 2005.

\bibitem{Huang_Jafar_Shamai_Vishwanath}
C.~Huang, S.~A. Jafar, S.~Shamai, and S.~Vishwanath, ``{On Degrees of Freedom
  Region of MIMO Networks without Channel State Information at Transmitters},''
  \emph{IEEE Transactions on Information Theory}, no.~2, pp. 849--857, Feb.
  2012.

\bibitem{Vaze_Varanasi_MIMOBC}
C.~S. Vaze and M.~K. Varanasi, ``The degree-of-freedom regions of {MIMO}
  broadcast, interference, and cognitive radio channels with no {CSIT},''
  \emph{IEEE Transactions on Information Theory}, vol.~58, no.~8, pp.
  5354--5374, 2012.

\bibitem{Arash_Jafar_TC}
A.~G. Davoodi and S.~A. Jafar, ``{Transmitter Cooperation under Finite
  Precision CSIT:A GDoF Perspective},'' \emph{IEEE Transactions on Information
  Theory}, 2016.

\bibitem{Arash_Bofeng_Jafar_BC}
A.~G. Davoodi, and S.~A. Jafar, ``{GDoF} of the {MISO BC}: Bridging
  the gap between finite precision and perfect {CSIT},''  \emph{IEEE Transactions on Information
  Theory}, pp. 1297--1301, 2016.


\bibitem{Yao_Varanasi_Hybrid}
W.~Yao and M.~K. Varanasi, ``Degrees of freedom of the two-user {MIMO}
  broadcast channel with private and common messages under hybrid {CSIT}
  models,'' \emph{IEEE Transactions on Information Theory}, May 2017.
%
%\bibitem{Bofeng_Arash_Jafar_ArXiv}
%B.~Yuan, A.~G. Davoodi, and S.~A. Jafar, ``{DoF} region of the {MIMO}
%  interference channel with partial {CSIT},'' \emph{Available on ArXiv}, April.
%  2019.
%
%\bibitem{Arash_Jafar_MIMOsym_ArXiv}
%A.~G. Davoodi and S.~A. Jafar, ``Aligned image sets and the {GDoF} of symmetric
%  {MIMO} interference channel with partial {CSIT},'' \emph{IEEE Global Communications Conference}, pp. 1--6, Dec. 2017.
%
%
%
%\bibitem{Arash_Jafar_IC}
%------, ``Generalized {D}egrees of {F}reedom of the {S}ymmetric {$K$}-{U}ser
%  {I}nterference {C}hannel under {F}inite {P}recision {CSIT},'' \emph{IEEE Transactions on Information Theory}, vol.~63,
%  no.~10, pp. 6561--6572, 2017.



\bibitem{Arash_Jafar_KMIMOIC}
------, ``$k$-user symmetric {$M\times N$} {MIMO} interference channel under
  finite precision {CSIT}: A {GDoF} perspective,'' \emph{IEEE Transactions on Information Theory}, July 2018.



\end{thebibliography}
\end{document}